\documentclass[twocolumn,showpacs,preprintnumbers,amsmath,amssymb,superscriptaddress]{revtex4-1}

\usepackage[utf8]{inputenc}
\usepackage[T1]{fontenc}
 
\usepackage{mathrsfs}
\usepackage{amsmath}
\usepackage{txfonts}
\usepackage{mathtools}
\usepackage{braket}
\usepackage{tensor}
\usepackage{xcolor}
\usepackage[abbreviations]{siunitx}
\usepackage{graphicx}
\usepackage{booktabs}

\usepackage[utf8]{inputenc}
\usepackage[T1]{fontenc}
\usepackage{CJKutf8}

\usepackage{float}

\usepackage{braket}

\usepackage{hyperref}
\hypersetup{
	colorlinks=true, 
	linktoc=all,     
	linkcolor=blue,  
}
\usepackage[normalem]{ulem}

\newcommand{\eMax}{\ensuremath{e_{\text{Max}}}}

\usepackage{hyperref}
\usepackage{cleveref}

\newcommand{\nord}[1]{\ensuremath{:\!#1:}}

\newcommand{\beq}{\begin{equation}}
\newcommand{\eeq}{\end{equation}}
\newcommand{\beqn}{\begin{eqnarray}}
\newcommand{\eeqn}{\end{eqnarray}}
\newcommand{\bsub}{\begin{subequations}}
\newcommand{\esub}{\end{subequations}}
\newcommand{\bpm}{\begin{pmatrix}}
\newcommand{\epm}{\end{pmatrix}}

\DeclareSIUnit{\fm}{\femto\meter}
\DeclareSIUnit{\MeVc}{\MeV\per\text{\ensuremath{c}}}

\allowdisplaybreaks[3]

\begin{document} 

\begin{CJK*}{UTF8}{gbsn}

\title{{\em Ab initio} mapping of the boundary of the $N=20$ island of inversion}

  \author{E. F. Zhou (周恩付)}
  \affiliation{School of Physics and Astronomy, Sun Yat-sen University, Zhuhai 519082, P.R. China}  
  \affiliation{Guangdong Provincial Key Laboratory of Quantum Metrology and Sensing, Sun Yat-Sen University, Zhuhai 519082, China } 

   \author{C. R. Ding (丁晨蓉)}    
  \affiliation{School of Physics and Astronomy, Sun Yat-sen University, Zhuhai 519082, P.R. China}  
  \affiliation{Guangdong Provincial Key Laboratory of Quantum Metrology and Sensing, Sun Yat-Sen University, Zhuhai 519082, China }

   \author{Q. Y. Luo (罗青杨)}    
  \affiliation{School of Physics and Astronomy, Sun Yat-sen University, Zhuhai 519082, P.R. China}  
  \affiliation{Guangdong Provincial Key Laboratory of Quantum Metrology and Sensing, Sun Yat-Sen University, Zhuhai 519082, China } 
  
  \author{J. M. Yao (尧江明)}     
  \email{Contact author: yaojm8@sysu.edu.cn}
  \affiliation{School of Physics and Astronomy, Sun Yat-sen University, Zhuhai 519082, P.R. China}    
  \affiliation{Guangdong Provincial Key Laboratory of Quantum Metrology and Sensing, Sun Yat-Sen University, Zhuhai 519082, China } 
 
\author{H.~Hergert}     
\affiliation{Facility for Rare Isotope Beams, Michigan State University, East Lansing, Michigan 48824-1321, USA.}
\affiliation{Department of Physics \& Astronomy, Michigan State University, East Lansing, Michigan 48824-1321, USA.}

\date{\today}

\begin{abstract}

 Starting from a chiral two- plus three-nucleon interaction, we perform a systematic study of the low-lying states of neutron-rich nuclei around $N=20$ using the in-medium generator coordinate method (IM-GCM), which combines the multi-reference in-medium similarity renormalization group (MR-IMSRG) with the quantum-number projected generator coordinate method (PGCM) defined in a full single-particle space. The main features of the energy spectra and electromagnetic properties of low-lying states in both even–even and odd-mass nuclei of this mass region are reasonably reproduced. The boundary of the $N=20$ island of inversion (IOI) is investigated, and the results indicate that $\nuclide[30]{Ne}$, $\nuclide[29,31,33]{Na}$, $\nuclide[31,32,33,34]{Mg}$, and $\nuclide[35]{Al}$ lie within the IOI, whereas $\nuclide[29]{F}$, $\nuclide[29]{Ne}$, $\nuclide[30]{Mg}$, $\nuclide[31,33]{Al}$, $\nuclide[34,35]{Si}$, and $\nuclide[35]{P}$ fall outside it. 
\end{abstract}
 
\maketitle

\section{Introduction}
\label{sec:introduction}

With the advent of radioactive ion beam facilities, numerous exotic nuclei far from the $\beta$-stability line have become accessible to nuclear experiments. Among these, neutron-rich nuclei with neutron numbers around $N=20$ have garnered significant research interest since the discovery of anomalies in their binding energies~\cite{Thibault:1975} and the unusually low excitation energy of the $2^+_1$ state in \nuclide[32]{Mg}~\cite{Detraz:1979}. Such a phenomenon deviates significantly from the patterns observed in magic nuclei, which are characterized by high excitation energies of $2^+_1$ states. The presence of the low-lying $2^+_1$ state in \nuclide[32]{Mg}, together with the later observation of strong electric quadrupole transition strengths~\cite{Motobayashi:1995Mg32}, indicates the onset of large quadrupole deformation. Similar phenomena were subsequently identified in neighboring isotopes with $Z=10-12$ and $N=20$~\cite{Pritychenko:1999Mg32,Yoneda:2001Mg34,Gade:2007Mg3436,Heyde1991Mg3230,Neyens:2005Mg31,Doornenbal:2009Ne32,Michimasa:2014Ne30,Doornenbal:2016Ne30,Pritychenko:2001Na31}. These nuclei constitute the first identified “island of inversion” (IOI) on the nuclear chart~\cite{Warburton:1990island}. Determining the boundary of this IOI has been the objective of numerous theoretical and experimental studies~\cite{Otsuka:2020}. 
    
The onset of large deformation in neutron-rich sodium isotopes around  $N=20$ was first explained using the deformation-constrained Skyrme-Hartree-Fock (SHF) approach. This approach highlighted the critical role of filling negative-parity $f_{7/2}$ levels instead of $d_{3/2}$ levels, resulting from level crossings driven by quadrupole deformation~\cite{Campi:1975HFM1}. This interpretation is supported by beyond-mean-field studies~\cite{Guzman:2000PRC,Rodriguez-Guzman:2002NPA,Niksic:2006kv,Yao:2011_Mg,Borrajo:2017PLB} based on modern energy-density functionals (EDFs) and shell model studies~\cite{Wildenthal:1983SM2,Poves:1987SM3,Warburton:1990island,Fukunishi:1992SM1}, which demonstrate that multi-particle multi-hole (mp-mh) excitations across the $N=20$ shell gap are essential for reproducing the pronounced quadrupole collectivity in the ground states of IOI nuclei. These studies have inspired extensive research on the evolution
of shell structure, collectivity, and the degradation of the traditional magic numbers in neutron-rich nuclei~\cite{Utsuno:2004SM4,Caurier:1998SM5,Heyde:2011,Caurier:2013sm6,Gade:2016,Otsuka:2020}. 

 Moreover, nuclei in the IOI around  $N=20$ exhibit pronounced shape coexistence~\cite{Wimmer:2010Mg32prl}, characterized by multiple low-lying states associated with distinct intrinsic shapes~\cite{Heyde:2011RMP}. In some nuclei within this mass region, strong shape mixing may occur, resulting in nuclear states that are quantum superpositions of configurations with distinct intrinsic shapes rather than nearly pure shape configurations. These two scenarios cannot be distinguished from the mean-field energy surface alone. A more detailed analysis requires decomposing the beyond-mean-field wave functions into contributions associated with different intrinsic shapes. Accurately capturing this shape-mixing effect is  challenging for nuclear {\em ab initio} methods.

The valence-space in-medium similarity renormalization group (VS-IMSRG)~\cite{Tsukiyama:2012PRC,Bogner:2014PRL,Stroberg:2019ARNPS} starting from chiral nuclear forces~\cite{Weinberg:1991,Epelbaum:2009RMP,Machleidt:2011PR} has been employed to describe the low-lying states of even-even isotopes with $Z=10-14$ and $N\simeq 20$~\cite{Miyagi:2020PRC}.  It was found that neutron excitations from the  $sd$ shell to the 
$fp$ shell play a prominent role in both the ground and excited states of these nuclei. Nevertheless, the excitation energies of the $2^+_1$ states and the $B(E2: 0^+_1\to 2^+_1)$ values are systematically overestimated and underestimated, respectively. This deficiency is expected to be addressed either by including the contributions of higher-body operators~\cite{Stroberg:2024} or by using a correlated reference state~\cite{Hergert:2020}. Consequently, the solution obtained within the deformed IMSRG framework does not conserve angular momentum, which necessitates the implementation of angular-momentum projection (AMP) to restore this symmetry. At present, no established protocol prescribes how AMP should be incorporated into the deformed IMSRG framework. Nevertheless, one can envisage employing the evolved operators together with explicitly symmetry-restored wave functions. AMP for beyond-mean-field wave functions has already been successfully implemented in coupled-cluster (CC) theory and subsequently applied to describe the low-lying states of IOI nuclei with great success~\cite{Hagen:2022PRC,Sun:2024_even,Sun:2025lk}.  Nevertheless, accurately predicting the energy ordering of competing states with different shapes remains a challenge~\cite{Sun:2024_even,Sun:2025lk}. 

The IMSRG and CC variants aim to incorporate the description of collectivity into frameworks that were aimed at capturing dynamical correlation, i.e., excitations involving only a few nucleons. Conversely, there have been recent attempts to add dynamical correlation on top of methods that are focused on collectivity and shape phenomena, in particular the quantum-number projected generator coordinate method (PGCM). The novel PGCM-PT(2), a second-order multi-reference perturbation theory based on PGCM wave functions, was recently developed and applied to study low-lying excited states in even-even neon isotopes~\cite{Frosini:2022_1,Frosini:2022_2,Frosini:2022_3}. While the low-lying states of most neon isotopes are well described~\cite{Frosini:2022_2}, accurately predicting the energy ordering of weakly and strongly deformed bands in \nuclide[30]{Ne} still remains a challenge for PGCM-PT(2). Last but not least, a recent valence-space PGCM study~\cite{Cao:2025} employing operators evolved within the VS-IMSRG framework~\cite{Stroberg:2017PRL} demonstrated that the low-lying states of $^{32}$Mg can be reproduced rather well by enlarging the valence space from the $sd$ shell to the $sdfp$ shell. 

In recent years, we have developed the in-medium generator coordinate method (IM-GCM)~\cite{Yao:2018PRC}, which combines the multi-reference in-medium similarity renormalization group (MR-IMSRG)~\cite{Hergert:2016PR} with the PGCM formulated in the full single-particle space. This method avoids the aforementioned issue of the deformed IMSRG by working with correlated reference states that have been explicit projected onto good quantum numbers, so that the evolving Hamiltonian remains a scalar operator and conserves the symmetries at all stages of our workflow. The IM-GCM has already been successfully applied to study the low-lying states of axially deformed \nuclide[48]{Ti}~\cite{Yao:2020PRL}, and triaxially deformed \nuclide[76]{Ge} and \nuclide[76]{Se}~\cite{Belley:2024PRL}. Recently, we extended this method to investigate shape coexistence and the weakening of the  $N=20$ magic number in both even–even and odd-mass neutron-rich nuclei starting from a chiral two- plus three-nucleon (NN+3N) interaction~\cite{Zhou:2024_short}. The approach reproduces the coexistence of weakly and strongly deformed states at comparable energies, although the spectra remain somewhat stretched. This study shows that the deformed ground state in IOI nuclei emerges through the systematic IMSRG evolution of the initial Hamiltonian into an effective one. As discussed in Refs.~\cite{Zhou:2024_short,Ding:2025}, increasing the IMSRG flow parameter embeds more residual correlations into the effective Hamiltonian, thereby improving the validity of mean-field–based approaches and the description of nuclear low-lying states. In the present work, we provide a  detailed description of this framework and applications to the nuclei of this region, with the goal of exploring the boundary of the IOI around $N=20$.  

The article is arranged as follows. In Sec.\ref{sec:framework}, we present the technical details of the methods employed in the present study, including the MR-IMSRG and PGCM. The expressions used to evaluate the transition strengths of tensor operators are also introduced in detail. In Sec.\ref{sec:Results and discussion}, we present results for the low-lying states of even-even and odd-mass nuclei around $N=20$, including magnetic dipole moments, electric quadrupole moments, and electromagnetic transitions. The main conclusions of this study are summarized in Sec.~\ref{sec:summary}.

\section{The theoretical framework}
\label{sec:framework} 
In this section, we provide a detailed description of the IM-GCM framework developed in Refs.~\cite{Yao:2018PRC,Yao:2020PRL}, with particular emphasis on its recent extension to odd-mass nuclei~\cite{Lin:2024arl,Zhou:2024_short}. An illustration of the procedure is shown in Fig.~\ref{fig:IMGCM}.

\subsection{The Hamiltonian}
\begin{figure}[tb]
 \centering
\includegraphics[width=8.6cm]{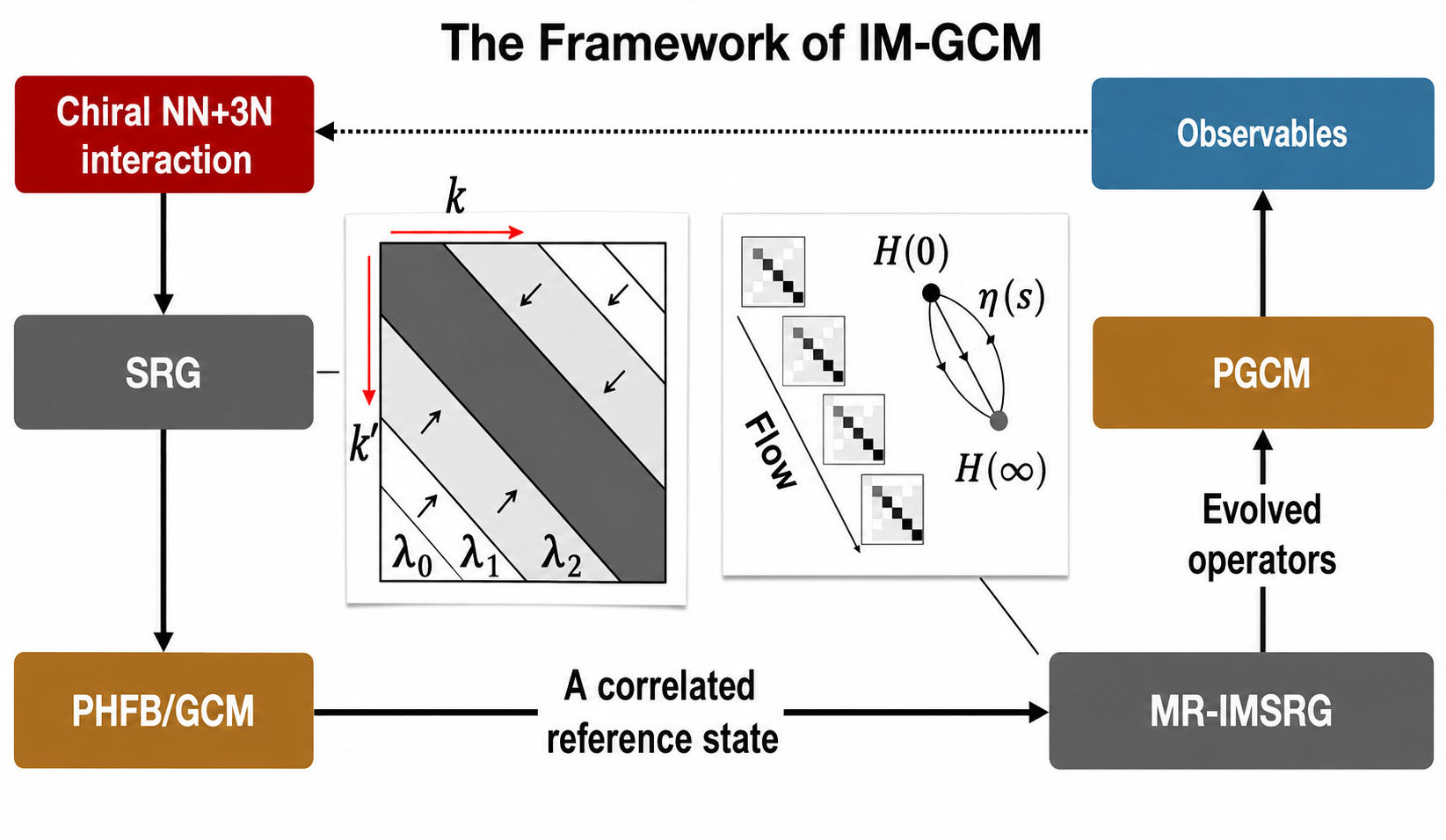} 
\caption{ (Color online) Flowchart of the IM-GCM framework \cite{Yao:2020PRL}. The central insets are schematic cartoons illustrating the evolution of the nuclear interaction in momentum space as a function of the SRG momentum-resolution scale (left) and the evolution of Hamiltonian matrix elements in configuration space with the MR-IMSRG flow parameter (right). }
\label{fig:IMGCM}
 \end{figure}

 In the IM-GCM,  we start from  an intrinsic nuclear $A$-body Hamiltonian containing both $NN$ and $3N$ interactions, 
\begin{equation}
\label{Eq:H0}
\hat H_0 = \left(1-\frac{1}{A}\right) T^{[1]}+\frac{1}{A} T^{[2]}
+   V^{[2]} +  W^{[3]},
\end{equation}
where the kinetic term is composed of one- and two-body pieces,
\beq 
T^{[1]} = \sum^A_{i=1} \frac{\mathbf{p}_i^2}{2 m_N}, \quad T^{[2]} =-\sum_{i<j} \frac{\mathbf{p}_i \cdot \mathbf{p}_j}{m_N},
\eeq 
$m_N$ is the mass of the nucleon and $\mathbf{p}_i$ the momentum of the $i$-th nucleon. This Hamiltonian is then evolved with the (free-space) SRG technique \cite{Bogner:2010PPNP}, which decouples the off-diagonal elements between the low- and high-momentum states,
\begin{equation}  
\label{Eq:FlowH}
  \frac{d\hat H_{\lambda}}{d\lambda}=-\frac{4}{\lambda^5}[[\hat T_{\rm rel}, \hat H_\lambda], \hat H_\lambda].
\end{equation}
Here, 
$\hat{T}_{\rm rel}$ is the relative kinetic energy of two nucleons. In this work, we employ the so-called {\em magic} chiral interaction EM1.8/2.0~\cite{Hebeler:2011PRC}, in which the $NN$ part \cite{Entem:2003PRC} is SRG-evolved to the resolution scale $\lambda = 1.8$ fm$^{-1}$, see an illustration in the central inset (left) of  Fig.\ref{fig:IMGCM}, while the low-energy constants (LECs) of the $3N$ interaction, $c_D$ and $c_E$, are directly fit to the \nuclide[3]{H} binding energy and the \nuclide[4]{He} matter radius while assuming a cutoff $\Lambda=2.0\,\mathrm{fm}^{-1}$.

\subsection{The MR-IMSRG flow} 

The basic idea of the MR-IMSRG is to use a flow equation to gradually decouple a
preselected \emph{correlated} reference state $\ket{\Psi_0}$ from all other states. As illustrated in Fig.\ref{fig:IMGCM}, the reference state is usually chosen as either a particle-number projected (PNP) HFB state~\cite{Hergert:2013PRL}, a deformed HFB state with both PNP and AMP~\cite{Yao:2018PRC}, a PGCM state \cite{Zhou:2024_short} or  other correlated states \cite{Gebrerufael:2017PRL}.  According to our previous studies~\cite{Belley:2023,Zhou:2024_short}, an ensemble of multiple symmetry-restored HFB states, determined through variation after particle-number projection (VAPNP) works better than  a PGCM state for the studies of nuclei with shape coexistence. Therefore, the ensemble reference state is employed subsequently. Specifically, following Ref.~\cite{Stroberg:2017PRL}, the ensemble reference state is defined by the density operator
$\hat{\rho}=c_1\ket{\Phi^{NZJ}_1}\bra{\Phi^{NZJ}_1}+c_2\ket{\Phi^{NZJ}_2}\bra{\Phi^{NZJ}_2}$,
where $\ket{\Phi^{JNZ}_{1,2}}$ denote the symmetry-restored $J^\pi=0^+$ states obtained by projecting the quasiparticle vacua associated with the spherical and energy-minimum configurations onto good angular momentum and particle number. In the present calculation, we take $c_1=c_2=0.5$ so that both configurations can be decoupled from excited configurations by the MR-IMSRG flow. 

The IMSRG-evolved Hamiltonian $\hat{H}(s)$ as a function of the continuous flow parameter $s$ is given by
\beq
\label{Hs}
 \hat H(s)=\hat U(s)\hat H_0\hat U^\dagger(s) \,.
\eeq 
where $\hat H_0$ is the initial Hamiltonian, i.e., the EM1.8/2.0, and $\hat U(0)=1$. The continuous unitary transformation $\hat U(s)$ drives $\hat H_0$ to
a specific form, e.g., by eliminating certain matrix elements or minimizing its
expectation value.  Taking the derivative $d/ds$ of both sides of Eq.\
\eqref{Hs} yields the flow equation 
\beqn
\label{flow-H}
\dfrac{d\hat H(s)}{ds} = [\hat \eta(s), \hat H(s)] \,,
\eeqn
where we have introduced the anti-Hermitian generator of the transformation,
\beq
\hat \eta(s)\equiv\dfrac{d\hat U(s)}{ds}\hat U^\dagger (s).
\eeq
 By introducing strings of creation and annihilation operators as 
\beq 
A^{pqr\ldots}_{stu\ldots} = a^\dagger_pa^\dagger_qa^\dagger_r\ldots a_u a_t a_s,
\eeq 
we rewrite the Hamiltonian  $\hat H_0$ in second quantization, 
\beqn
\hat H_0 
\label{eq:Hamiltonian_sq}
&=&  \sum_{pq} \bar{t}^p_{q} A^p_q  + \dfrac{1}{4}\sum_{pqrs}  \bar{v}^{pq}_{rs} 
A^{pq}_{rs}
+\dfrac{1}{36}\sum_{pqrstu}\bar{w}^{pqr}_{stu} A^{pqr}_{stu},
\eeqn 
where the symbols hatted with bars are introduced for the sake of brevity,
\beq
\bar{t} \equiv \left(1-\frac{1}{A}\right)T^{[1]},\quad 
\bar{v} \equiv \frac{1}{A}T^{[2]}  + V^{[2]},\quad
 \bar{w}\equiv W^{[3]}.  
\eeq 
The above Hamiltonian is normal-ordered with respect to the previously chosen reference state $\ket{\Psi_0}$,
\beqn
\label{eq:normal-ordered-H}
\hat H_0 &=& E_0 + \sum_{pq}f^{p}_{q} \nord{A^p_q}
+ \dfrac{1}{4}\sum_{pqrs} \Gamma^{pq}_{rs} \nord{A^{pq}_{rs}}\nonumber\\
&& + \dfrac{1}{36}\sum_{pqrstu} \bar{w}^{pqr}_{stu}  \nord{A^{pqr}_{stu}}.
\eeqn 
By definition, the expectation values of normal-ordered operators, indicated by
$\nord{A^{p\ldots}_{q\ldots}}$, with respect to the reference state are zero. For  the ensemble reference state, it corresponds to ${\rm Tr[\rho \nord{A^{p\ldots}_{q\ldots}}]}=0$ \cite{Stroberg:2017PRL}.
Thus, the normal-ordered zero-body term corresponds to the reference-state
energy $E_0$, which is given by
\beqn
\label{H:0b}
 E_0
&=&\sum_{pq} \bar{t}^p_{q}\rho^p_q
+\dfrac{1}{4}\sum_{pqrs}  \bar{v}^{pq}_{rs} \rho^{pq}_{rs}  +\dfrac{1}{36}\sum_{pqrstu} \bar{w}^{pqr}_{stu}\rho^{pqr}_{stu} \,.
\eeqn
The normal-ordered one-body and two-body terms are
\beqn
\label{H:1b}
 f^{p}_{q}&=& \bar{t}^{p}_{q}  +  \sum_{rs} \bar{v}^{pr}_{qs} \rho^r_s
+\dfrac{1}{4}\sum_{rstu} \bar{w}^{prs}_{qtu}\rho^{rs}_{tu} \,,\\
\label{H:2b}
 \Gamma^{pq}_{rs}  &=& \bar{v}^{pq}_{rs}+ \sum_{tu} \bar{w}^{pqt}_{rstu}\rho^{t}_{u}\,.
\eeqn
The density matrices are defined as
\beqn 
\rho^{pqr\cdots}_{stu\cdots} &=& \bra{ \Psi_0} A^{pqr\cdots}_{stu\cdots}\ket{\Psi_0},
\eeqn
and the corresponding
\emph{irreducible} density matrices  
\bsub
\label{cumulants}
\beqn
\lambda^p_q &=&  \rho^p_q\,, \\
\lambda^{pq}_{rs} &=& \rho^{pq}_{rs}  - {\cal A}(\lambda^p_r\lambda^q_s)
 = \rho^{pq}_{rs}  - \lambda^p_r\lambda^q_s +  \lambda^p_s\lambda^q_r\,,\\
\lambda^{pqr}_{stu} &=& \rho^{pqr}_{stu} - {\cal
A}(\lambda^p_s\lambda^{qr}_{tu}+\lambda^p_s\lambda^{q}_{t}\lambda^{r}_{u}) \,,
\eeqn
\esub
where the antisymmetrization operator ${\cal A}$ generates all possible \emph{unique} permutations of upper indices and lower indices. The irreducible density matrices encode the correlations in the reference state. For independent particle states, the $n$-body ($n\geq2$) irreducible densities vanish.

As we evolve $\hat{H}(s)$ towards $s\to\infty$, the expectation value of the evolved Hamiltonian  with respect to the chosen reference state $\ket{\Psi_0}$ converges to the exact ground-state energy. Since an ensemble reference state is used in this work, additional post-processing, such as a PGCM diagonalization, is employed to extract the energies of nuclear low-lying states (see below). Moreover, in practical implementations, all operators are truncated to include only normal-ordered two-body terms—a scheme known as the NO2B approximation. This approximation has demonstrated success for both closed-shell nuclei~\cite{Tsukiyama:2011PRL} and open-shell systems~\cite{Hergert:2016PR}, provided that collective multi-particle–multi-hole correlations are encoded in the reference state~\cite{Hergert:2020}. 

The (MR-)IMSRG can be reformulated by using the so-called Magnus expansion \cite{Magnus:1954}. It stipulates that if certain convergence conditions are satisfied, the unitary transformation $\hat U(s)$ can be written as a proper exponential of an anti-Hermitian operator $\hat \Omega(s)$\cite{Morris:2015}
\beq
\label{eq:transformation}
\hat U(s)\equiv e^{\hat \Omega(s)}\,.
\eeq
Instead of solving the flow equation for $\hat{H}(s)$, one can implement the (MR-)IMSRG evolution as a flow equation for $\hat
\Omega$: 
\beq
\label{flow_omega}
\dfrac{d\hat \Omega(s)}{ds}=\sum^\infty_{n=0} \dfrac{B_n}{n!} [\hat \Omega(s),\hat \eta(s)]^{(n)} \,, 
\eeq
where we define nested commutators as
\bsub\beqn
  \left[\hat \Omega(s),\hat \eta(s)\right]^{(0)} &=& \hat \eta(s)\,,\\ 
  \left[\hat \Omega(s),\hat \eta(s)\right]^{(n)} &=& \left[\hat \Omega(s), \left[\hat \Omega(s),\hat \eta(s)\right]^{(n-1)}\right]\,,
\eeqn\esub
and $B_{n=0,1,2,3,\cdots}$ are the Bernoulli numbers $\{ 1, -1/2, 1/6, 0, \cdots
\}$~\cite{Bernoulli}.   
The advantage of the Magnus formulation is that $\Omega(s)$ can now be used to construct arbitrary observables at any given time, while the direct integration approach requires all relevant observables to be co-evolved with the Hamiltonian \cite{Morris:2015,Hergert:2016PR}.

In the present applications, we use the Brillouin generator for $\eta(s)$ \cite{Hergert:2017PS}, 
\bsub
\label{eq:Brillouin}
\beqn
\eta^p_q &\equiv& \bra{\Psi_0} \left[ \hat H, \nord{A^p_q} \right] \ket{\Psi_0}\,,\\
\eta^{pq}_{rs} &\equiv& \bra{\Psi_0} \left[ \hat H, \nord{A^{pq}_{rs}} \right] \ket{\Psi_0},
\eeqn
\esub
which is essentially the gradient of the energy under a general unitary transformation.

In the present work, all operators $\hat{O}$ of interest are evolved consistently with the Hamiltonian using the unitary transformation $\hat{U}(s)$ in Eq.~(\ref{eq:transformation}),
\begin{equation}
\hat{O}(s) = \hat{U}(s) \hat{O}(0) \hat{U}^\dagger(s) = 
e^{\hat{\Omega}(s)}\hat{O}(0)e^{-\hat{\Omega}(s)}\,,
\end{equation}
and evaluated via the Baker-Campbell-Hausdorff (BCH) expansion. The explicit expressions for a general tensor operator are provided in Appendix~\ref{app:tensor_operator}.

\subsection{The PGCM for low-lying states} 

A direct application of the IMSRG framework to target nuclear excited states presents significant challenges \cite{Hergert:2017bc}. As a result, it is usually combined with other many-body methods for nuclear excitations \cite{Tsukiyama:2012PRC,Stroberg:2019ARNPS,Gebrerufael:2017PRL,Parzuchowski:2017yq,Yao:2018PRC}. In this work, the wave functions of nuclear low-lying states are derived through a PGCM calculation based on the evolved Hamiltonian $\hat{H}(s)$,
\begin{equation}
\label{eq:gcmwf}
\vert \Psi^{NZJM\pi}_\alpha\rangle
=\sum_{c} f^{NZJM\pi}_{\alpha c} \hat P^J_{MK} \hat P^N\hat P^Z \ket{\Phi_\kappa(\mathbf{q})}.
\end{equation}
Here, $\alpha$ labels states with the same quantum numbers, while $c=\{K,\kappa,\mathbf{q}\}$ denotes a collective index for different deformed configurations, with $\mathbf{q}$ representing the shape parameters. Unless otherwise specified, we assume axial symmetry for these configurations.
The operators $\hat P^{J}_{MK}$ and $\hat{P}^{N, Z}$  are projectors that select components with the angular momentum $J$, neutron number $N$ and proton number $Z$~\cite{Ring:1980},
\begin{subequations}
\begin{align}
\label{eq:Euler_angles}
\hat P^{J}_{MK} &\equiv \frac{2J+1}{8\pi^2}\int d\Omega D^{J\ast}_{MK}(\Omega) \hat R(\Omega),\\
\label{eq:PNP_gauge}
\hat P^{N_\tau} &\equiv \frac{1}{2\pi}\int^{2\pi}_0 d\varphi_{\tau} e^{i\varphi_{\tau}(\hat N_\tau-N_\tau)}.
\end{align}
\end{subequations}
The operator $\hat P^J_{MK}$ extracts the component of angular momentum along the intrinsic axis $z$, which is conventionally labeled by $K$. The Wigner $D$-function is defined as $D^{J}_{MK}(\Omega)\equiv\bra{JM}\hat R(\Omega)\ket{JK}=\bra{JM}e^{i\phi\hat J_z}e^{i\theta\hat J_y}e^{i\psi\hat J_z}\ket{JK}$, where $\Omega=(\phi, \theta, \psi)$ represents the three Euler angles. The particle-number operator is defined as usual, $\hat N=\sum_k a^\dagger_k a_k$. 

For an even-even nucleus, the mean-field configuration is chosen as a quasiparticle vacuum~\cite{Yao:2020PRL}.  For an odd-mass nucleus, the mean-field configuration,  labeled as $\ket{\Phi^{\rm (OA)}_\kappa(\mathbf{q})}$, is constructed as a one-quasiparticle excitation of a quasiparticle vacuum~\cite{Ring:1980,Lin:2024arl}, 
\begin{eqnarray}
\label{eq:odd-mass-wfs}
 \ket{\Phi^{\rm (OA)}_\kappa(\mathbf{q})}  =\alpha^\dagger_\kappa \ket{\Phi_{(\kappa)}(\mathbf{q})},\quad  \alpha_\kappa \ket{\Phi_{(\kappa)}(\mathbf{q})}=0,
\end{eqnarray} 
where $\ket{\Phi_{(\kappa)}(\mathbf{q})}$ is also labeled with the collective coordinate $\mathbf{q}$, and has an even-number parity. The quasiparticle operators $(\alpha, \alpha^\dagger)$ are connected to the original single-particle operators $(a, a^\dagger)$ via the Bogoliubov transformation~\cite{Ring:1980}, 
\begin{eqnarray}
\label{eq:Bogoliubov_transformation}
\left(
\begin{array}{cc}
\alpha\\
\alpha^{\dag}\\
\end{array}
\right)=\left(
\begin{array}{cc}
U^\dag&V^{\dag}\\
V^T&U^T\\
\end{array}
\right)\left(
\begin{array}{cc}
a\\
a^\dag\\
\end{array}
\right),
\end{eqnarray}
where the $U, V$ matrices  are determined from the VAPNP+HFB calculation~\cite{Lin:2024arl,Bally:2021_EPJA}, 
\beq 
\delta \frac{\bra{\Phi^{\rm (OA)}_\kappa(\mathbf{q})} \hat H(s) \hat P^N\hat P^Z \ket{\Phi^{\rm (OA)}_\kappa(\mathbf{q})}}
{\bra{\Phi^{\rm (OA)}_\kappa(\mathbf{q})}  \hat P^N\hat P^Z \ket{\Phi^{\rm (OA)}_\kappa(\mathbf{q})}}=0.
\eeq 
The configuration of the $k$-th one-quasiparticle state with odd-number parity is obtained self-consistently by exchanging the $k$-th column of the $U$ and $V$ matrices in the HFB wave function~\cite{Ring:1980}:
\begin{equation}
(U_{pk},V_{pk}) \longleftrightarrow (V^\ast_{pk},U^\ast_{pk}),
\end{equation}
where the index $p=(\tau n\ell j\Omega)_p\equiv (n_p, \xi_p)$ is a label for the spherical harmonic oscillator basis, and $k$ the label for a quasiparticle state.  In this case, quasiparticle states are labeled with quantum numbers $K^\pi$, where $K=\Omega_p$ is the projection of the angular momentum $j_p$ along the $z$-axis, and parity $\pi=(-1)^{\ell_p}$. The collective coordinate $\mathbf{q}$ is replaced with the dimensionless quadrupole deformation $\beta_2$,
\beq 
\beta_2=\frac{4\pi}{3AR^2}\bra{\Phi^{\rm (OA)}_\kappa(\mathbf{q})} r^2Y_{20}\ket{\Phi^{\rm (OA)}_\kappa(\mathbf{q})},
\eeq 
 with $R=1.2A^{1/3}$ fm.   We note that the Kramers' degeneracy is lifted due to the breaking of time-reversal invariance in the self-consistent HFB calculations for odd-mass nuclei.

The weight functions $f^{NZJM\pi}_{\alpha c}$ are determined by varying the ground-state energy for the ansatz \eqref{eq:gcmwf}, which leads to the so-called Hill-Wheeler-Griffin (HWG) equation~\cite{Hill:1953,Ring:1980}
\begin{eqnarray}
\label{eq:HWG}
\sum_{c'}
\Bigg[\mathscr{H}^{NZJ\pi}_{cc'}
-E_\alpha^{J }\mathscr{N}^{NZJ\pi }_{cc'} \Bigg]
f^{NZJM\pi }_{\alpha c'}=0,
\end{eqnarray}
where the Hamiltonian kernel  and norm kernel are defined by
\begin{eqnarray}
\label{eq:kernel}
 &&\mathscr{O}^{NZJ\pi}_{cc'}\nonumber\\
 &=&\bra{NZ J\pi; c}  \hat O \ket{NZ J\pi; c'} \nonumber\\ 
 &=&  \frac{2J+1}{8\pi^2} \int d\Omega D_{KK'}^{J\ast}(\Omega)
 \int_0^{2\pi}d\varphi_n \frac{e^{-iN\varphi_n}}{2\pi} 
  \int_0^{2\pi}d\varphi_p \frac{e^{-iZ\varphi_p}}{2\pi} \nonumber\\
&&\times \bra{\Phi^{\rm (OA)}_{\kappa}(\mathbf{q})}
    \hat O \hat R(\Omega)  e^{i\hat Z\varphi_p}e^{i\hat N\varphi_n} \ket{\Phi^{\rm (OA)}_{\kappa'}(\mathbf{q}')},
\end{eqnarray}
with the operator $\hat O$ representing $\hat H(s)$ and $\hat{1}$, respectively. The parity $\pi$ is defined by the quasiparticle configurations $\ket{\Phi^{\rm (OA)}_{\kappa}(\mathbf{q})}$. The norm overlap $\bra{\Phi^{\rm (OA)}_{\kappa}(\mathbf{q})}  \hat R(\Omega)  e^{i\hat Z\varphi_p}e^{i\hat N\varphi_n} \ket{\Phi^{\rm (OA)}_{\kappa'}(\mathbf{q}')}$ of the HFB wave functions with odd-number parity is computed with the Pfaffian formula introduced in Ref.~\cite{Avez:2012PRC}.

The HWG equation (\ref{eq:HWG}) for a given set of quantum numbers $(NZJ\pi)$ is solved  in the standard way, as discussed in Refs. \cite{Ring:1980,Yao:2010}. First, we diagonalize the norm kernel $\mathscr{N}^{NZJ\pi }_{cc'}$, and construnct a new basis by using the  eigenfunctions with eigenvalue above a pre-chosen cutoff value to remove possible redundancy in the original basis. The Hamiltonian is then diagonalized in this new basis. In this way, one is able to obtain the energies $E_\alpha^{J}$ and
the mixing weights $f^{J\alpha}_{c}$ of nuclear states $\vert \Psi^{J}_\alpha\rangle$. Since the basis functions $\ket{NZ J; c}$ are nonorthogonal, one usually introduces the collective wave function 
\begin{equation}
\label{eq:coll_wf}
g^{J\pi}_\alpha(K, \mathbf{q})=\sum_{c'} (\mathscr{N}^{1/2})^{NZJ\pi}_{c,c'} f^{NZJM\pi }_{\alpha c'},
 \end{equation}
 which is properly normalized. The distribution of $g^{J\pi}_\alpha(K, \mathbf{q})$ over $K$ and $\mathbf{q}$ reflects the contribution of each basis function to the nuclear state $\vert \Psi^{NZJM\pi}_\alpha\rangle$.

\subsection{Electromagnetic observables}

The electromagnetic multipole transition  strength from an initial state with spin parity $J^{\pi_i}_{i, \alpha_i}$ to a final state  with spin parity $J^{\pi_f}_{f, \alpha_f}$  is determined by
\begin{eqnarray}
&& B(T\lambda; J_i\alpha_i\pi_i\rightarrow J_f\alpha_f\pi_f) \nonumber\\
&=& \hat{J}^{-2}_i   \left| \sum_{c_f, c_i} f^{J_i\alpha_i \pi_i}_{c_i}f^{J_f\alpha_f \pi_f}_{c_f} 
    \langle NZJ_f\pi_f,c_f||\hat T_\lambda ||NZJ_i\pi_i,c_i\rangle
    \right|^2,
\end{eqnarray} 
where the configuration-dependent reduced matrix element is given by  
\begin{eqnarray}
\label{eq:reduced_matrix_element}
    &&\langle NZ J_f\pi_f; c_f ||\hat  T_\lambda|| NZ J_i\pi_i; c_i\rangle\nonumber\\
    =&& \delta_{\pi_f\pi_i,(-1)^L}(-1)^{J_f-K_f}
    \cfrac{\hat{J}_i^2\hat{J}_f^2}{8\pi^2} \sum_{\nu M} \left(
 \begin{array}{ccc}
J_f&\lambda&J_i\\
-K_f &\nu&M\\
\end{array}
    \right) \nonumber\\
   &&\times  \int d\Omega D_{MK_i}^{J_i\ast}(\Omega)
\bra{\Phi^{\rm (OA)}_{\kappa_f}(\mathbf{q}_f)}
    \hat T_{\lambda\nu}
  \hat R(\Omega) \hat P^Z\hat P^N\hat P^{\pi_i}
   \ket{\Phi^{\rm (OA)}_{\kappa_i}(\mathbf{q}_i)}.\nonumber\\
\end{eqnarray}
Here, $\hat T_{\lambda\nu}$ stands for the tensor operator of either magnetic $(L=\lambda+1)$  or  electric ($L=\lambda$) multipole moments. The magnetic dipole and electric quadrupole moment are defined, respectively, by~\cite{Ring:1980},
\beqn
 \hat{\mathbf{M}}
&=& \sqrt{\frac{3}{4\pi}} \Bigg(g_s  \hat{\mathbf{s}} +  g_\ell  \hat{\boldsymbol{\ell}}\Bigg)\mu_N, \\ 
\hat{Q}_{2\nu} &=& e r^2 Y_{2\nu},
\eeqn  
with $(g_s, g_\ell)$ being the spin and orbital angular-momentum $g$-factors with bare values ($g_s=5.586,~g_\ell=1.0$) for protons and ($g_s=-3.826,~g_\ell=0$) for neutrons. The bare charges $e_\pi=1$ for protons and $e_\nu=0$ for neutrons are used for the operator $\hat{Q}_{e\nu}$. Since  the full single-particle space is employed, and we aim to describe microscopic correlations explicitly through our many-body methods, there is no need to introduce effective charges or $g$-factors  for neutrons and protons in the calculation of nuclear electromagnetic properties.

The reduced matrix element of a one-body tensor operator $\hat{T}_{\lambda\mu}$ between nuclear states can be expressed as a product of single-particle reduced matrix elements and the corresponding transition densities~\cite{Yao:2015TD,Gao:PRC2023},
\beqn
&&\langle NZ J_f\pi_f; c_f ||\hat  T_\lambda|| NZ J_i\pi_i; c_i\rangle\nonumber\\
&=& \sum_{ab} \hat{\lambda}^{-1}\langle n_a l_a j_a\vert\vert \hat T_{\lambda}\vert\vert n_b l_b j_b\rangle
\rho^{\lambda; J_i\to J_f}_{ab},
\eeqn
where the one-body transition density is determined by 
\beqn
 \rho^{\lambda; J_i\to J_f}_{ab}  
&\equiv&  \langle NZ J_f\pi_f; c_f ||[a^\dagger_a \tilde a_b]_{\lambda}|| NZ J_i\pi_i; c_i\rangle\nonumber\\
  &=&\hat{J}_f \sum_\nu  \langle J_i K_f-\nu, \lambda \nu\vert J_f K_f\rangle  \dfrac{\hat{J}^2_i}{8\pi^2}
\int d\Omega D^{J_i \ast}_{K_f-\nu,K_i}(\Omega) \nonumber\\
&&\times \sum_{m_am_b} \langle j_am_a j_b -m_b\vert \lambda\nu\rangle
(-1)^{j_b-m_b} \nonumber\\
&&  \times\bra{\Phi^{\rm (OA)}_{\kappa_f}(\mathbf{q}_f)}  a^\dagger_a a_b \hat R(\Omega) \hat P^N\hat P^Z\ket{\Phi^{\rm (OA)}_{\kappa_i}(\mathbf{q}_i)}.
\eeqn
Similarly, it is straightforward to compute the magnetic dipole moment $\mu(J^\pi_\alpha)$ of the state $|\Psi^{JJ}_\alpha\rangle$, 
\beqn
\mu\left(J_{\alpha}^{\pi}\right)
&\equiv & \left\langle \Psi^{J\pi}_\alpha\left|\hat{\mu}_{10}\right| \Psi^{J\pi}_\alpha\right\rangle \nonumber\\
&= &\sqrt{\frac{4 \pi}{3}} \left(\begin{array}{ccc}
J & 1 & J \\
-J & 0 & J
\end{array}\right) \nonumber \\
&& \times \sum_{c_{f}, c_{i}} f_{c_i}^{J \alpha \pi} f_{c_f}^{J \alpha \pi} \left\langle N Z J \pi ; c_{f}\left\|\hat{M}_{1}\right\| N Z J\pi ; c_{i}\right\rangle,
\eeqn
and the spectroscopic quadrupole moment  
\beqn 
\label{eq:Q_s}
Q_{s}\left(J_{\alpha}^{\pi}\right)
&\equiv & \sqrt{\frac{16 \pi}{5}}\left\langle \Psi^{J\pi}_\alpha\left|\hat{Q}_{20}\right| \Psi^{J\pi}_\alpha\right\rangle \nonumber \\
&=&  \sqrt{\frac{16 \pi}{5}}\left(\begin{array}{ccc}
J & 2 & J \\
-J & 0 & J
\end{array}\right) \sum_{c_{f}, c_{i}} f_{c_i}^{J \alpha \pi} f_{c_f}^{J \alpha \pi}  \nonumber \\
&& \times\left\langle N Z J \pi ; c_{f}\left\| \hat Q_2\right\| N Z J \pi; c_{i}\right\rangle.
\eeqn 

As discussed above, within the IMSRG framework all observables must be evolved consistently with the Hamiltonian. The IMSRG transformation induces many-body components in transition operators, including higher-body terms (see Appendix~\ref{app:tensor_operator}). In the present work, we neglect contributions from genuine two-body electromagnetic currents~\cite{Miyagi:2024}, but retain the induced two-body transition operators generated by the IMSRG evolution. Their contributions are evaluated by contracting the evolved operators with the corresponding two-body transition densities, whose expressions are given in Appendix~\ref{app:Two_body_density}.

 \section{Results and discussion}
\label{sec:Results and discussion}
  
 We perform IM-GCM calculations for the low-lying states of nuclei around the $N=20$, employing a spherical harmonic-oscillator basis $\ket{nljm}$ with $e = 2n + l \leq \eMax = 8$ (i.e., 9 major oscillator shells)  and $\hbar \omega = 16$ MeV. For the $3N$ interaction, we discard all matrix elements involving states with $e_1+e_2+e_3>14$.  The calculations are based on the low-resolution chiral NN+3N interaction EM1.8/2.0~\cite{Hebeler:2011PRC}. To assess the uncertainty associated with the finite model space, we take $^{32}$Mg as a representative example and extrapolate the results to the limit $\eMax \to \infty$ according to  $O(\eMax)=O(\infty)+a\exp(-b\eMax)$, where $O(\infty)$, $a$, and $b$ are fitted to the results obtained with $\eMax=6$, 8, and 10. The extrapolated values indicate an increase of approximately 1\% in the excitation energy of the $2_1^+$ state and 6\% in the $B(E2;2_1^+ \to 0_1^+)$ transition strength, whereas the excitation energy of the $0_2^+$ state increases by up to 25\%. Varying $\hbar\omega$ between 12 and 16 MeV does not change the level ordering, but modifies the excitation energies of the $2^+_1$ and $0^+_2$ states by approximately 7\% and the spectroscopic quadrupole moment by about 5\%.  Despite these uncertainties, the ordering of the low-lying states and the classification of nuclei as inside or outside the IOI in the subsequent study remain largely unchanged.  

 We note that strict decoupling of the reference state through Eq.~(\ref{flow_omega})  is generally  not guaranteed within the truncated MR-IMSRG(2) framework. We therefore employ the Brillouin generator (\ref{eq:Brillouin}), which has a well-posed definition as the gradient of the energy under unitary variations and efficiently suppresses the dominant off-diagonal couplings in practical applications. At large flow parameters, the evolution can become slow as it attempts to decouple nearly degenerate low-lying states, while uncertainties from omitted induced higher-body operators may accumulate. Following Ref.~\cite{Yao:2020PRL}, we therefore terminate the flow at $s=0.16~\mathrm{MeV}^{-1}$ and  treat the remaining correlations explicitly through the subsequent GCM diagonalization.  To estimate the possible uncertainty associated with terminating the flow at a finite value of $s$, we perform calculations for $^{32}$Mg using Hamiltonians evolved to several different flow parameters and extrapolate the results to the limit $s\to\infty$. The ground-state energy decreases from $-245.8$ MeV to $-246.9$ MeV, while the excitation energy of the $0^+_2$ state increases from 2.0 MeV to 2.7 MeV.  Further details can be found in Ref.~\cite{Zhou:2024_short}.

\subsection{Low-lying states of even-even nuclei}  
 
\begin{figure}[tb]
 \centering
\includegraphics[width=8.4cm]{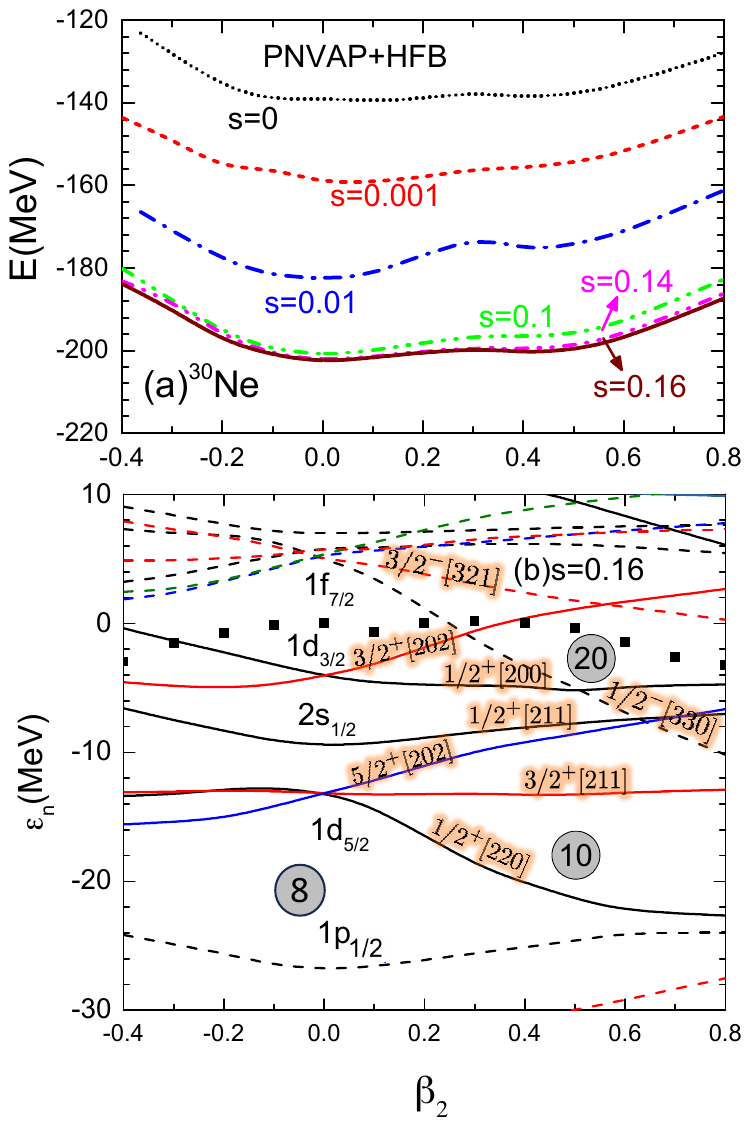} 
\caption{ (Color online) (a) Deformation energy curves of \nuclide[30]{Ne} obtained from VAPNP+HFB calculations using IMSRG-evolved Hamiltonians with various values of the flow parameter $s$,  as a function of the axial quadrupole deformation $\beta_2$.
(b) Nilsson diagram for neutrons in \nuclide[30]{Ne} calculated with the evolved Hamiltonian $\hat{H}(s)$ at $s = 0.16$ MeV$^{-1}$, showing single-particle energies as functions of $\beta_2$. The Nilsson asymptotic quantum numbers $\Omega^\pi[Nn_z\Lambda]$ are shown for several relevant orbitals. Here, $N$ denotes the principal oscillator quantum number, while $n_z$, $\Omega$, and $\Lambda$ represent the projections of $N$, $j$, and $\ell$ onto the symmetry axis, respectively. Black squares mark the neutron Fermi levels.}
\label{fig:Ne30s}
 \end{figure}

Figure~\ref{fig:Ne30s}(a) shows the energy curves of \nuclide[30]{Ne} as a function of the quadrupole deformation parameter $\beta_2$ obtained from the VAPNP+HFB calculation using a set of IMSRG-evolved Hamiltonians $\hat{H}(s)$ for different values of the flow parameter $s$. It is evident that with the IMSRG-evolved Hamiltonian $\hat{H}(s)$, the energy surface shifts systematically downward as the flow parameter $s$ increases. This behavior indicates that an increasing amount of many-body correlations is incorporated into the mean-field description, thereby reducing the contribution of residual correlations for $\hat{H}(s)$ at larger values of  $s$. A similar phenomenon has been observed in PGCM-PT calculations with IMSRG-evolved Hamiltonians~\cite{Frosini:2022_1,Frosini:2022_2,Frosini:2022_3}.  The energy curve stabilizes reasonably well around $s=0.16$ MeV$^{-1}$, where it exhibits two energy minima with distinct shapes: the lower one corresponds to a spherical shape, while the other is located at a prolate shape with $\beta_2 = 0.5$. 

As discussed in Refs.~\cite{Zhou:2024_short,Ding:2025}, the observables of atomic nuclei should be invariant under a unitary transformation like the IMSRG (aside from truncation errors). However, the single-particle energies derived from  the $\hat H(s)$ are not observables and will also evolve with the flow parameter $s$. The Nilsson diagram of neutrons at $s=0.16$ MeV$^{-1}$ is displayed in Fig.~\ref{fig:Ne30s}(b). It is seen that the onset of the strongly deformed prolate shape is attributed to the level crossing of the downward-sloping $\Omega^\pi=1/2^-$ component of the $\nu f_{7/2}$ orbital and the upward-sloping $\Omega^\pi=3/2^+$ component of the $\nu d_{3/2}$ orbital, which results in a significant $N=20$ shell gap around $\beta_2 = 0.5$.  This result is consistent with studies using phenomenological  interactions \cite{Stevenson:2002,Kimura:2002,Rodriguez-Guzman:2002NPA}.

 \begin{figure}[bt]
 \centering
\includegraphics[width=8.4cm]{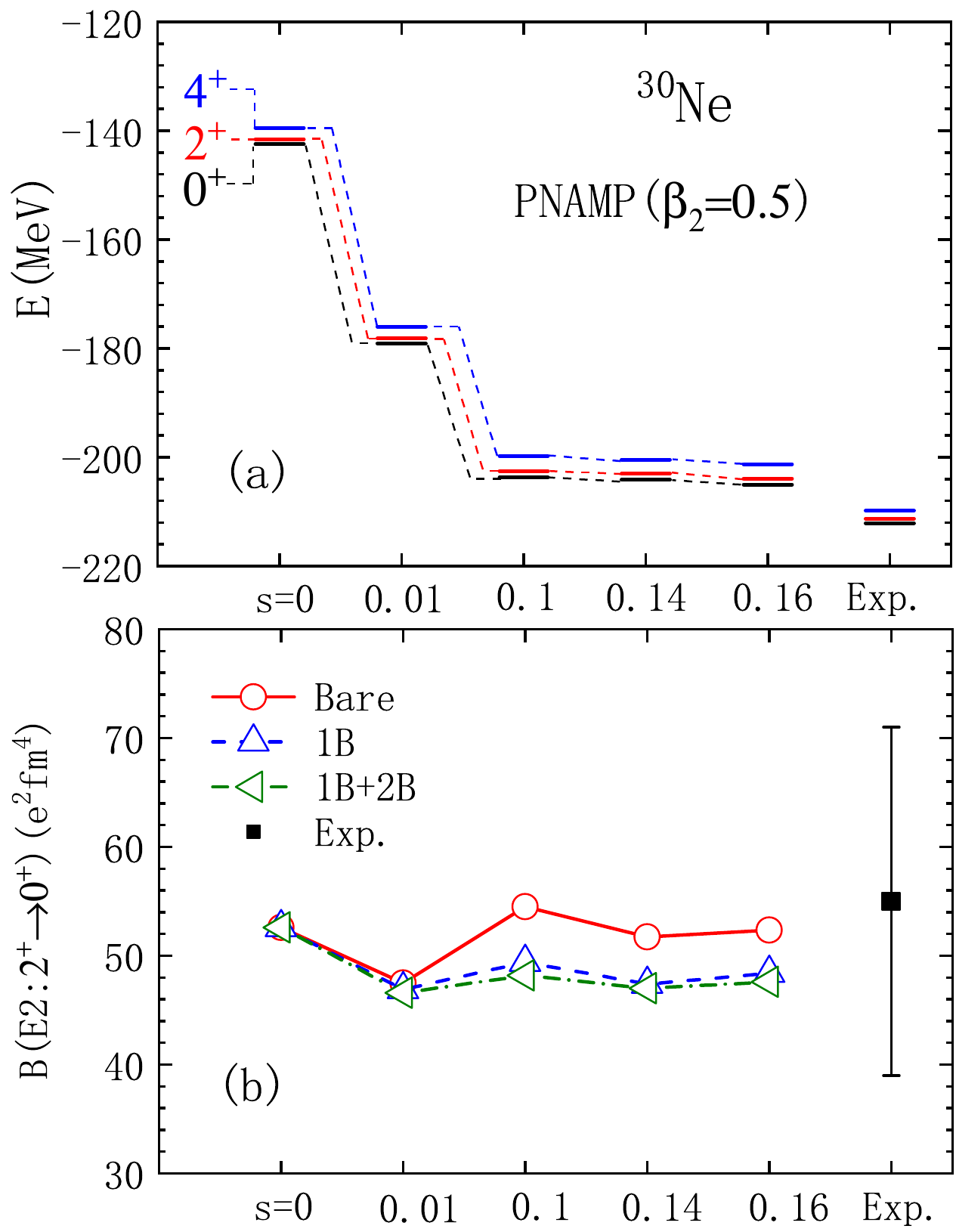}
\caption{ (Color online) (a) The low-energy spectra and (b) the $B(E2: 2^+ \rightarrow 0^+)$ values of \nuclide[30]{Ne} from the IM-PHFB calculation using different $\hat{H}(s)$ based on the single HFB state with $\beta_2 = 0.5$. In panel (b), either the bare one-body transition operator (Bare) or the evolved one-body operator with (1B+2B) and without (1B) evolved two-body operators is used. The experimental data are taken from Refs.~\cite{Ne30E2Doornenbal:2016,NNDC}.}
\label{fig:Ne30_1B+2B}
 \end{figure}

 \begin{figure*}[bt]
 \centering
\includegraphics[width=14cm]{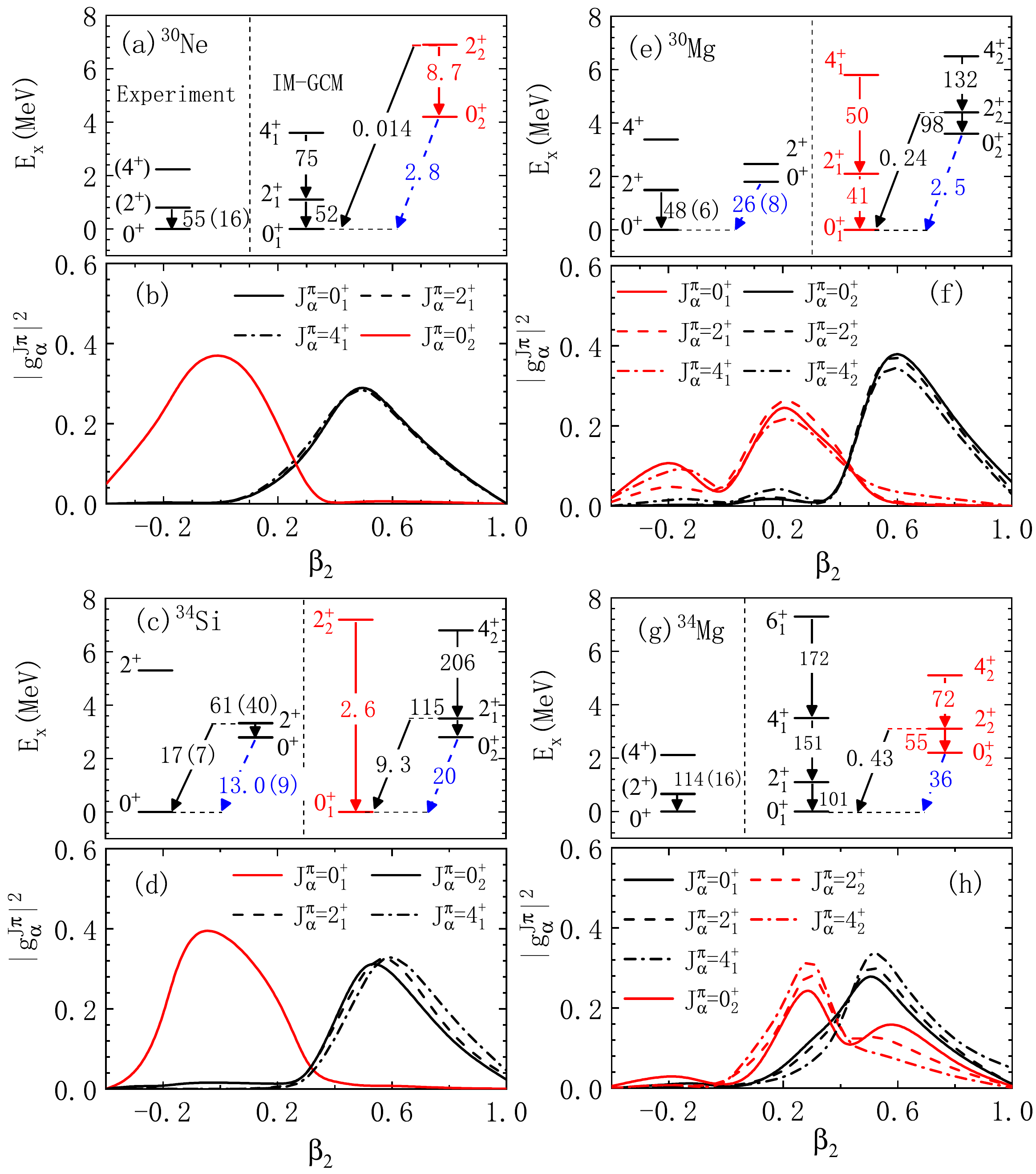}
\caption{ (Color online)  Energy spectra and distributions of collective wave functions $|g_\alpha^{J\pi}|^2$ [Eq.~(\ref{eq:coll_wf})] for the low-lying states of $\nuclide[30]{Ne}$ (a, b), $\nuclide[34]{Si}$ (c, d), $\nuclide[30]{Mg}$ (e, f), and $\nuclide[34]{Mg}$ (g, h), obtained from IM-GCM calculations and compared with available data~\cite{NNDC,sheeta=30,Rotaru:2012Si34isomer}. Solid arrows represent the calculated $B(E2)$ transition strengths (in units of e$^2$fm$^4$), while dashed arrows denote the $10^{3}\rho^2(E0)$ values. The collective wave functions corresponding to weakly deformed states are shown in red, and those for strongly deformed states are shown in black.}
\label{fig:evenlevel}
 \end{figure*}

Figure~\ref{fig:Ne30_1B+2B}(a) displays the excitation spectra of the $0^+$, $2^+$, and $4^+$ states obtained from particle-number and angular-momentum projected (PNAMP)  calculations based on the prolate HFB configuration with $\beta_2=0.5$ and different evolved Hamiltonians $\hat{H}(s)$.  This method is hereafter referred to as IMSRG+PHFB (IM-PHFB). In a perfect calculation, the energies should be independent of $s$, but the NO2B truncation in the (MR-)IMSRG and the approximations inherent to the IM-PHFB calculation, e.g., due to the basis truncation or selection of generator coordinates, violate unitarity in the many-body system. Thus, the variation of the energies with $s$ in Fig.~\ref{fig:Ne30_1B+2B}(a) indicates the effect of missing many-body correlations in the IM-PHFB. It is important to note that the impact of the \emph{many-body truncations} becomes progressively smaller as $\hat{H}(s)$ is evolved, and the overall variation with $s$ decreases if we improve on the IMSRG and IM-PHFB truncations, as demonstrated in Ref.\cite{Frosini:2022_3}.

In our present calculations, we see that the states ``stabilize'' around $s=0.16$ MeV$^{-1}$, reflecting the behavior seen in Fig.~\ref{fig:Ne30s}. The absolute energies are systematically higher than the data. This overestimation of the energies is expected to be improved by enlarging $\eMax$. Indeed, we find that extrapolating the $\eMax$ value to $\infty$ lowers down the entire spectrum.  See for instance the supplementary file of Ref.~\cite{Zhou:2024_short}. 
Figure~\ref{fig:Ne30_1B+2B}(b) compares the $B(E2:2^+ \rightarrow 0^+)$ values for \nuclide[30]{Ne}, for the calculation  based on the same configuration with  $\beta_2 = 0.5$, by using the transition operators with different flow  parameter $s$. We see that the $B(E2)$ value obtained with the evolved one-body transition operator at $s=0.16$ MeV$^{-1}$ is about 8\% smaller than that by the bare one. In contrast, the evolved two-body transition operator only contributes about 2\%.

An atomic nucleus with multiple minima on its potential-energy surface may exhibit either shape coexistence or shape mixing. We are examining this phenomenon in the low-lying states of \nuclide[30]{Ne} and its neighboring even-even nuclei, including \nuclide[34]{Si} and \nuclide[30,34]{Mg}. The corresponding energy spectra and collective wave functions are displayed in Fig.~\ref{fig:evenlevel} (also see Ref.~\cite{Zhou:2024_short}). Overall, both the energy spectra and the electric quadrupole transition strengths are reasonably reproduced.
According to the distribution of the collective wave functions, the low-energy structure of \nuclide[30]{Ne} is dominated by a strongly deformed rotational band, along with a weakly deformed $0^+_2$ state. In contrast, for \nuclide[34]{Si}, which has four more protons than \nuclide[30]{Ne}, the energy ordering of the strongly and weakly deformed $0^+$ states is inverted.  both \nuclide[30]{Ne} and \nuclide[34]{Si} therefore exhibit shape coexistence.   For \nuclide[30]{Mg}, the IM-GCM predicts a strong mixture of weakly deformed prolate and oblate configurations with $\beta_2\simeq\pm0.2$ in the ground state. The excited states of \nuclide[30]{Mg} belonging to the rotational band are predominantly characterized by a strongly prolate deformed shape with $\beta_2\simeq 0.6$. With the addition of four more neutrons to \nuclide[30]{Mg}, the energy ordering of the two shapes in \nuclide[34]{Mg} is inverted. Both \nuclide[30]{Mg} and \nuclide[34]{Mg} thus display features of both shape coexistence and shape mixing.

 In addition to examining the distribution of the collective wave function, these phenomena can also be indicated from the electric monopole ($E0$) transition strength $\rho^2(E0)$. This strength, for transitions between the lowest $0^+$ states via the emission of conversion electrons, is defined as
\beq
\rho^2(E0; 0^+_2 \to 0^+_1)
=\left|\frac{ \bra{0^+_1}e \sum_{p} \hat{r}_{p}^{2}\ket{0^+_2}}{e R^{2}}\right|^{2}.
\eeq
The $\rho^2(E0)$ value is sensitive to both the degree of deformation and the extent of mixing between configurations associated with different shapes~\cite{Heyde:2011RMP}. As shown in Fig.~\ref{fig:evenlevel}, the $\rho^2(E0)$ strengths are generally weak in all four nuclei, indicating only limited mixing between the two lowest $0^+$ states with different radii. Among them, \nuclide[34]{Mg} exhibits the strongest configuration mixing and, consequently, the largest $E0$ transition strength.

 \begin{figure}[tb]
 \centering
\includegraphics[width=7cm]{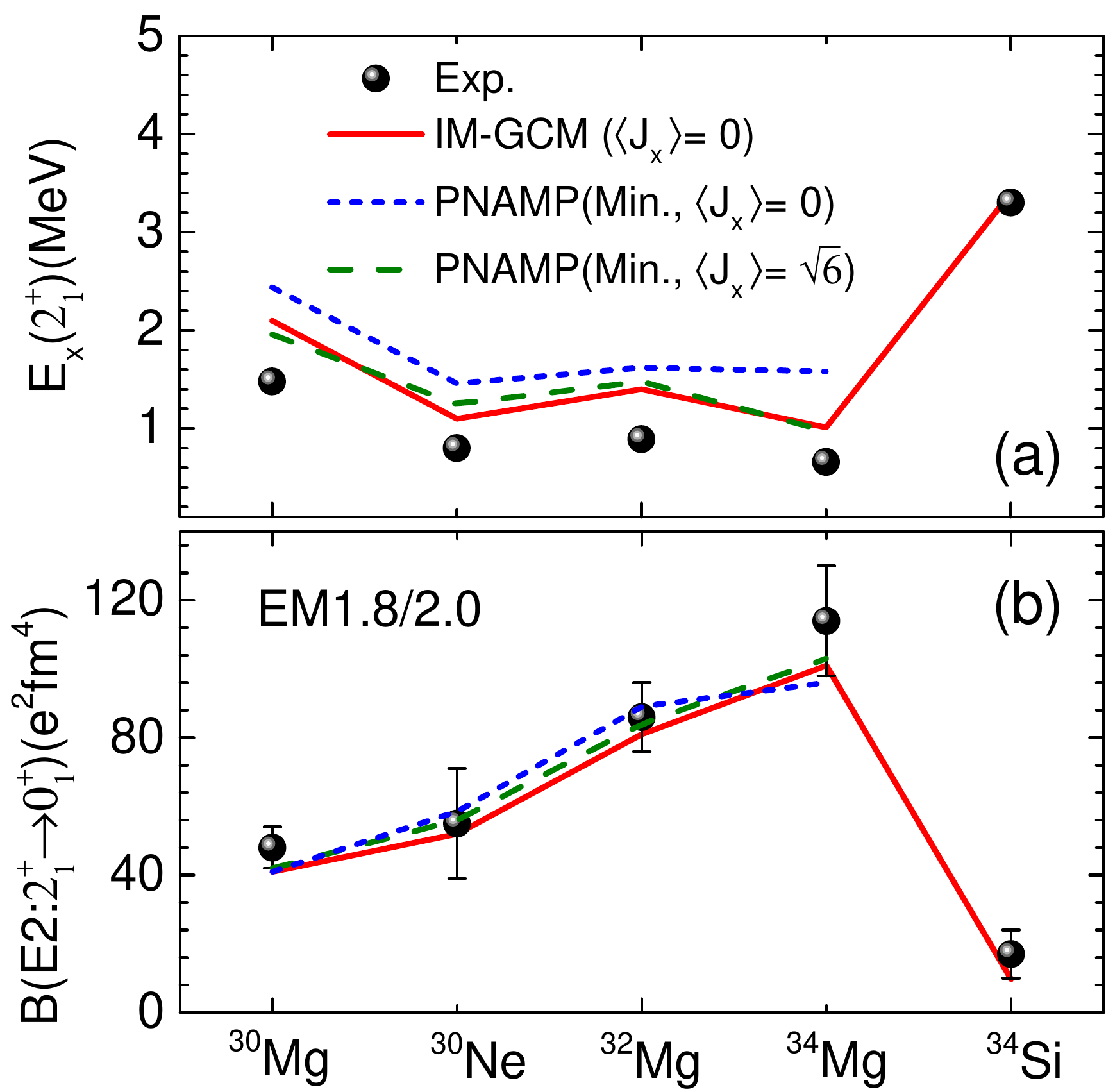}
\caption{(Color online) (a) Excitation energies of the $2_1^+$ state and (b) transition probabilities $B(E2: 2_1^+ \rightarrow 0_1^+)$ for \nuclide[30,32,34]{Mg}, \nuclide[30]{Ne}, and \nuclide[34]{Si}. The bullets and red solid lines represent the experimental data and the IM-GCM calculations using the configuration with zero cranking frequency ($\langle J_x \rangle = 0$). Results from PNAMP calculations based on the lowest-energy configuration with zero and nonzero cranking frequencies ($\langle J_x \rangle = 0$ and $\langle J_x \rangle = \sqrt{6}$, respectively) are also provided for comparison. The three-dimensional AMP is used to restore the angular momentum of the configurations with nonzero cranking frequency.
 }
  \label{fig:even_e2} 
 \end{figure}

Figure~\ref{fig:even_e2} displays the systematic trends in the excitation energies of the $2^+_1$ states and the $B(E2: 2_1^+\to 0_1^+)$ transition strengths for the five nuclei \nuclide[30]{Ne}, \nuclide[30,32,34]{Mg}, and \nuclide[34]{Si} from the IM-GCM calculation. The results show that the excitation energies $E_x(2^+_1)$ for \nuclide[30]{Ne} and \nuclide[32,34]{Mg} are significantly lower than those for \nuclide[30]{Mg} and \nuclide[34]{Si}. Correspondingly, the $B(E2: 2_1^+\to 0_1^+)$ values for the former three nuclei are generally higher than those for the latter two. We note that the $E_x(2^+_1)$ values are generally overestimated, which is expected to improve by incorporating cranked configurations with non-zero frequency~\cite{Zhou:2024_short,Belley:2024PRL}. For illustration, we compare the results of PNAMP calculations based on the cranked configuration with $\langle J_x\rangle=\sqrt{6}$ and those based on the non-cranked configuration. The comparison clearly shows that the excitation energy of the $2^+$ state decreases significantly in the PNAMP calculation with the cranked configuration. In short, our IM-GCM calculations indicate that shape coexistence is prevalent in nuclei around \nuclide[32]{Mg}. Specifically, the ground states of \nuclide[30]{Mg} and \nuclide[34]{Si} are weakly deformed, while those of \nuclide[30]{Ne} and \nuclide[34]{Mg} are strongly deformed. This places the latter two nuclei within the IOI.

\subsection{Low-lying states of odd-mass nuclei}

 \begin{figure}[tb]
 \centering
\includegraphics[width=8.5cm]{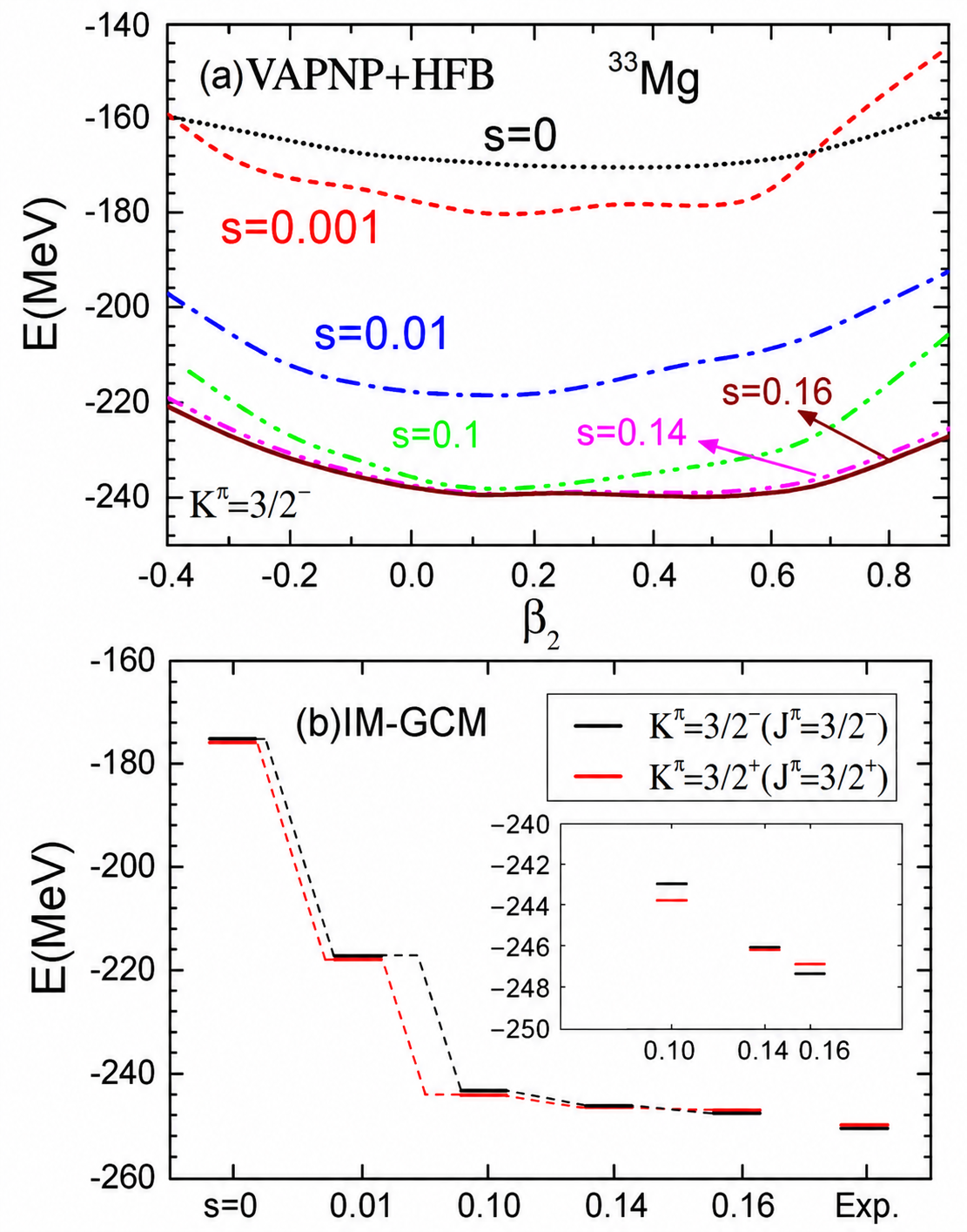}
\caption{(Color online) (a) Energy curves of one-quasiparticle states with $K^\pi = 3/2^-$ from the VAPNP+HFB calculation for \nuclide[33]{Mg} as a function of the quadrupole deformation $\beta_2$, obtained using IMSRG-evolved Hamiltonians with different values of the flow parameter $s$. (b) Evolution of the lowest $J^\pi = 3/2^\pm$ states with $K^\pi = 3/2^\pm$ in \nuclide[33]{Mg} obtained from the IM-GCM calculation as a function of the flow parameter $s$. Experimental data are taken from Ref.~\cite{Bazin:2021nti}. }
  \label{fig:Mg33_flow} 
 \end{figure}

The low-lying states of odd-mass nuclei are generally more complex than those of neighboring even-even nuclei. This complexity is particularly evident in odd-mass nuclei around $N=20$ with shape coexistence. In Ref.~\cite{Zhou:2024_short}, we demonstrated the crucial role of dynamical correlations in accurately reproducing the spin-parity of the ground state of \nuclide[33]{Mg}. Here, we provide additional details on the results for this nucleus.  Figure \ref{fig:Mg33_flow}(a) shows the energy curves of the one-quasiparticle state of \nuclide[33]{Mg} with $K^\pi = 3/2^-$ from the VAPNP+HFB calculation using the IMSRG-evolved $\hat{H}(s)$ for \nuclide[32]{Mg}. Similar to the case of \nuclide[30]{Ne} in Fig.\ref{fig:Ne30s}, the energy curves stabilize at $s = 0.16$ MeV$^{-1}$.  Figure~\ref{fig:Mg33_flow}(b) shows that the energy ordering of the $J^\pi = 3/2^-$ and $3/2^+$ states in \nuclide[33]{Mg} reverses beyond $s \approx 0.14$ MeV$^{-1}$, yielding a $3/2^-$ ground state consistent with experiment~\cite{Bazin:2021nti}. This behavior reflects the evolution of Nilsson levels in \nuclide[32]{Mg}~\cite{Zhou:2024_short}, where a crossing between the $\Omega^\pi = 3/2^+$ component of the $\nu d_{3/2}$ orbital and the $\Omega^\pi = 3/2^-$ component of the $\nu f_{7/2}$ orbital shifts from $\beta_2 \approx 0.5$ to 0.4 with increasing $s$. As a result, the unpaired neutron occupies the $\Omega^\pi =3/2^-$ component instead of $3/2^+$ component,  explaining the emergence of a strongly deformed energy minimum in the energy surface obtained using the IMSRG-evolved Hamiltonian.

 \begin{figure}[tb]
 \centering
\includegraphics[width=7cm]{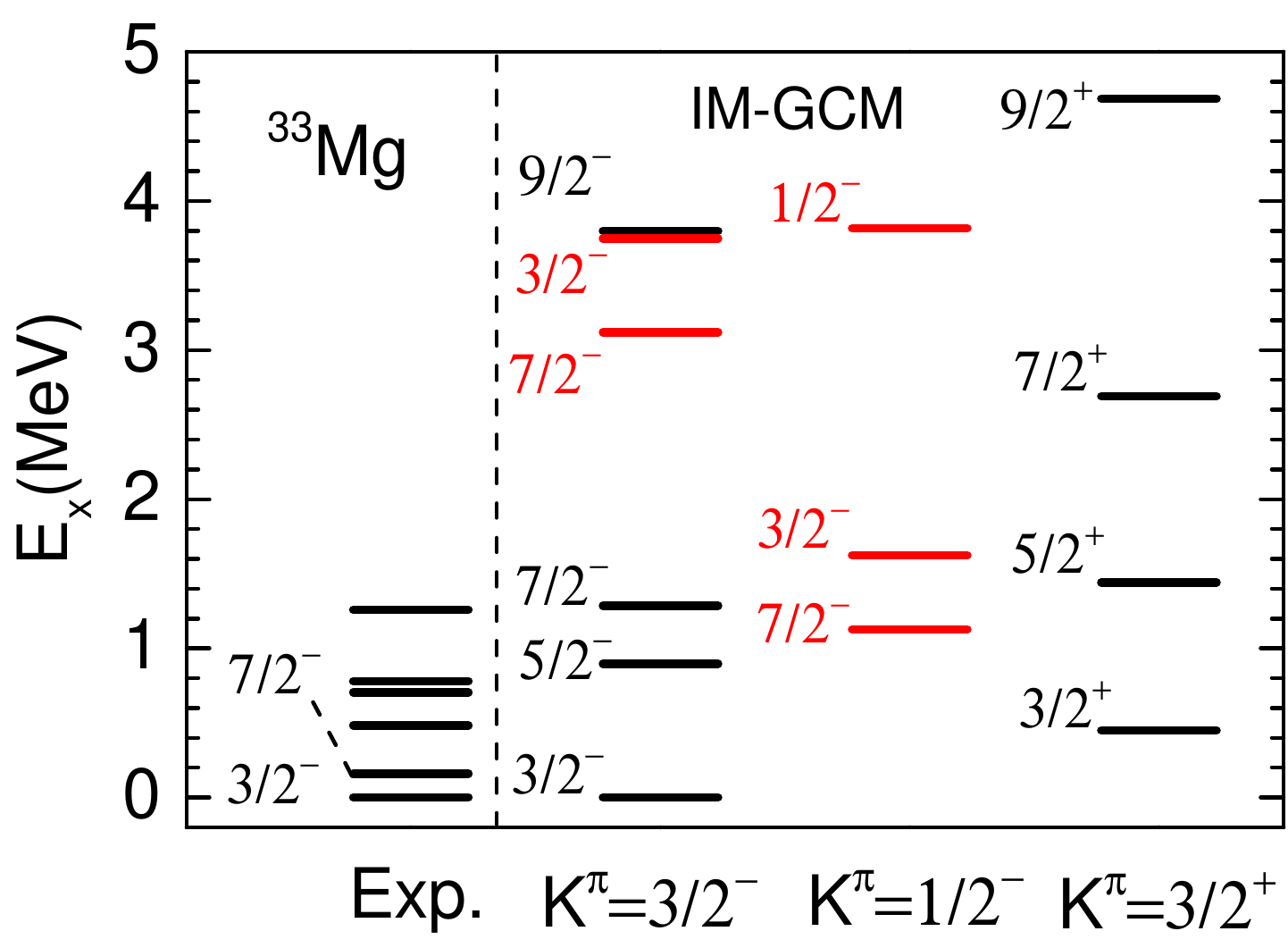}
\caption{(Color online) Low-lying states of \nuclide[33]{Mg} with $K^\pi=3/2^-,1/2^-$ and $3/2^+$  from  the IM-GCM calculation using $H(s=0.16)$, in comparison with experimental data~\cite{Bazin:2021nti}. The states dominated by the weakly and strongly deformed configurations are indicated with red and  black colors, respectively.  The energy spectra after $K$-mixing have been reported in Ref. \cite{Zhou:2024_short}. }
  \label{fig:Mg33_IMGCM_no_Kmixing} 
 \end{figure}

 \begin{figure}[tb]
 \centering
\includegraphics[width=8.5cm]{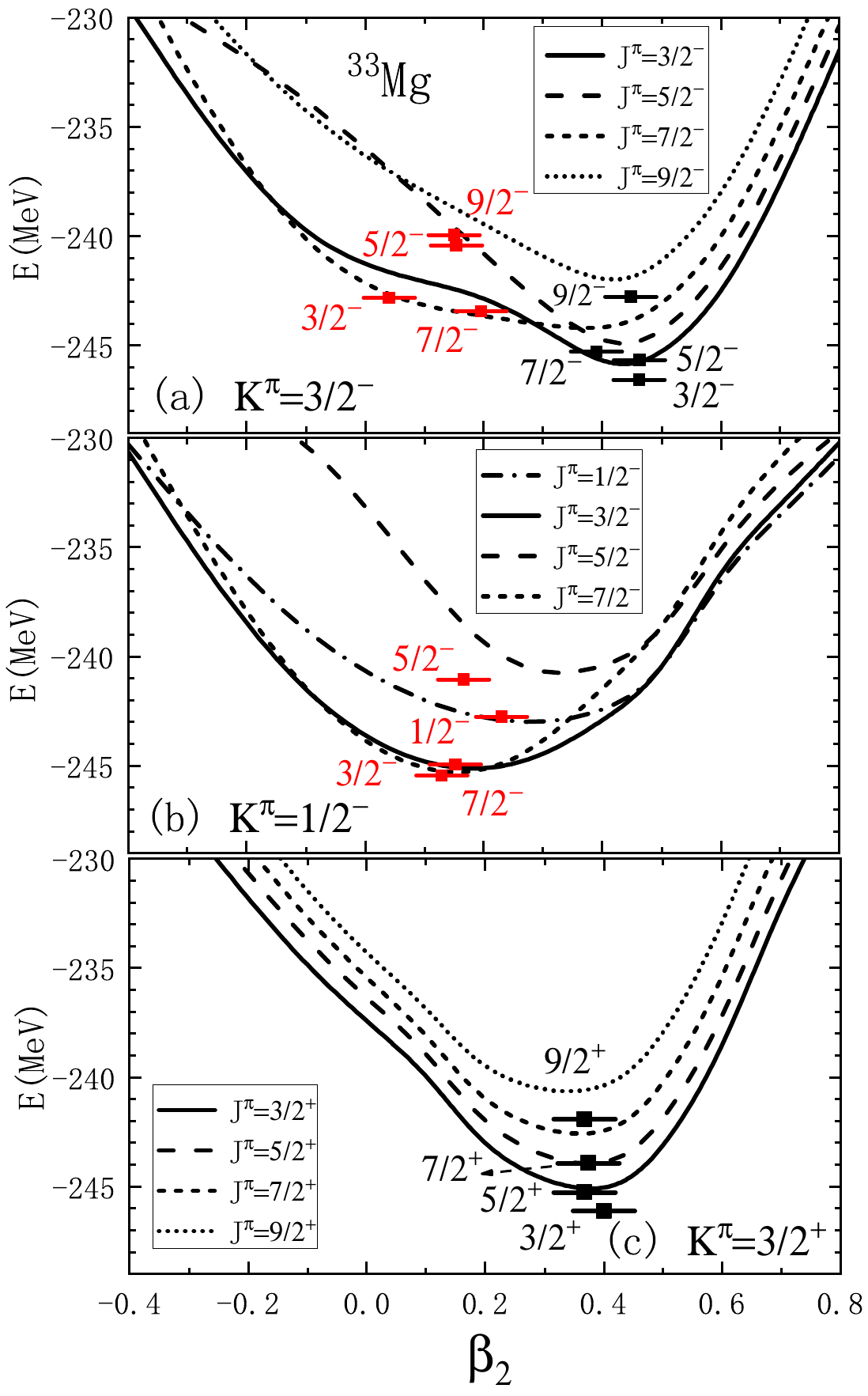}
\caption{(Color online) Energies of one-quasiparticle states with (a) $K^\pi = 3/2^-$, (b) $K^\pi = 1/2^-$, and (c) $K^\pi = 3/2^+$, from the PHFB calculation using the Hamiltonian $\hat{H}(s=0.16)$, with projection onto correct particle numbers and different angular momentum $J^\pi$, as a function of the quadrupole deformation parameter $\beta_2$. The discrete low-lying states $J^\pi_\alpha$, obtained by mixing different axially deformed one-quasiparticle states using the PGCM, are plotted at their mean quadrupole deformation $\bar{\beta}_2^{J\pi\alpha}$, as defined in Eq.~(\ref{eq:mean_beta2}).}
  \label{fig:Mg33_AMPE} 
 \end{figure}

Figure~\ref{fig:Mg33_IMGCM_no_Kmixing} shows the low-lying spectrum of \nuclide[33]{Mg} for $K^\pi = 3/2^-, 1/2^-$, and $3/2^+$ from IM-GCM calculations using $\hat{H}(s = 0.16)$. A clear rotational band is seen for the $K^\pi = 3/2^+$ sequence ($J^\pi = 3/2^+, 5/2^+, 7/2^+, 9/2^+$), while no such structure emerges for $K^\pi = 1/2^-$ and $3/2^-$. This behavior is clarified in Fig.~\ref{fig:Mg33_AMPE}, which presents the energy curves of one-quasiparticle states projected onto particle number and angular momentum $J^\pi$ as functions of quadrupole deformation $\beta_2$. The low-lying states $J^\pi_\alpha$ in Fig.~\ref{fig:Mg33_IMGCM_no_Kmixing} result from configuration mixing in the PGCM and are plotted at their mean deformation $\bar{\beta}_2^{J\pi\alpha}$, as defined in Ref.~\cite{Yao:2022_HB}
\beq
\label{eq:mean_beta2}
\bar{\beta}_2^{J\alpha}(K^\pi)
=\sum_{\beta_2} |g^{J\pi}_\alpha(K, \beta_2)|^2\beta_2.
\eeq 
As shown in Fig.~\ref{fig:Mg33_AMPE}(c), the energy curves with $K^\pi = 3/2^+$ but different $J^\pi$ have well-defined energy minima around the same quadrupole deformation, $\beta_2 = 0.4$. The observed rotational band ($J^\pi = 3/2^+, 5/2^+, \ldots$) with $\Delta J = 1$ is primarily built on this strongly deformed energy-minimal configuration, consistent with the particle-rotor model in the strong coupling limit \cite{Ring:1980}. As shown in Fig.~\ref{fig:Mg33_AMPE}(a), the energy curves for $K^\pi = 3/2^-$ exhibit shallow minima around $\beta_2 = 0.5$, in contrast to the deeper minima for $K^\pi = 3/2^+$. This leads to enhanced shape mixing, particularly in the $J^\pi = 7/2^-$ state, and explains why the $K^\pi = 3/2^-$ sequence ($J^\pi = 3/2^-, 5/2^-, \ldots$) in Fig.\ref{fig:Mg33_IMGCM_no_Kmixing} deviates from a rotational band. In comparison, Fig.\ref{fig:Mg33_AMPE}(b) shows that the energy minima for $J^\pi = 7/2^-, 3/2^-$ with $K^\pi = 1/2^-$ lie in weakly deformed regions, consistent with the weak-coupling limit of the particle-rotor model, where the lowest state has $J \simeq j = 7/2$\cite{Ring:1980}. Thus, the $J^\pi = 7/2^-$ and $3/2^-$ states arising from $K^\pi = 1/2^-$ configurations in Fig.~\ref{fig:Mg33_IMGCM_no_Kmixing} are predominantly associated with weak deformation. This conclusion is consistent with our previous analysis of the collective wave functions $|g_\alpha^{J\pi}(K,q)|^2$ after $K$-mixing~\cite{Zhou:2024_short}, which shows that the $3/2^-$ state is dominated by a strongly prolate-deformed configuration with $K^\pi = 3/2^-$, while the $7/2^-$ state primarily consists of the weakly deformed configurations with different $K$ values. Notably, the $3/2^+$ state shares a similar dominant configuration with the $3/2^-$ state.

 \begin{figure*}[tb]
 \centering
\includegraphics[width=17cm]{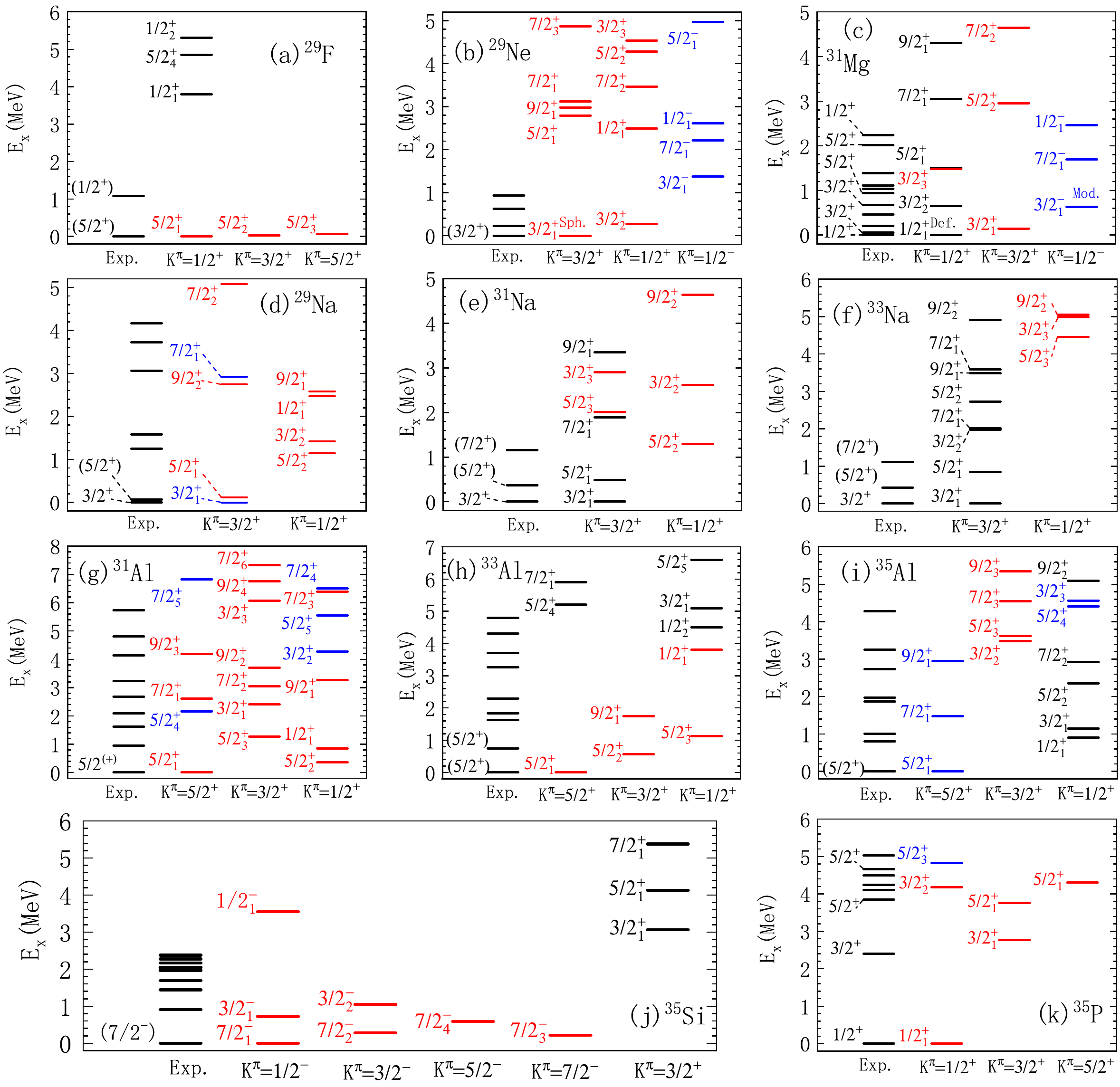}
\caption{(Color online) The energy spectra of the low-lying states of \nuclide[29]{Ne}, \nuclide[29,31,33]{Na}, \nuclide[31]{Mg}, \nuclide[31,33,35]{Al}, \nuclide[35]{Si} and \nuclide[35]{P} around $N=20$ and $Z=12$ from the IM-GCM calculation based on the configurations with different values of  $K^\pi$.  The states dominated by the weakly ($\bar{\beta}_2<0.2$), moderately  ($0.2<\bar{\beta}_2<0.4$), and  strongly ($\bar{\beta}_2>0.4$) deformed configurations are indicated with red, blue, and black colors, respectively. The  data are taken  from Refs.\cite{NNDC,Doornenbal:2017} }. 
\label{fig:spectra_odd_mass} 
 \end{figure*}

Figure~\ref{fig:spectra_odd_mass} presents the low-lying IM-GCM energy spectra of eleven odd-mass nuclei around $N=20$ and $Z=12$, including \nuclide[29]{F}, \nuclide[29]{Ne}, \nuclide[29,31,33]{Na}, \nuclide[31]{Mg}, \nuclide[31,33,35]{Al}, \nuclide[35]{Si}, and \nuclide[35]{P}, using evolved Hamiltonians $\hat{H}(s = 0.16)$ for each  nucleus. The experimental spin-parities of the low-lying states, including ground states, are well reproduced across all nuclei. Notably, the ground states of \nuclide[29]{F}, \nuclide[29]{Ne}, \nuclide[31,33]{Al}, \nuclide[35]{Si}, and \nuclide[35]{P} are dominated by weakly deformed configurations, while \nuclide[29]{Na} and \nuclide[35]{Al} show moderate deformation. Strong quadrupole deformation is found only in the ground states of \nuclide[31]{Mg} and \nuclide[31,33]{Na}.

 \begin{figure}[tb]
 \centering
\includegraphics[width=7cm]{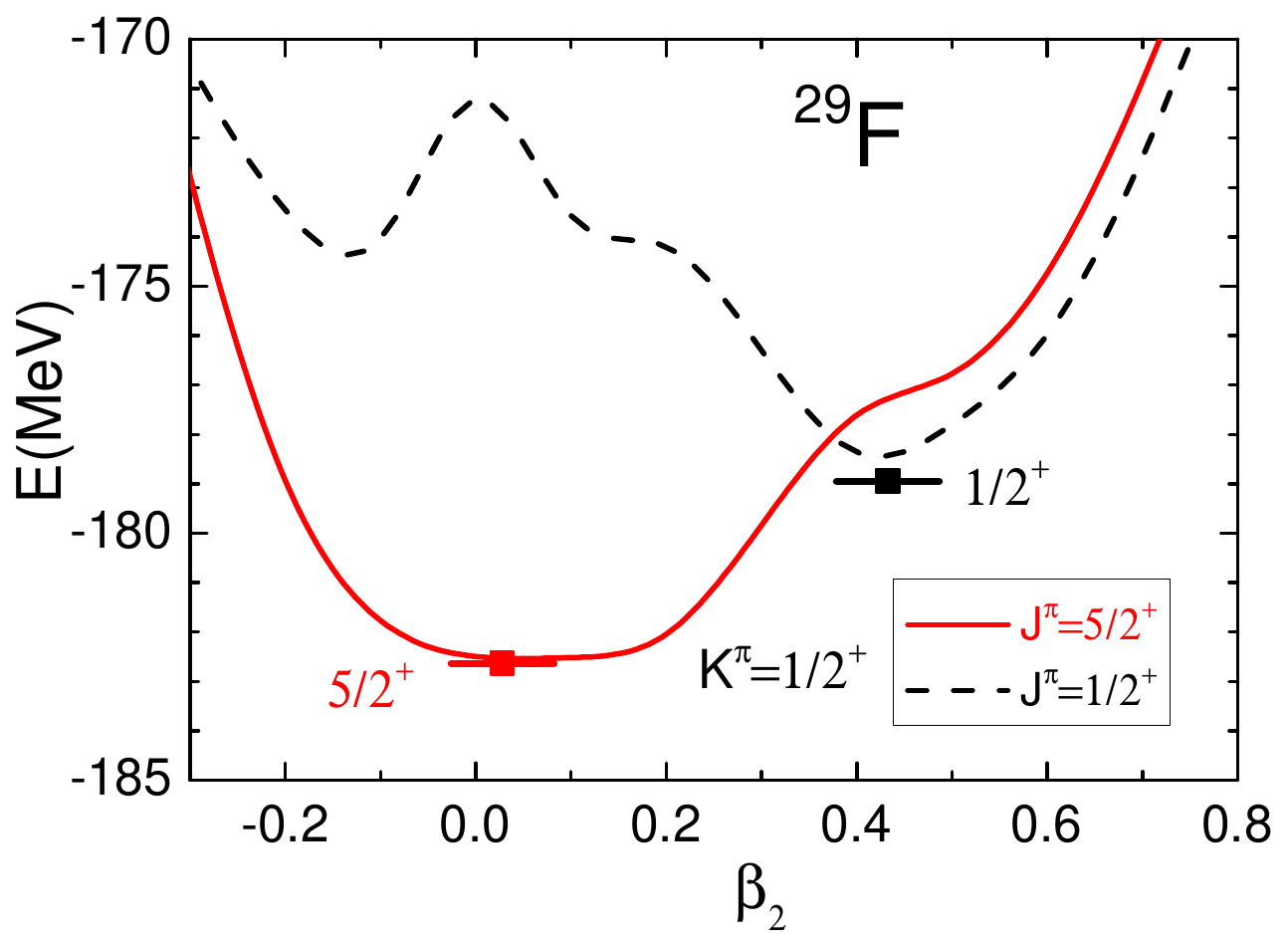}
\caption{(Color online)  Same as Fig.~\ref{fig:Mg33_AMPE}, but for  the one-quasiparticle states of \nuclide[29]{F} with  $K^\pi = 1/2^+$. }
  \label{fig:PES_F29_K1/2}
 \end{figure}
 
 \begin{figure*}[tb]
 \centering
\includegraphics[width=14cm]{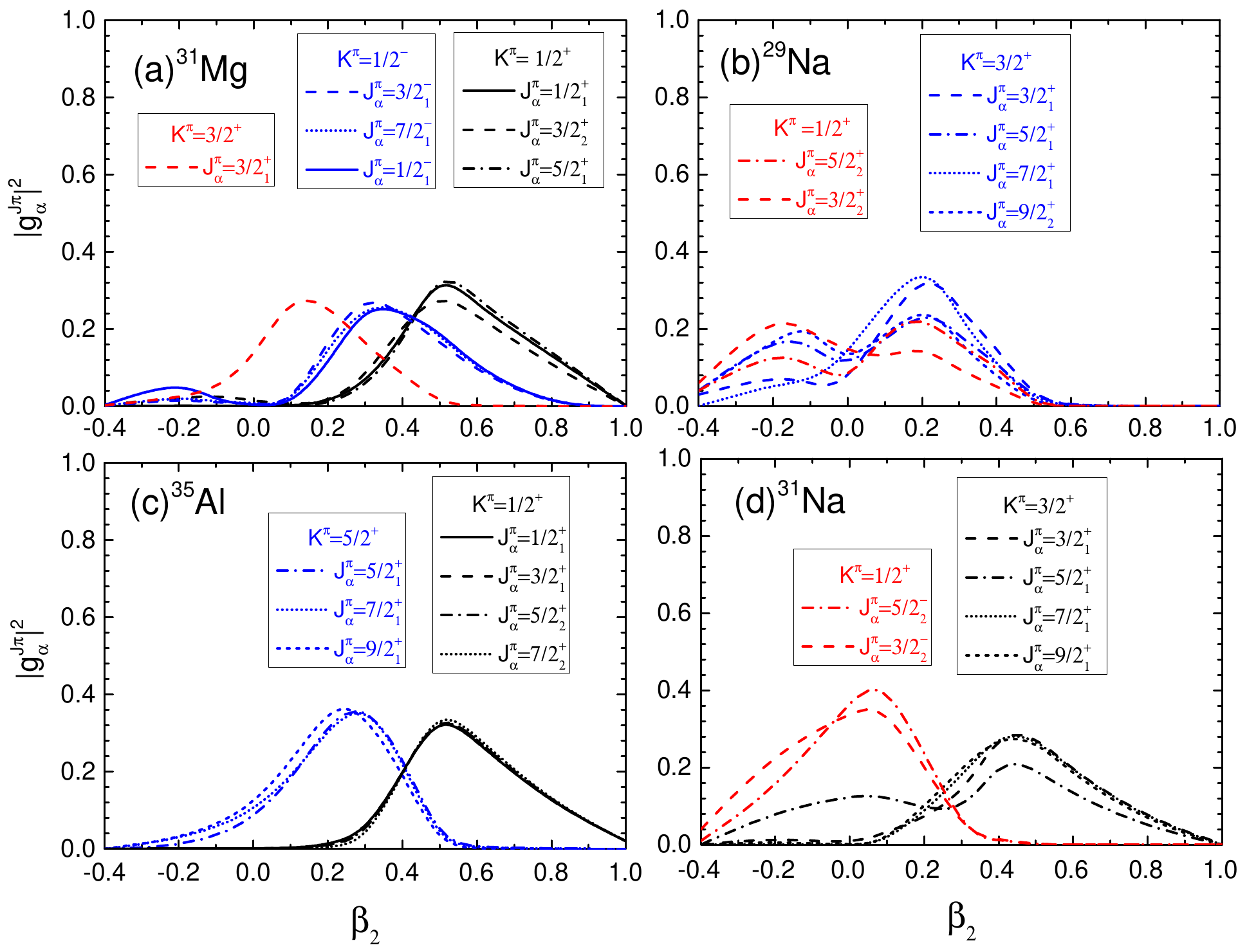}
\caption{(Color online) The collective wave functions of  states  $J_\alpha^\pi$ with different $K^\pi$ in (a) \nuclide[31]{Mg}, (b) \nuclide[29]{Na}, (c) \nuclide[35]{Al}, and (d) \nuclide[31]{Na} from the IM-GCM calculation as a function of the quadrupole deformation parameter $\beta_2$.  
 }
  \label{fig:oddacoll} 
 \end{figure*}  

The Nilsson diagram of neutrons in \nuclide[30]{Ne}, shown in Fig.~\ref{fig:Ne30s}(b), helps elucidate the low-energy structures of \nuclide[29]{F} and \nuclide[29]{Ne}. In \nuclide[29]{F}, the unpaired proton occupies the $1/2^+[220]$ component of the $d_{5/2}$ orbital, yielding a near-spherical ground state with $J^\pi = 5/2^+$. The first excited state, by contrast, is strongly deformed with $J^\pi = 1/2^+$, as shown in Fig.\ref{fig:PES_F29_K1/2}.   These results align with experimental data and shell-model calculations~\cite{Doornenbal:2017}, despite the exclusion of continuum coupling effects. This suggests that the continuum effect contributes only moderately to the ground state of \nuclide[29]{F}~\cite{Kahlbow:2024}. Similarly, the ground state of \nuclide[29]{Ne} is nearly spherical, with the unpaired neutron in the $3/2^+[202]$ component of the $d_{3/2}$ orbital.  In contrast, \nuclide[31]{Mg}, with two additional protons compared to \nuclide[29]{Ne}, exhibits a strongly deformed ground state with spin-parity $1/2^+$, built on configurations with $K^\pi=1/2^+$ (see Fig.~\ref{fig:oddacoll} for the collective wave function distribution as a function of quadrupole deformation). The predominant configuration in the ground state of \nuclide[31]{Mg} occurs around $\beta_2 \approx 0.5$. At this deformation, the downward-sloping Nilsson orbital $1/2^-[330]$ drops below the upward-sloping orbitals $3/2^+[202]$ and $1/2^+[200]$, corresponding—relative to the spherical configuration—to the promotion of two neutrons from the $1d_{3/2}$ orbital to the $1f_{7/2}$ orbital. This dominant configuration is commonly referred to as a $2p$–$3h$ configuration with respect to the $N=20$ shell gap~\cite{Kimura:2007}. Moderately deformed negative-parity states arise from the occupation of the $1/2^-[330]$ component of the $1f_{7/2}$ orbital, instead of the $3/2^+[202]$ Nilsson orbital. However, the $3/2^+_1$ state in \nuclide[31]{Mg} is primarily characterized by a spherical (or weakly deformed) configuration, with the valence neutron occupying the $3/2^+[202]$ component of the $1d{3/2}$ orbital. It has been found experimentally that the  $3/2^+_1$ state is actually an isomer with a lifetime of 12 $n$s~\cite{NNDC}. According to our IM-GCM calculations with the $K$-mixing, we find $B(E2: 3/2_1^+ \to 1/2_1^+) = 5.2$ e$^2$fm$^4$ and $B(M1: 3/2_1^+ \to 1/2_1^+) = 3.6 \times 10^{-3} \mu_N^2$. The  corresponding transition probability $T$ (in units of $s^{-1}$)  is defined as \cite{Ring:1980},
\bsub
\begin{eqnarray}
    T(E2) & = & 1.223 \times 10^9 \cdot E^5_\gamma \cdot B(E2), \\
    T(M1) & = & 1.779 \times 10^{13} \cdot E^3_\gamma
    \cdot B(M1),
 \end{eqnarray}
 \esub
 where $E_\gamma$ is the energy difference (in MeV) between the two states. Using the transition strengths and energies from the IM-GCM calculation, we obtain $T^{-1}(E2) = 2.1 \ \mu s$ and $T^{-1}(M1) = 4.6 \ ns$ for the $3/2_1^+$ state. When using the experimental data for the transition energy, we find $T^{-1}(E2) = 0.5 \ ms$ and $T^{-1}(M1) = 125 \ ns$ for the same state. In both cases, the results indicate that the primary decay mode of the $3/2_1^+$ state is the $M1$ transition rather than the $E2$ transition. 
 
 \begin{table}[]
 \centering
     \tabcolsep=16pt
     \caption{The $B(E2: 5/2^+_1\to 3/2^+_1)$ (e$^2$fm$^4$) values in  \nuclide[29,31,33]{Na} from the IM-GCM calculations.  The results are compared  with those of shell model calculations based on the USDB interaction using the effective charges $e_\pi=1.5 e, e_\nu=0.5 e$ for protons and neutrons, respectively, as well as available experimental data, taken from Refs.~\cite{Naisoptop2023,Na31qs}.  }
     \begin{tabular}{lccc}
     \hline  
  \hline \\
       Isotopes  & USDB &  IM-GCM &   Exp. \\   \hline  
     \nuclide[29]{Na} & 96  & 63  &  93(16)  \\ 
         \nuclide[31]{Na} & 76 & 79  &   - \\ 
         \nuclide[33]{Na} & - & 97  &   - \\         
     \hline
  \hline           
     \end{tabular}
     \label{tab:Na_qmuE2}
 \end{table}
 
The low-lying energy spectra of \nuclide[29,31,33]{Na} present a significant challenge for many nuclear models, including {\em ab initio} coupled-cluster theory~\cite{Sun:2025lk} and valence-space shell models~\cite{Naisoptop2023} starting from nuclear chiral interactions. Figure~\ref{fig:spectra_odd_mass} displays the low-lying states of these three neutron-rich sodium isotopes from our IM-GCM calculations. The ground states of these isotopes share the same spin-parity $J^\pi=3/2^+$, indicating that they are primarily configurations with $K^\pi=3/2^+$. However, the quadrupole deformation of the dominant configuration varies, as shown in Fig.\ref{fig:oddacoll}. With increasing neutron number, the predominant configuration of the $3/2^+_1$ state shifts toward larger deformation, illustrating a shape transition from moderate to strong deformation. This shape transition is further reflected in the systematics of the $B(E2: 5/2^+_1 \to 3/2^+_1)$ values, as provided in Table~\ref{tab:Na_qmuE2}. Conversely, states with $K^\pi=1/2^+$ are generally dominated by weakly deformed configurations.
Among the three sodium isotopes, \nuclide[31]{Na} exhibits a low-energy structure characterized by coexisting distinct shapes with competing energies. The ground-state spin-parity $J^\pi=3/2^+$ of \nuclide[29,31,33]{Na} is primarily determined by the unpaired proton occupying the $3/2^+[211]$ component of the $1d_{5/2}$ orbital. 

Similarly, the ground states of \nuclide[31,33,35]{Al} share the same spin-parity, but with $J^\pi=5/2^+$ and $K^\pi=5/2^+$, corresponding to the valence neutron occupying the $5/2^+[202]$ component of the spherical $1d_{5/2}$ orbital. Additionally, the dominant configuration of the ground state shifts toward larger deformation regions. As shown in Fig.~\ref{fig:oddacoll}, the peak of the collective wave function for the ground state of \nuclide[35]{Al} is located around $\beta_2=0.5$.

\begin{figure}[tb]
 \centering
\includegraphics[width=8.5cm]{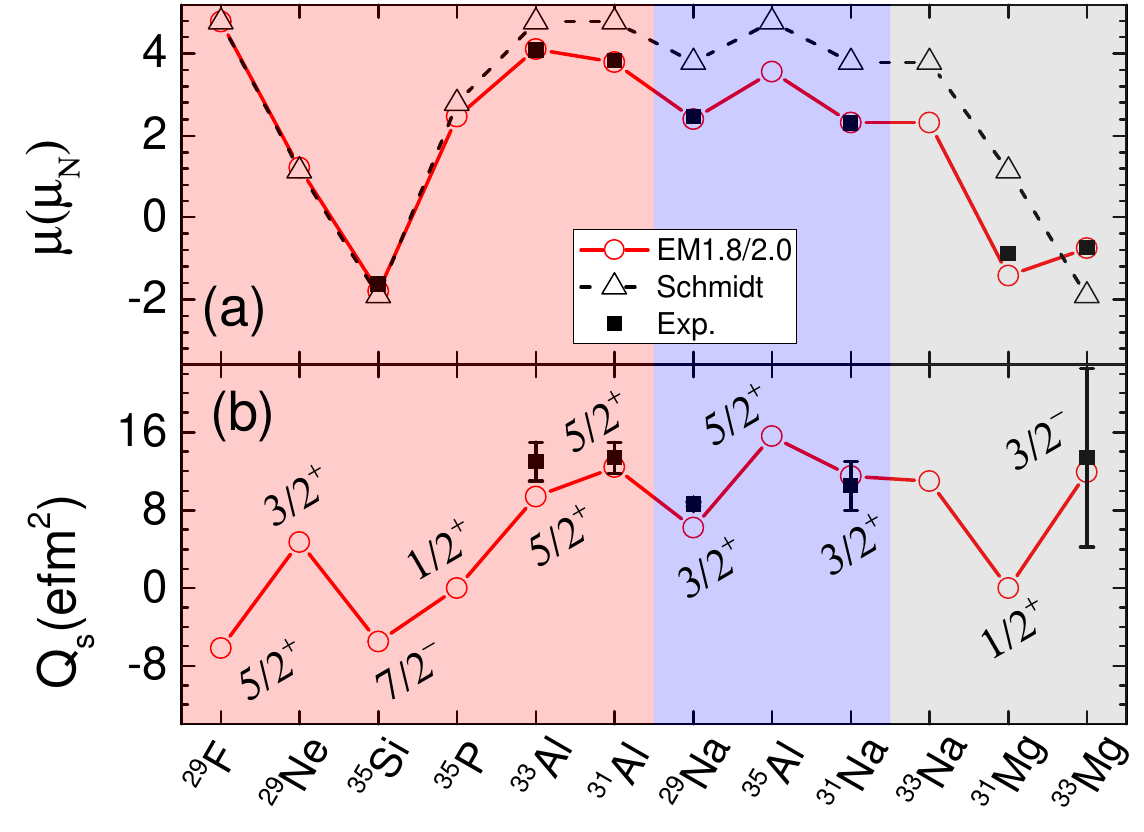}
\caption{(Color online)  The magnetic dipole moments $\mu$ (a) and spectroscopic quadrupole moments $Q_s$ (b)
of the ground states of odd-mass nuclei around $N=20$ from the IM-GCM calculation, in comparison with available data. The contribution of the induced two-body magnetic dipole moment is generally less than 1\%. The regions with red, blue, and gray  colors indicate the nuclei with weakly, moderate and strong deformations, respectively. The data are taken from Refs.~\cite{SheetA=31,sheeta=29,sheetA=33,sheeta=35,qsAl33,qsMg33,Na31qs}}
\label{fig:odda_qmu} 
 \end{figure}

Figure~\ref{fig:odda_qmu} compares the magnetic dipole moments $\mu$ and spectroscopic quadrupole moments $Q_s$  for the ground states of  odd-mass nuclei around $N=20$.   Both $Q_s$ and $\mu$ are surprisingly well reproduced by our IM-GCM calculation, considering the
uncertainties in $Q_s$ and $\mu$ arising from the choice of model parameters ($\eMax, \hbar\omega$ and flow parameter $s$) are estimated to be within 10\% and 1\%, respectively \cite{Zhou:2024_short}. In particular, one can see that the  Schmidt formula works rather well for  the magnetic dipole moments of weakly deformed nuclei, but fails for strongly deformed nuclei, highlighting the importance of configuration mixing beyond a single-particle picture. The $Q_s$ is highly sensitive to nuclear deformation. For even–even nuclei, the sign of $Q_s$ provides a straightforward indication of the intrinsic shape: a negative $Q_s$ corresponds to a prolate shape ($K=0$), while a positive $Q_s$ indicates an oblate shape ($K=J$). This simple correspondence, however, does not generally hold for odd-mass nuclei, where the sign of $Q_s$ depends on the specific combination of quantum numbers $(J,K)$. For the ground state of $\nuclide[33]{Mg}$ with $J=K=3/2$, the relevant Wigner $3j$-symbol in Eq.~(\ref{eq:Q_s}) ~\cite{Varshalovich:1988},
\beq
\begin{pmatrix}
J & 2 & J \\
-K & 0 & K
\end{pmatrix}
= \frac{3K^2 - J(J+1)}{\sqrt{(2J+1)(2J+3)(2J-1)J(J+1)}},
\eeq
is positive. Consequently, the spectroscopic quadrupole moment $Q_s$ has the same sign as the intrinsic quadrupole moment. In other words, the observed positive value of $Q_s$ for the ground state of $\nuclide[33]{Mg}$ reflects a predominantly prolate rather than oblate shape. The same reasoning applies to \nuclide[29,31,33]{Na} and \nuclide[35]{Al}, consistent with the conclusion obtained from the recent heavy-ion inelastic scattering measurement \cite{Salinas:2026}. The reasonable agreement with the corresponding data indicates that the  IM-GCM approach captures both single-particle and collective effects across the $N=20$ IOI region.

\subsection{The boundary of the $N=20$ island of inversion}

 \begin{figure}[tb]
 \centering
 \includegraphics[width=8.4cm]{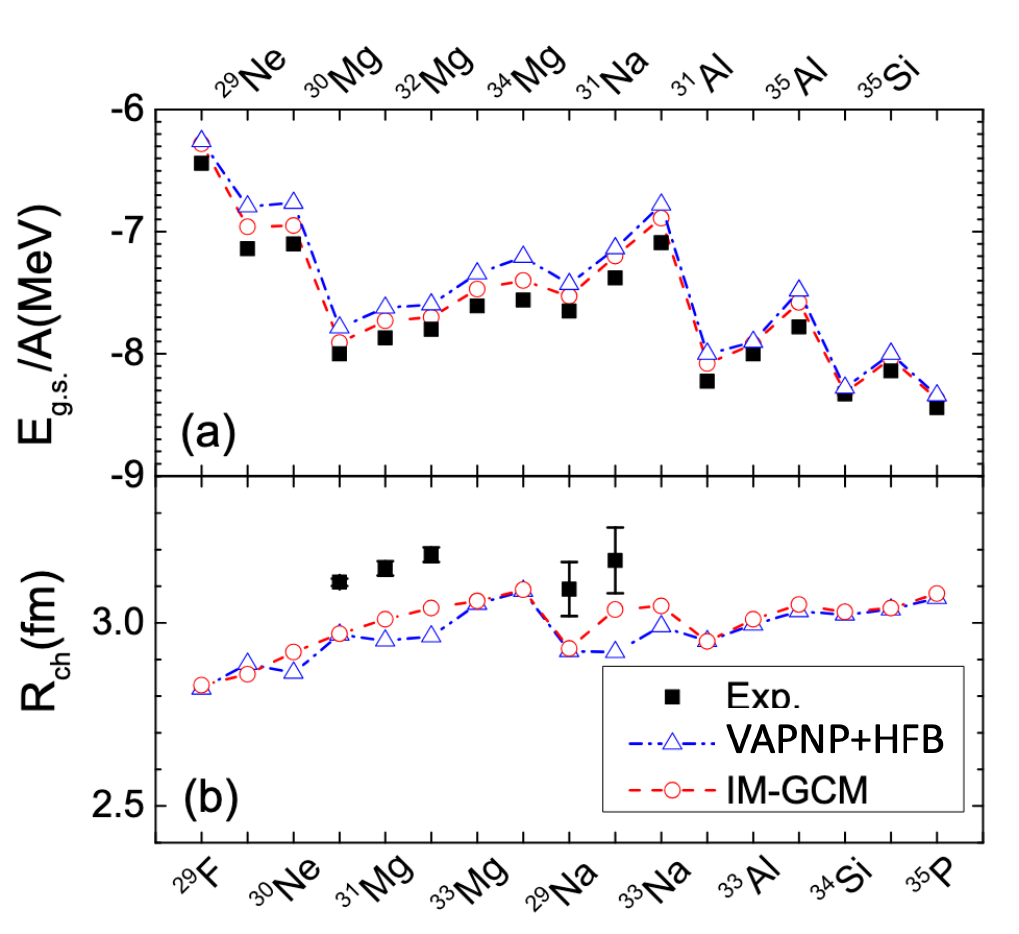} 
\caption{(Color online) (a) The average binding energy $E/A$ and (b) the rms charge radii $R_{\rm ch}$ for the ground states of nuclei around $N=20$ from  the calculations of  VAPNP+HFB and IM-GCM methods, in comparison with the corresponding data from Ref.\cite{sheetformass} and Refs.\cite{radii,Yordanov:2012}, respectively. } 
\label{fig:island} 
 \end{figure}

Figure~\ref{fig:island} displays the binding energy per nucleon 
$E/A$ and root-mean-square (rms) charge radii $R_{\rm ch}$ of the 17 nuclei around $N=20$.  The charge radii $R_{\rm ch}$ are evaluated in the following way~\cite{radiiexp}
\begin{eqnarray}
   R^2_{\rm ch}=  \langle r_{\rm pp}^2 \rangle + \langle R_{p}^2 \rangle +\cfrac{N}{Z}~\langle R_{n}^2 \rangle + \cfrac{3 \hbar^2}{4m_p^2c^2},
\end{eqnarray}
where  $3 \hbar^2/4m_p^2c^2 = 0.033$~fm$^2$, ~$\langle R_{p}^2 \rangle = 0.77$~fm$^2$
and  $\langle R_{n}^2 \rangle = -0.1149$ fm$^2$. The expectation value $\langle r_{\rm pp}^2 \rangle$ is the mean-square radius of the proton distribution inside a nucleus from VAPNP+HFB and IM-GCM calculations, respectively.   The systematic behaviors of $E/A$ and $R_{\rm ch}$ are reasonably reproduced, but the quantities are systematically over- and underestimated, respectively. The rms error for $E/A$ in the IM-GCM method is approximately 0.12 MeV, which is expected to decrease with a larger $\eMax$ value~\cite{Zhou:2024_short}.
Recently, the masses of $^{29,31,33}$Na were measured~\cite{Lykiardopoulou:2025}, from which the empirical shell gap can be extracted. The shell gap is defined as the difference between two-neutron separation energies separated by two neutrons, $
\Delta_{2n}(Z,N)=S_{2n}(Z,N)-S_{2n}(Z,N+2)$.
Our calculations yield $\Delta_{2n}(^{32}\mathrm{Mg}) = 4.469$ MeV and $\Delta_{2n}(^{31}\mathrm{Na}) = 3.142$ MeV, systematically larger than the experimental values of 1.100 MeV and 1.967 MeV, respectively. These calculated values should, however, be interpreted with caution. They remain sensitive to the choice of reference states and to the size of the model space. In particular, the use of different reference states for neighboring nuclei may introduce residual uncertainties in the finite differences, while the present ($e_{\max}=8$) model space is insufficient to fully converge their absolute binding energies. These effects may modify the extracted shell gaps by several MeV.

The charge radii are systematically underestimated by approximately 0.10 fm, mainly due to deficiencies in the chiral interaction EM1.8/2.0~\cite{Hebeler:2011PRC}. We note that the charge radii of Ne and Mg isotopes~\cite{Novario:2020} are better reproduced with the chiral interaction $\Delta$NNLO$_{\rm GO}$(450)~\cite{Jiang:2020}. Despite this overall offset, the systematic trends of the charge radii are well captured. Interestingly, a significant difference is observed between the predictions of VAPNP+HFB and IM-GCM for \nuclide[31]{Na}. This discrepancy is attributed to the coexistence of two nearly degenerate states with weakly and strongly deformed shapes. Compared to VAPNP+HFB, the IM-GCM calculation includes AMP, which allows the configuration with a large prolate deformed shape to gain more energy than the weakly deformed configuration. This leads to a shift of the ground state from the weakly deformed state to the strongly deformed state, resulting in an increase in the charge radius for \nuclide[31]{Na} that is consistent with the experimental trend.

 \begin{figure}[tb]
 \centering
 \includegraphics[width=8.4cm]{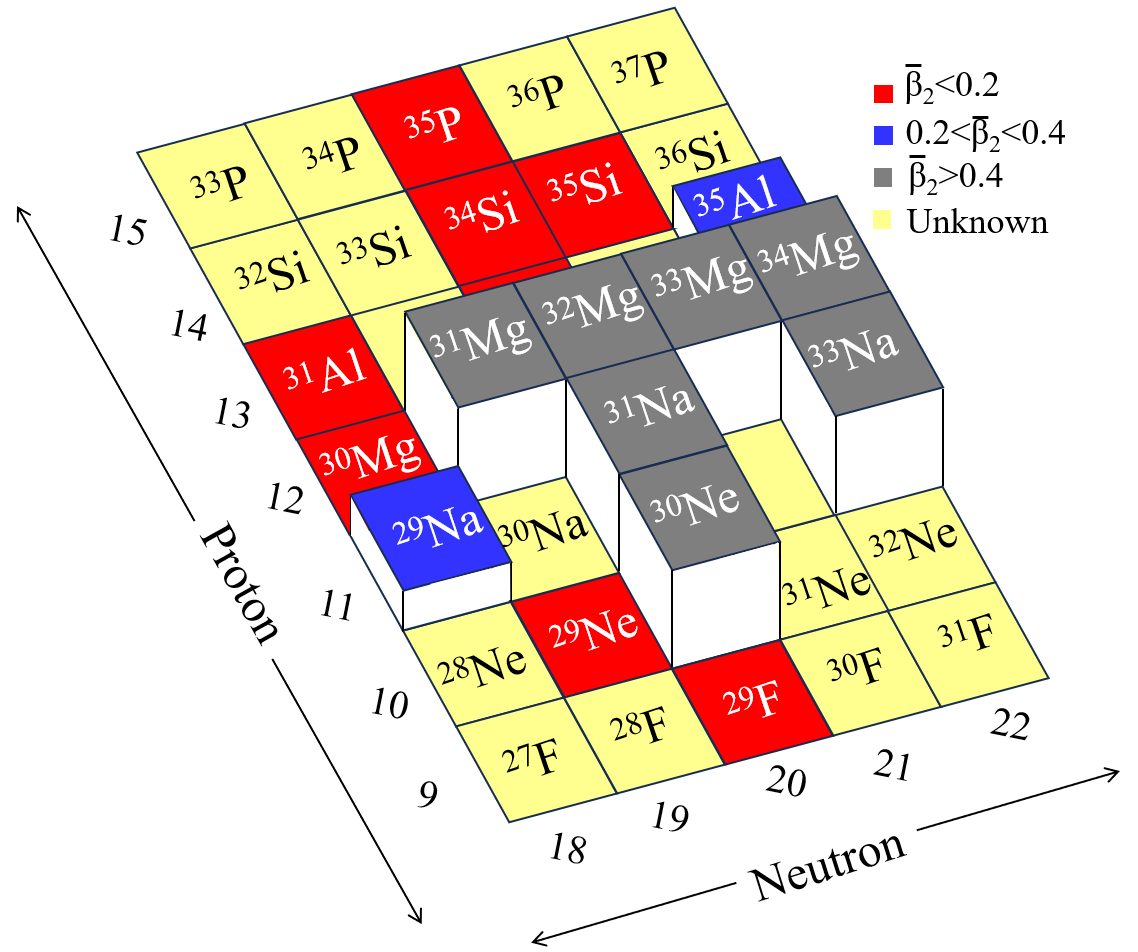} 
\caption{(Color online) A cartoon depiction of the mean quadrupole deformations (\ref{eq:mean_beta2}) of the ground states of nuclei around $N=20$. The nuclei predicted to be weakly, moderately, and strongly deformed are indicated in red, green, and gray, respectively. Nuclei not studied in this work, mostly odd-odd nuclei, are marked in yellow.} 
\label{fig:boundary_IOI} 
 \end{figure}

  Figure~\ref{fig:boundary_IOI} summarizes the ground-state deformations for the nuclei studied around $N=20$, classified into three regions: weak deformation ($\bar{\beta}_2 < 0.2$), moderate deformation ($0.2 < \bar{\beta}_2 < 0.4$), and strong deformation ($\bar{\beta}_2 > 0.4$).  The division of the regions is guided by the neutron Nilsson diagram  for \nuclide[30]{Ne}, as shown in Fig.~\ref{fig:Ne30s}. The $1/2^-[330]$ component of the $\nu f{7/2}$ orbital crosses with the $3/2^+[202]$ component of the $\nu d{3/2}$ orbital around $\beta_2 = 0.3$. Configurations with $\beta_2 > 0.3$ are usually interpreted as particle-hole excitations across the $N=20$ shell. Based on this classification, \nuclide[30]{Ne}, \nuclide[29,31,33]{Na}, \nuclide[31,32,33,34]{Mg}, and \nuclide[35]{Al} are within the IOI, while \nuclide[29]{F}, \nuclide[29]{Ne}, \nuclide[30]{Mg}, \nuclide[31,33]{Al}, \nuclide[34,35]{Si}, and \nuclide[35]{P} lie outside it.

\section{Conclusions}
\label{sec:summary}
 The in-medium generator coordinate method (IM-GCM) is an \emph{ab initio} technique that combines the In-Medium Similarity Renormalization Group (IMSRG) and quantum-number projected GCM to achieve a microscopic description of intrinsically deformed nuclei. In recent years, we have applied it to  describe low-lying states of medium-mass deformed nuclei, as well as nuclear matrix elements for neutrinoless double-beta decay, starting from chiral two- plus three-nucleon interactions. In this work, we present a comprehensive introduction to the IM-GCM framework, including expressions for consistently evolved transition operators, and demonstrate its capabilities by applying it to low-lying states in both even--even and odd-mass neutron-rich nuclei.

As a case study, we apply the method to nuclei in the vicinity of the $N=20$ island of inversion (IOI). A wide range of observables, including excitation spectra, electric quadrupole and magnetic dipole transition strengths, spectroscopic quadrupole moments, and magnetic dipole moments, are calculated and found to be in reasonable agreement with available experimental data, underscoring the robustness and predictive power of the IM-GCM approach. Our results further highlight the importance of both dynamical and static correlations in reproducing the low-energy structure of nuclei in this region. In the present framework, these correlations are incorporated efficiently through the IMSRG evolution and the PGCM configuration mixing, respectively. Based on a systematic analysis of the calculated spectra and electromagnetic properties, we determine the extent of the IOI and conclude that $\nuclide[30]{Ne}$, $\nuclide[29,31,33]{Na}$, $\nuclide[31,32,33,34]{Mg}$, and $\nuclide[35]{Al}$ lie within the island, whereas $\nuclide[29]{F}$, $\nuclide[29]{Ne}$, $\nuclide[30]{Mg}$, $\nuclide[31,33]{Al}$, $\nuclide[34,35]{Si}$, and $\nuclide[35]{P}$ are located outside it.
 
This study highlights the strong potential of the IM-GCM for describing low-lying states in nuclei exhibiting pronounced deformation and shape coexistence. Odd-odd nuclei, whose structures are expected to be even more intricate, were not considered in the present work. Work to extend the IM-GCM to such systems is in progress, and it will be essential for a more complete and definitive determination of the IOI boundary. In addition, continuum effects are likely to play an important role for nuclei along the south-eastern edge of the IOI, and incorporating these effects represents an important and challenging direction for future investigations. Finally, one of the major strengths of ab initio nuclear theory is its capability to quantify theoretical uncertainties in a systematic manner. Such an analysis has not yet been fully carried out in the present work and remains a challenging task for future investigations. Nevertheless, the development of robust uncertainty quantification frameworks for nuclear low-lying states \cite{Zhang:2024nqr} will be crucial for establishing predictive, high-precision descriptions of nuclei around the IOI and for fully exploiting the potential of ab initio approaches in this region.

 \section*{Data availability statement}
The data that support the findings of this article are openly available \cite{Data}.

\section*{acknowledgments} 
We thank B. Bally, C. F. Jiao and T. R. Rodríguez for helpful discussions. This work is supported in part by the National Natural Science Foundation of China (Grants No. 125B2108 and 12375119).  H.H. ackowledges the support of the U.S. Department of Energy, Office of Science, Office of Nuclear Physics under Award Nos. DE-SC0023516 and DE-SC0023175 (SciDAC-5 NUCLEI Collaboration).  This work was supported in part through computational resources and services provided by the Institute for Cyber-Enabled Research at Michigan State University, as well as the Beijing Super Cloud Computing Center (BSCC).

\appendix

\section{MR-IMSRG evolution of a tensor operator }
 \label{app:tensor_operator}

The expressions for evolving a tensor operator $\hat{T}^L_\mu(s)$ of rank $L$ in the single-reference IMSRG have been derived in Ref.~\cite{Parzuchowski:2017}. Here we extend the formalism to the multi-reference (MR)-IMSRG case, which contains  several additional terms depending on higher-body irreducible density matrices $\lambda^{[k]}$ of the preselected correlated reference state~\cite{Yao:2018PRC}. The transformed tensor operator $\hat{T}^L_\mu(s)$  is computed using the BCH formula 
\beqn
\hat T^L_\mu(s) &=& e^{\hat\Omega(s)}\hat{T}^L_\mu e^{-\hat\Omega(s)}
=\sum_{n=0}^\infty \frac{1}{n!} \hat{T}^{L(n)}_\mu \nonumber\\
&=&\hat T^L_\mu
+ [\hat\Omega(s), \hat{T}^L_\mu] 
+\frac{1}{2!}[\hat\Omega(s), [\hat\Omega(s), \hat{T}^L_\mu]]+\cdots
\eeqn

\begin{figure}[tb]
\centering
\includegraphics[width=8.5cm]{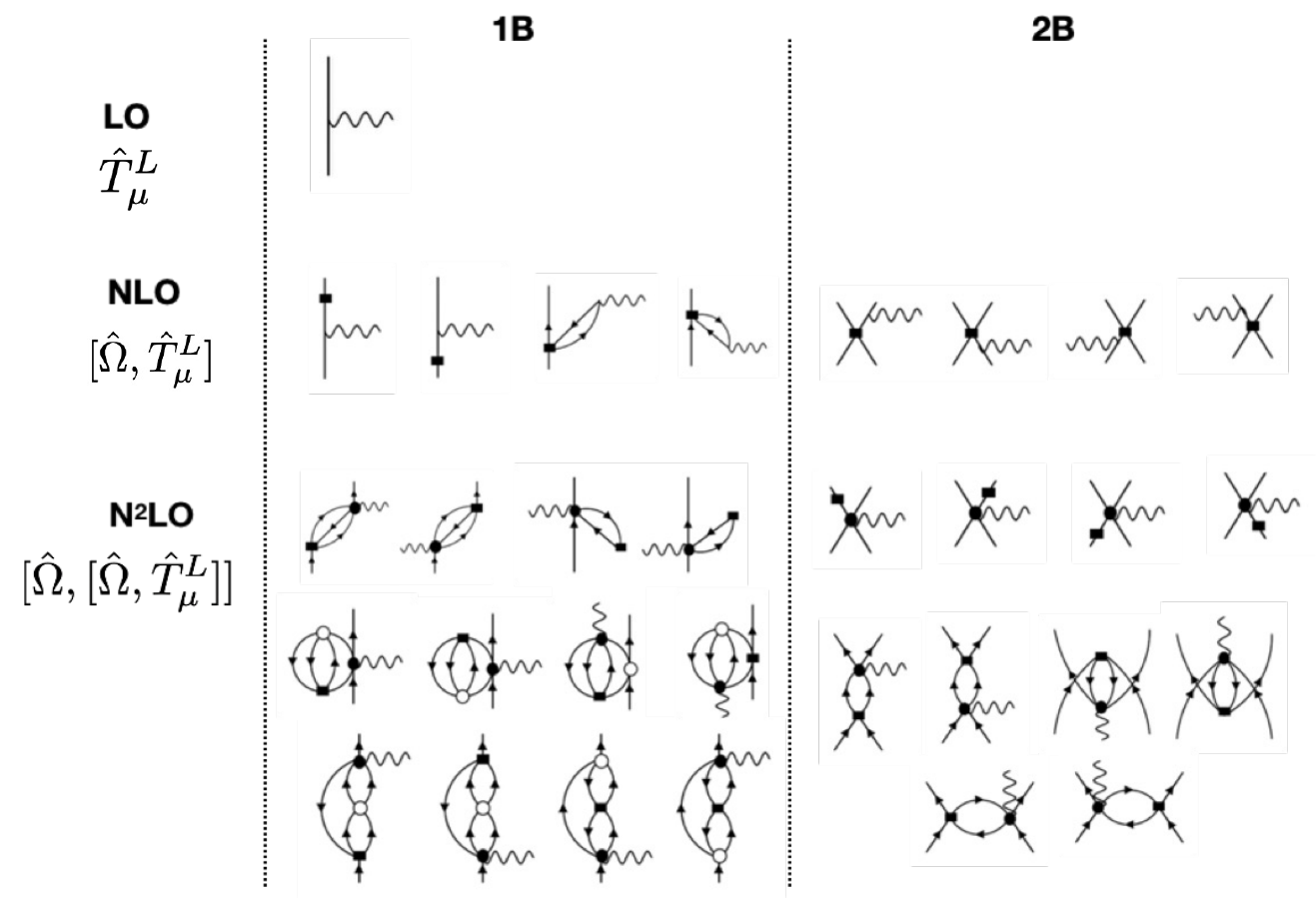}  
\caption{ (Color online) Antisymmetrized Goldstone diagrams of a tensor operator $\hat{T}^{L}_{\mu}(s)$ of rank $L$ under the NO2B approximation in the BCH expansion.  The filled squares are for one and two-body parts of $\hat\Omega$, hollow dots mark insertions of  two-body irreducible density $\lambda^{[2]}$,  wavy lines correspond to a tensor operator $\hat{T}^L_\mu$.  }
\label{Fig:tensor_operator_imsrg}
\end{figure}

Each term of this expression is depicted schematically in Fig.~\ref{Fig:tensor_operator_imsrg}. The Magnus operator $\hat
\Omega$ is of scalar type and has been determined by solving the flow equation (\ref{flow_omega}). The  tensor operator $\hat{T}^{L(n)}_{\mu}(s)$  of rank $L$   is defined as the nested commutators,
 \beqn
\hat{T}^{L(n)}_{\mu}(s) 
= [\hat\Omega(s),   \hat{T}^{L(n-1)}_{\mu}(s)],
\eeqn 
with $\hat{T}^{L(0)}_{\mu}(s)=\hat{T}^L_\mu$. Like all other many-body operators, the tensor operator $\hat{T}^{L(n)}_{\mu}(s)$ is also truncated up to normal-ordered two-body terms (NO2B), where the normal-ordered one-body term is 
\beqn
 \hat{T}^{L (n)}_\mu(1B)
=\sum_{ac} \hat L^{-1} \langle a\vert\vert T^{L (n)}\vert\vert c\rangle  [a^\dagger_a \tilde a_c]^{L}_{\mu}\,,
\eeqn
where the reduced one-body matrix element can be decomposed into terms depending on different ranks of irreducible densities, with the truncation of $k\leq2$,
\beqn
  \langle a \vert\vert T^{L(n)} \vert\vert c\rangle 
 &=&\sum_{k=0,1,2}\langle a \vert\vert T^{L(n)} \vert\vert c\rangle(\lambda^{[k]}),
 \eeqn
where $\lambda^{[k]}$ denotes the irreducible density of rank $k$, defined in Eq.(\ref{cumulants}).

 Introducing the following notation for the reduced matrix elements of the tensor operator, 
\beqn
T^L_{pq} &\equiv& \langle p \vert\vert T^L \vert\vert q\rangle,\\
T^{(J_1J_2)L}_{pq,rs} &\equiv& \langle (pq) J_1 \vert\vert T^L\vert \vert (rs)J_2\rangle,
\eeqn
we find the following expression for the leading term, which does not depend on the density:
 \beqn 
 &&\langle a \vert\vert T^{L(n)} \vert\vert c\rangle(\lambda^{[0]})\nonumber\\
&=&\sum_{b} \left(  \delta_{j_aj_b}      \Omega_{ab}  T^{L (n-1)}_{bc}  
   -  \delta_{j_bj_c} \Omega_{bc} T^{L(n-1)}_{ab}   \right).
\eeqn
The second term depends on the one-body density $\lambda^{[1]}$, which is diagonalized to obtain the (fractional) occupation numbers $n_p$ and $\bar{n}_p = 1-n_p$:
\beqn 
&& \langle a \vert\vert T^{L(n)} \vert\vert c\rangle(\lambda^{[1]})\nonumber\\
&=&-\sum_{bd}(n_b-n_d) \left[\Omega^L_{a\bar cb\bar d} T^{L(n-1)}_{bd}
 - \hat j_b T^{(L 0)L (n-1)}_{a\bar cb\bar d}  \Omega_{bd} \right]\nonumber\\
 &&+\dfrac{1}{2} 
   \sum_{ebd}\sum_{J_1J_2} (n_d\bar n_e\bar n_b +\bar n_d n_en_b) 
    (-1)^{j_c+j_d+L-J_1}  \hat J_1\hat J_2  \nonumber\\
 &&\times\left\{\begin{array}{ccc}
J_1 & J_2 & L \\
 j_c & j_a & j_d \\
 \end{array} 
 \right\}   
 \left[ \Omega^{J_1}_{daeb}   T^{(J_1J_2)L(n-1)}_{ebdc}
 - T^{(J_1J_2)L (n-1)}_{daeb}  \Omega^{J_2}_{ebdc} \right].\nonumber\\
\eeqn
The final term contains the two-body irreducible density $\lambda^{[2]}$, and involves a significantly more complex calculation,
\beqn
 && \langle a \vert\vert T^{L(n)} \vert\vert c\rangle(\lambda^{[2]})\nonumber\\
 &=&
  \sum_{J_1J_2} \sum_{a}   
    \left[  T^{(J_1J_2)L (n-1)} \cdot   \lambda^{J_2} \cdot
    \Omega^{J_2} \right]_{a\bar 2a \bar 1} 
    \nonumber\\
    &&\times (-1)^{J_1+j_a-j_2+1}\hat J_1\hat J_2 
     \left\{\begin{array}{ccc}
 j_1 & j_a & J_2  \\
 J_1 & L  & j_2 \\
 \end{array}
 \right\}  \nonumber\\
    &&  +   \sum_{J_1J_2}   \sum_{a}  
     \left[  T^{(J_1J_2)L(n-1)} \cdot   \lambda^{J_2} \cdot
    \Omega^{J_2} \right]_{1\bar a 2 \bar a} 
 \nonumber\\
    &&\times (-1)^{J_2+L+j_a+j_1} \hat J_1 \hat J_2 
     \left\{\begin{array}{ccc}
 j_2 & j_a & J_2  \\
 J_1 & L  & j_1 \\
 \end{array}
 \right\}  \nonumber\\
 &&+ \sum_{J_1J_2} \sum_{a}   
    \left[  T^{(J_1J_2)L(n-1)} \cdot   \lambda^{J_2} \cdot
    \Omega^{J_2} \right]_{a\bar 2a \bar 1} 
  \nonumber\\
    &&\times  (-1)^{J_1+j_a-j_2+1}\hat J_1\hat J_2 
     \left\{\begin{array}{ccc}
 j_1 & j_a & J_2  \\
 J_1 & L  & j_2 \\
 \end{array}
 \right\}  \nonumber\\
    &&  +   \sum_{J_1J_2}   \sum_{a}  
     \left[  T^{(J_1J_2)L(n-1)} \cdot   \lambda^{J_2} \cdot
    \Omega^{J_2} \right]_{1\bar a 2 \bar a} 
  \nonumber\\
    &&\times(-1)^{J_2+L+j_a+j_1} \hat J_1 \hat J_2 
     \left\{\begin{array}{ccc}
 j_2 & j_a & J_2  \\
 J_1 & L  & j_1 \\
 \end{array}
 \right\}  \nonumber\\
 &&- \dfrac{1}{2}    \sum_{abcdq}    \sum_{J} 
    \left[ \Omega^{J} \cdot \lambda^J \right]_{abbq} 
    T^{(L0)L(n-1)}_{1\bar 2a\bar q}    
    (-1)^{j_b+j_q-J}   \hat J^2  \hat j^{-1}_{a}\delta_{j_aj_q} 
      \nonumber\\
 &&
 -  \dfrac{1}{2}      \sum_{abcds} 
          \sum_{J}   \left[\Omega^{J}\cdot  \lambda^J\right]_{cdds}  T^{(0L)L(n-1)}_{c\bar s2\bar 1}  
          (-1)^{j_c+j_d+J} \hat J^2 \hat j^{-1}_c   \delta_{j_c j_s}
    \nonumber\\
    &&
 - \dfrac{1}{2}    \sum_{adpqs}\sum_{J_3J_4} 
  \left[\lambda^{J_3}\cdot T^{(J_3J_4)L(n-1)}\right]_{sdas}   \Omega^{L}_{a\bar d 2\bar 1} \nonumber\\
    &&\times 
    (-1)^{J_3+L+J_4+j_a+j_d+j_1+j_2}\hat J_3  \hat J_4
   \left\{\begin{array}{ccc}
 j_a & j_s & J_4 \\
 J_3 & L & j_d \\
 \end{array} 
 \right\}   \nonumber\\
 &&
 - \dfrac{1}{2}   \sum_{bcqrs}\sum_{J_3J_4}
       \left[T^{(J_3J_4)L(n-1)}\cdot  \lambda^{J_4}\right]_{cqqb}  \Omega^{L}_{c\bar b1\bar 2 } 
         (-1)^{L+j_c+j_b} \nonumber\\
    &&\times \hat J_3   \hat J_4  
   \left\{\begin{array}{ccc}
 j_c & j_q & J_3 \\
 J_4 & L & j_b \\
 \end{array} 
 \right\}, 
\eeqn
 where the symbol ``$\cdot$'' represents the matrix product of two matrices of two-body operators,
 \beq 
 (A \cdot B)_{abcd} = \sum_{ef}A_{abef}B_{efcd}.
 \eeq
 The expression for the $J$-coupled irreducible two-body density $\lambda^J_{abcd}$ has been given in Eq.(C2) of Ref.~\cite{Yao:2018PRC}. The 
Pandya transformation  for the reduced matrix element of the scalar $\Omega$,
 \beqn
 \bar\Omega^J_{1\bar 2 3\bar 4}
 =-\sum_{J'} \hat J^{'2}
   \left\{\begin{array}{ccc}
 j_1 & j_2 & J  \\
 j_3 & j_4  & J' \\
 \end{array}
 \right\}
 \Omega^{J'}_{14 32}\,,
 \eeqn
 and that for the reduced matrix elements of a two-body tensor operator of rank $\lambda$,
 \beqn
 T^{(J_3J_4)L}_{p\bar s r\bar q}
 &=& -\sum_{J_1J_2} (-1)^{J_2+J_4+j_q+j_s}
 \hat J_1 \hat J_2 \hat J_3 \hat J_4
  \left\{\begin{array}{ccc}
 j_p & j_s & J_3  \\
 j_q & j_r  & J_4\\
 J_1 & J_2 & L
 \end{array}
 \right\}
 T^{(J_1J_2)L}_{pqrs}.\nonumber\\
 \eeqn
  are used in the calculation. The expression for the NO2B tensor operator is identical to the previously derived expression in Ref.~\cite{Parzuchowski:2017}, since only up to one-body density matrices can appear in the commutators. The only difference is that occupation numbers can take any value between 0 and 1 in our case (cf. Ref.~ \cite{Hergert:2017PS}).

\section{Transition strength of a two-body tensor operator }
 \label{app:Two_body_density}
A general two-body tensor operator $T^{[2]}_{\lambda\mu}$ can be written in $J$-coupled form and in second quantization as
\beqn
\hat{T}_{\lambda\mu}^{[2]} & = & -\cfrac{1}{4} \sum_{J_1J_2}\sum_{pqrs}\hat{\lambda}^{-1}\bar{T}_{pqrs}^{(J_1J_2)\lambda}
\left[ [a_p^\dag a_q^\dag]^{J_1} [\tilde{a}_r \tilde{a}_s]^{J_2}\right]_\mu^\lambda\nonumber\\
&\equiv& \cfrac{1}{4} \sum_{J_1J_2}\sum_{pqrs}\hat{\lambda}^{-1}\bar{T}_{pqrs}^{(J_1J_2)\lambda} \hat{\rho}_{pqrs}^{(J_1J_2)\lambda \mu}
\eeqn
 where we define the two-body transition density operator as
 \beqn
\hat{\rho}_{pqrs}^{(J_1J_2)\lambda \mu} & \equiv & 
- \left[ [a_p^\dag a_q^\dag]^{J_1} [\tilde{a}_r \tilde{a}_s]^{J_2}\right]_\mu^\lambda = \sum_{M_1M_2} (-1)^{J_2-M_2}C_{M_1-M_2\mu}^{J_1J_2\lambda} \nonumber\\
& &\times \sum_{m_pm_qm_rm_s}C_{m_pm_qM_1}^{j_pj_qJ_1}C_{m_rm_sM_2}^{j_rj_sJ_2}a_p^\dag a_q^\dag a_s a_r.
\eeqn

 In PGCM calculations,  the  matrix element for the transition by the two-body tensor operator $\hat{T}^{[2]}_{\lambda\mu}$  from an initial state $J_i$ to the final state $J_f$ can be written in terms of its reduced matrix element,
\beqn
&&\langle J_f M_f\pi_f ,\alpha_f |\hat{T}^{[2]}_{\lambda\mu}| J_i M_i\pi_i ,\alpha_i \rangle \nonumber\\
&=& \cfrac{(-1)^{2\lambda}}{\sqrt{2J_f+1}}\langle J_i M_i \lambda \mu|J_fM_f\rangle \langle J_f\pi_f,\alpha_f||\hat{T}^{[2]}_\lambda||J_i\pi_i,\alpha_i\rangle,
\eeqn
 and $E\lambda$ transition strength is finally given by
\beqn
 B(T\lambda: J_i\alpha_i\pi_i\to  J_f\alpha_f\pi_f)
=\cfrac{1}{2J_i+1}
\left |
\langle J_f\pi_f,\alpha_f||\hat{T}^{[2]}_\lambda||J_i\pi_i,\alpha_i\rangle
\right|^2.
\eeqn
For an odd-mass nucleus, the reduced matrix element  reads
\beqn
&&\langle J_f\pi_f,\alpha_f||\hat{T}^{[2]}_\lambda||J_i\pi_i,\alpha_i\rangle \nonumber\\
&=&\sum_{c_ic_f}f_{\alpha_f}^{J_f\pi_f}(c_f)f_{\alpha_i}^{J_i\pi_i}(c_i) 
\langle NZJ_f\pi_f,c_f||\hat{T}^{[2]}_\lambda||NZJ_i\pi_i,c_i\rangle
\eeqn
with the configuration-dependent one determined by
\beqn
&&\langle NZJ_f\pi_f,c_f||\hat{T}^{[2]}_\lambda||NZJ_i\pi_i,c_i\rangle\nonumber\\
&=&\cfrac{1}{4}\sum_{J_1J_2}\sum_{pqrs}\hat{\lambda}^{-1} \bar{T}_{pqrs}^{(J_1J_2)\lambda}\rho_{pqrs}^{(J_1J_2)\lambda}(NZJ_f\pi_fc_f,J_i\pi_ic_i).
\eeqn
The two-body transition density is given by
\beqn
&&\rho_{pqrs}^{J_1J_2\lambda}(NZJ_f\pi_fc_f,J_i\pi_ic_i)\nonumber\\
&=& \cfrac{\hat{J}_f^2\hat{J}_i^2}{8\pi^2}(-1)^{J_f-K_f} \sum_{\mu K_i'}\left(\begin{array}{ccc}
 J_f & \lambda & J_i  \\
 -K_f & \mu  & K_i' \\
 \end{array}\right)\nonumber\\
 &&\times \int d\Omega D_{K_i'K_i}^{J_i*}(\Omega) \langle \Phi^{(OA)}_{\kappa_f}(\mathbf{q}_f)|\hat \rho_{pqrs}^{(J_1J_2)\lambda\mu} \hat R(\Omega)\hat P^N \hat P^Z |\Phi^{(OA)}_{\kappa_i}(\mathbf{q}_i)\rangle\nonumber\\
\eeqn
with
\beqn
&&\langle \Phi^{(OA)}_{\kappa_f}(\mathbf{q}_f)|\hat \rho_{pqrs}^{(J_1J_2)\lambda\mu} \hat R(\Omega)\hat P^N \hat P^Z |\Phi^{(OA)}_{\kappa_i}(\mathbf{q}_i)\rangle\nonumber \\
&=& \sum_{M_1M_2}(-1)^{J_2-M_2}C_{M_1-M_2\mu}^{J_1J_2\lambda}\sum_{m_pm_qm_rm_s}C_{m_pm_qM_1}^{j_pj_qJ_1}C_{m_rm_sM_2}^{j_rj_sJ_2}\nonumber\\
&&\times\langle \Phi^{(OA)}_{\kappa_f}(\mathbf{q}_f)|a_p^\dag a_q^\dag a_s a_r \hat R(\Omega)\hat P^N \hat P^Z |\Phi^{(OA)}_{\kappa_i}(\mathbf{q}_i)\rangle.
\eeqn


\begin{thebibliography}{106}%
\makeatletter
\providecommand \@ifxundefined [1]{%
 \@ifx{#1\undefined}
}%
\providecommand \@ifnum [1]{%
 \ifnum #1\expandafter \@firstoftwo
 \else \expandafter \@secondoftwo
 \fi
}%
\providecommand \@ifx [1]{%
 \ifx #1\expandafter \@firstoftwo
 \else \expandafter \@secondoftwo
 \fi
}%
\providecommand \natexlab [1]{#1}%
\providecommand \enquote  [1]{``#1''}%
\providecommand \bibnamefont  [1]{#1}%
\providecommand \bibfnamefont [1]{#1}%
\providecommand \citenamefont [1]{#1}%
\providecommand \href@noop [0]{\@secondoftwo}%
\providecommand \href [0]{\begingroup \@sanitize@url \@href}%
\providecommand \@href[1]{\@@startlink{#1}\@@href}%
\providecommand \@@href[1]{\endgroup#1\@@endlink}%
\providecommand \@sanitize@url [0]{\catcode `\\12\catcode `\$12\catcode
  `\&12\catcode `\#12\catcode `\^12\catcode `\_12\catcode `\%12\relax}%
\providecommand \@@startlink[1]{}%
\providecommand \@@endlink[0]{}%
\providecommand \url  [0]{\begingroup\@sanitize@url \@url }%
\providecommand \@url [1]{\endgroup\@href {#1}{\urlprefix }}%
\providecommand \urlprefix  [0]{URL }%
\providecommand \Eprint [0]{\href }%
\providecommand \doibase [0]{http://dx.doi.org/}%
\providecommand \selectlanguage [0]{\@gobble}%
\providecommand \bibinfo  [0]{\@secondoftwo}%
\providecommand \bibfield  [0]{\@secondoftwo}%
\providecommand \translation [1]{[#1]}%
\providecommand \BibitemOpen [0]{}%
\providecommand \bibitemStop [0]{}%
\providecommand \bibitemNoStop [0]{.\EOS\space}%
\providecommand \EOS [0]{\spacefactor3000\relax}%
\providecommand \BibitemShut  [1]{\csname bibitem#1\endcsname}%
\let\auto@bib@innerbib\@empty
\bibitem [{\citenamefont {Thibault}\ \emph {et~al.}(1975)\citenamefont
  {Thibault}, \citenamefont {Klapisch}, \citenamefont {Rigaud}, \citenamefont
  {Poskanzer}, \citenamefont {Prieels}, \citenamefont {Lessard},\ and\
  \citenamefont {Reisdorf}}]{Thibault:1975}%
  \BibitemOpen
  \bibfield  {author} {\bibinfo {author} {\bibfnamefont {C.}~\bibnamefont
  {Thibault}}, \bibinfo {author} {\bibfnamefont {R.}~\bibnamefont {Klapisch}},
  \bibinfo {author} {\bibfnamefont {C.}~\bibnamefont {Rigaud}}, \bibinfo
  {author} {\bibfnamefont {A.~M.}\ \bibnamefont {Poskanzer}}, \bibinfo {author}
  {\bibfnamefont {R.}~\bibnamefont {Prieels}}, \bibinfo {author} {\bibfnamefont
  {L.}~\bibnamefont {Lessard}}, \ and\ \bibinfo {author} {\bibfnamefont
  {W.}~\bibnamefont {Reisdorf}},\ }\href {\doibase 10.1103/PhysRevC.12.644}
  {\bibfield  {journal} {\bibinfo  {journal} {Phys. Rev. C}\ }\textbf {\bibinfo
  {volume} {12}},\ \bibinfo {pages} {644} (\bibinfo {year} {1975})}\BibitemShut
  {NoStop}%
\bibitem [{\citenamefont {Detraz}\ \emph {et~al.}(1979)\citenamefont {Detraz},
  \citenamefont {Guillemaud}, \citenamefont {Huber}, \citenamefont {Klapisch},
  \citenamefont {Langevin}, \citenamefont {Naulin}, \citenamefont {Thibault},
  \citenamefont {Carraz},\ and\ \citenamefont {Touchard}}]{Detraz:1979}%
  \BibitemOpen
  \bibfield  {author} {\bibinfo {author} {\bibfnamefont {C.}~\bibnamefont
  {Detraz}}, \bibinfo {author} {\bibfnamefont {D.}~\bibnamefont {Guillemaud}},
  \bibinfo {author} {\bibfnamefont {G.}~\bibnamefont {Huber}}, \bibinfo
  {author} {\bibfnamefont {R.}~\bibnamefont {Klapisch}}, \bibinfo {author}
  {\bibfnamefont {M.}~\bibnamefont {Langevin}}, \bibinfo {author}
  {\bibfnamefont {F.}~\bibnamefont {Naulin}}, \bibinfo {author} {\bibfnamefont
  {C.}~\bibnamefont {Thibault}}, \bibinfo {author} {\bibfnamefont {L.~C.}\
  \bibnamefont {Carraz}}, \ and\ \bibinfo {author} {\bibfnamefont
  {F.}~\bibnamefont {Touchard}},\ }\href {\doibase 10.1103/PhysRevC.19.164}
  {\bibfield  {journal} {\bibinfo  {journal} {Phys. Rev. C}\ }\textbf {\bibinfo
  {volume} {19}},\ \bibinfo {pages} {164} (\bibinfo {year} {1979})}\BibitemShut
  {NoStop}%
\bibitem [{\citenamefont {Motobayashi}\ \emph {et~al.}(1995)\citenamefont
  {Motobayashi} \emph {et~al.}}]{Motobayashi:1995Mg32}%
  \BibitemOpen
  \bibfield  {author} {\bibinfo {author} {\bibfnamefont {T.}~\bibnamefont
  {Motobayashi}} \emph {et~al.},\ }\href {\doibase
  10.1016/0370-2693(95)00012-A} {\bibfield  {journal} {\bibinfo  {journal}
  {Phys. Lett. B}\ }\textbf {\bibinfo {volume} {346}},\ \bibinfo {pages} {9}
  (\bibinfo {year} {1995})}\BibitemShut {NoStop}%
\bibitem [{\citenamefont {Pritychenko}\ \emph {et~al.}(1999)\citenamefont
  {Pritychenko} \emph {et~al.}}]{Pritychenko:1999Mg32}%
  \BibitemOpen
  \bibfield  {author} {\bibinfo {author} {\bibfnamefont {B.~V.}\ \bibnamefont
  {Pritychenko}} \emph {et~al.},\ }\href {\doibase
  10.1016/S0370-2693(99)00850-3} {\bibfield  {journal} {\bibinfo  {journal}
  {Phys. Lett. B}\ }\textbf {\bibinfo {volume} {461}},\ \bibinfo {pages} {322}
  (\bibinfo {year} {1999})},\ \bibinfo {note} {[Erratum: Phys.Lett.B 467,
  309--309 (1999)]}\BibitemShut {NoStop}%
\bibitem [{\citenamefont {Yoneda}\ \emph {et~al.}(2001)\citenamefont {Yoneda}
  \emph {et~al.}}]{Yoneda:2001Mg34}%
  \BibitemOpen
  \bibfield  {author} {\bibinfo {author} {\bibfnamefont {K.}~\bibnamefont
  {Yoneda}} \emph {et~al.},\ }\href {\doibase 10.1016/S0370-2693(01)00025-9}
  {\bibfield  {journal} {\bibinfo  {journal} {Phys. Lett. B}\ }\textbf
  {\bibinfo {volume} {499}},\ \bibinfo {pages} {233} (\bibinfo {year}
  {2001})}\BibitemShut {NoStop}%
\bibitem [{\citenamefont {Gade}\ \emph {et~al.}(2007)\citenamefont {Gade} \emph
  {et~al.}}]{Gade:2007Mg3436}%
  \BibitemOpen
  \bibfield  {author} {\bibinfo {author} {\bibfnamefont {A.}~\bibnamefont
  {Gade}} \emph {et~al.},\ }\href {\doibase 10.1103/PhysRevLett.99.072502}
  {\bibfield  {journal} {\bibinfo  {journal} {Phys. Rev. Lett.}\ }\textbf
  {\bibinfo {volume} {99}},\ \bibinfo {pages} {072502} (\bibinfo {year}
  {2007})}\BibitemShut {NoStop}%
\bibitem [{\citenamefont {Heyde}\ and\ \citenamefont
  {Wood}(1991)}]{Heyde1991Mg3230}%
  \BibitemOpen
  \bibfield  {author} {\bibinfo {author} {\bibfnamefont {K.}~\bibnamefont
  {Heyde}}\ and\ \bibinfo {author} {\bibfnamefont {J.~L.}\ \bibnamefont
  {Wood}},\ }\href {https://api.semanticscholar.org/CorpusID:121947991}
  {\bibfield  {journal} {\bibinfo  {journal} {Journal of Physics G}\ }\textbf
  {\bibinfo {volume} {17}},\ \bibinfo {pages} {135} (\bibinfo {year}
  {1991})}\BibitemShut {NoStop}%
\bibitem [{\citenamefont {Neyens}\ \emph {et~al.}(2005)\citenamefont {Neyens}
  \emph {et~al.}}]{Neyens:2005Mg31}%
  \BibitemOpen
  \bibfield  {author} {\bibinfo {author} {\bibfnamefont {G.}~\bibnamefont
  {Neyens}} \emph {et~al.},\ }\href {\doibase 10.1103/PhysRevLett.94.022501}
  {\bibfield  {journal} {\bibinfo  {journal} {Phys. Rev. Lett.}\ }\textbf
  {\bibinfo {volume} {94}},\ \bibinfo {pages} {022501} (\bibinfo {year}
  {2005})}\BibitemShut {NoStop}%
\bibitem [{\citenamefont {Doornenbal}\ \emph {et~al.}(2009)\citenamefont
  {Doornenbal} \emph {et~al.}}]{Doornenbal:2009Ne32}%
  \BibitemOpen
  \bibfield  {author} {\bibinfo {author} {\bibfnamefont {P.}~\bibnamefont
  {Doornenbal}} \emph {et~al.},\ }\href {\doibase
  10.1103/PhysRevLett.103.032501} {\bibfield  {journal} {\bibinfo  {journal}
  {Phys. Rev. Lett.}\ }\textbf {\bibinfo {volume} {103}},\ \bibinfo {pages}
  {032501} (\bibinfo {year} {2009})},\ \Eprint {http://arxiv.org/abs/0906.3775}
  {arXiv:0906.3775 [nucl-ex]} \BibitemShut {NoStop}%
\bibitem [{\citenamefont {Michimasa}\ \emph {et~al.}(2014)\citenamefont
  {Michimasa} \emph {et~al.}}]{Michimasa:2014Ne30}%
  \BibitemOpen
  \bibfield  {author} {\bibinfo {author} {\bibfnamefont {S.}~\bibnamefont
  {Michimasa}} \emph {et~al.},\ }\href {\doibase 10.1103/PhysRevC.89.054307}
  {\bibfield  {journal} {\bibinfo  {journal} {Phys. Rev. C}\ }\textbf {\bibinfo
  {volume} {89}},\ \bibinfo {pages} {054307} (\bibinfo {year}
  {2014})}\BibitemShut {NoStop}%
\bibitem [{\citenamefont {Doornenbal}\ \emph
  {et~al.}(2016{\natexlab{a}})\citenamefont {Doornenbal} \emph
  {et~al.}}]{Doornenbal:2016Ne30}%
  \BibitemOpen
  \bibfield  {author} {\bibinfo {author} {\bibfnamefont {P.}~\bibnamefont
  {Doornenbal}} \emph {et~al.},\ }\href {\doibase 10.1103/PhysRevC.93.044306}
  {\bibfield  {journal} {\bibinfo  {journal} {Phys. Rev. C}\ }\textbf {\bibinfo
  {volume} {93}},\ \bibinfo {pages} {044306} (\bibinfo {year}
  {2016}{\natexlab{a}})}\BibitemShut {NoStop}%
\bibitem [{\citenamefont {Pritychenko}\ \emph {et~al.}(2001)\citenamefont
  {Pritychenko}, \citenamefont {Glasmacher}, \citenamefont {Brown},
  \citenamefont {Cottle}, \citenamefont {Ibbotson}, \citenamefont {Kemper},
  \citenamefont {Riley},\ and\ \citenamefont {Scheit}}]{Pritychenko:2001Na31}%
  \BibitemOpen
  \bibfield  {author} {\bibinfo {author} {\bibfnamefont {B.~V.}\ \bibnamefont
  {Pritychenko}}, \bibinfo {author} {\bibfnamefont {T.}~\bibnamefont
  {Glasmacher}}, \bibinfo {author} {\bibfnamefont {B.~A.}\ \bibnamefont
  {Brown}}, \bibinfo {author} {\bibfnamefont {P.~D.}\ \bibnamefont {Cottle}},
  \bibinfo {author} {\bibfnamefont {R.~W.}\ \bibnamefont {Ibbotson}}, \bibinfo
  {author} {\bibfnamefont {K.~W.}\ \bibnamefont {Kemper}}, \bibinfo {author}
  {\bibfnamefont {L.~A.}\ \bibnamefont {Riley}}, \ and\ \bibinfo {author}
  {\bibfnamefont {H.}~\bibnamefont {Scheit}},\ }\href {\doibase
  10.1103/PhysRevC.63.011305} {\bibfield  {journal} {\bibinfo  {journal} {Phys.
  Rev. C}\ }\textbf {\bibinfo {volume} {63}},\ \bibinfo {pages} {011305}
  (\bibinfo {year} {2001})}\BibitemShut {NoStop}%
\bibitem [{\citenamefont {Warburton}\ \emph {et~al.}(1990)\citenamefont
  {Warburton}, \citenamefont {Becker},\ and\ \citenamefont
  {Brown}}]{Warburton:1990island}%
  \BibitemOpen
  \bibfield  {author} {\bibinfo {author} {\bibfnamefont {E.~K.}\ \bibnamefont
  {Warburton}}, \bibinfo {author} {\bibfnamefont {J.~A.}\ \bibnamefont
  {Becker}}, \ and\ \bibinfo {author} {\bibfnamefont {B.~A.}\ \bibnamefont
  {Brown}},\ }\href {\doibase 10.1103/PhysRevC.41.1147} {\bibfield  {journal}
  {\bibinfo  {journal} {Phys. Rev. C}\ }\textbf {\bibinfo {volume} {41}},\
  \bibinfo {pages} {1147} (\bibinfo {year} {1990})}\BibitemShut {NoStop}%
\bibitem [{\citenamefont {Otsuka}\ \emph {et~al.}(2020)\citenamefont {Otsuka},
  \citenamefont {Gade}, \citenamefont {Sorlin}, \citenamefont {Suzuki},\ and\
  \citenamefont {Utsuno}}]{Otsuka:2020}%
  \BibitemOpen
  \bibfield  {author} {\bibinfo {author} {\bibfnamefont {T.}~\bibnamefont
  {Otsuka}}, \bibinfo {author} {\bibfnamefont {A.}~\bibnamefont {Gade}},
  \bibinfo {author} {\bibfnamefont {O.}~\bibnamefont {Sorlin}}, \bibinfo
  {author} {\bibfnamefont {T.}~\bibnamefont {Suzuki}}, \ and\ \bibinfo {author}
  {\bibfnamefont {Y.}~\bibnamefont {Utsuno}},\ }\href {\doibase
  10.1103/RevModPhys.92.015002} {\bibfield  {journal} {\bibinfo  {journal}
  {Rev. Mod. Phys.}\ }\textbf {\bibinfo {volume} {92}},\ \bibinfo {pages}
  {015002} (\bibinfo {year} {2020})},\ \Eprint
  {http://arxiv.org/abs/1805.06501} {arXiv:1805.06501 [nucl-th]} \BibitemShut
  {NoStop}%
\bibitem [{\citenamefont {Campi}\ \emph {et~al.}(1975)\citenamefont {Campi},
  \citenamefont {Flocard}, \citenamefont {Kerman},\ and\ \citenamefont
  {Koonin}}]{Campi:1975HFM1}%
  \BibitemOpen
  \bibfield  {author} {\bibinfo {author} {\bibfnamefont {X.}~\bibnamefont
  {Campi}}, \bibinfo {author} {\bibfnamefont {H.}~\bibnamefont {Flocard}},
  \bibinfo {author} {\bibfnamefont {A.~K.}\ \bibnamefont {Kerman}}, \ and\
  \bibinfo {author} {\bibfnamefont {S.}~\bibnamefont {Koonin}},\ }\href
  {\doibase 10.1016/0375-9474(75)90065-2} {\bibfield  {journal} {\bibinfo
  {journal} {Nucl. Phys. A}\ }\textbf {\bibinfo {volume} {251}},\ \bibinfo
  {pages} {193} (\bibinfo {year} {1975})}\BibitemShut {NoStop}%
\bibitem [{\citenamefont {Rodr\'{\i}guez-Guzm\'an}\ \emph
  {et~al.}(2000)\citenamefont {Rodr\'{\i}guez-Guzm\'an}, \citenamefont
  {Egido},\ and\ \citenamefont {Robledo}}]{Guzman:2000PRC}%
  \BibitemOpen
  \bibfield  {author} {\bibinfo {author} {\bibfnamefont {R.~R.}\ \bibnamefont
  {Rodr\'{\i}guez-Guzm\'an}}, \bibinfo {author} {\bibfnamefont {J.~L.}\
  \bibnamefont {Egido}}, \ and\ \bibinfo {author} {\bibfnamefont {L.~M.}\
  \bibnamefont {Robledo}},\ }\href {\doibase 10.1103/PhysRevC.62.054319}
  {\bibfield  {journal} {\bibinfo  {journal} {Phys. Rev. C}\ }\textbf {\bibinfo
  {volume} {62}},\ \bibinfo {pages} {054319} (\bibinfo {year}
  {2000})}\BibitemShut {NoStop}%
\bibitem [{\citenamefont {Rodriguez-Guzman}\ \emph {et~al.}(2002)\citenamefont
  {Rodriguez-Guzman}, \citenamefont {Egido},\ and\ \citenamefont
  {Robledo}}]{Rodriguez-Guzman:2002NPA}%
  \BibitemOpen
  \bibfield  {author} {\bibinfo {author} {\bibfnamefont {R.}~\bibnamefont
  {Rodriguez-Guzman}}, \bibinfo {author} {\bibfnamefont {J.~L.}\ \bibnamefont
  {Egido}}, \ and\ \bibinfo {author} {\bibfnamefont {L.~M.}\ \bibnamefont
  {Robledo}},\ }\href {\doibase 10.1016/S0375-9474(02)01019-9} {\bibfield
  {journal} {\bibinfo  {journal} {Nucl. Phys. A}\ }\textbf {\bibinfo {volume}
  {709}},\ \bibinfo {pages} {201} (\bibinfo {year} {2002})},\ \Eprint
  {http://arxiv.org/abs/nucl-th/0204074} {arXiv:nucl-th/0204074} \BibitemShut
  {NoStop}%
\bibitem [{\citenamefont {Niksic}\ \emph {et~al.}(2006)\citenamefont {Niksic},
  \citenamefont {Vretenar},\ and\ \citenamefont {Ring}}]{Niksic:2006kv}%
  \BibitemOpen
  \bibfield  {author} {\bibinfo {author} {\bibfnamefont {T.}~\bibnamefont
  {Niksic}}, \bibinfo {author} {\bibfnamefont {D.}~\bibnamefont {Vretenar}}, \
  and\ \bibinfo {author} {\bibfnamefont {P.}~\bibnamefont {Ring}},\ }\href
  {\doibase 10.1103/PhysRevC.73.034308} {\bibfield  {journal} {\bibinfo
  {journal} {Phys. Rev. C}\ }\textbf {\bibinfo {volume} {73}},\ \bibinfo
  {pages} {034308} (\bibinfo {year} {2006})},\ \Eprint
  {http://arxiv.org/abs/nucl-th/0601005} {arXiv:nucl-th/0601005} \BibitemShut
  {NoStop}%
\bibitem [{\citenamefont {Yao}\ \emph {et~al.}(2011)\citenamefont {Yao},
  \citenamefont {Mei}, \citenamefont {Chen}, \citenamefont {Meng},
  \citenamefont {Ring},\ and\ \citenamefont {Vretenar}}]{Yao:2011_Mg}%
  \BibitemOpen
  \bibfield  {author} {\bibinfo {author} {\bibfnamefont {J.~M.}\ \bibnamefont
  {Yao}}, \bibinfo {author} {\bibfnamefont {H.}~\bibnamefont {Mei}}, \bibinfo
  {author} {\bibfnamefont {H.}~\bibnamefont {Chen}}, \bibinfo {author}
  {\bibfnamefont {J.}~\bibnamefont {Meng}}, \bibinfo {author} {\bibfnamefont
  {P.}~\bibnamefont {Ring}}, \ and\ \bibinfo {author} {\bibfnamefont
  {D.}~\bibnamefont {Vretenar}},\ }\href {\doibase 10.1103/PhysRevC.83.014308}
  {\bibfield  {journal} {\bibinfo  {journal} {Phys. Rev. C}\ }\textbf {\bibinfo
  {volume} {83}},\ \bibinfo {pages} {014308} (\bibinfo {year} {2011})},\
  \Eprint {http://arxiv.org/abs/1006.1400} {arXiv:1006.1400 [nucl-th]}
  \BibitemShut {NoStop}%
\bibitem [{\citenamefont {Borrajo}\ and\ \citenamefont
  {Egido}(2017)}]{Borrajo:2017PLB}%
  \BibitemOpen
  \bibfield  {author} {\bibinfo {author} {\bibfnamefont {M.}~\bibnamefont
  {Borrajo}}\ and\ \bibinfo {author} {\bibfnamefont {J.~L.}\ \bibnamefont
  {Egido}},\ }\href {\doibase 10.1016/j.physletb.2016.11.037} {\bibfield
  {journal} {\bibinfo  {journal} {Phys. Lett. B}\ }\textbf {\bibinfo {volume}
  {764}},\ \bibinfo {pages} {328} (\bibinfo {year} {2017})},\ \Eprint
  {http://arxiv.org/abs/1611.06982} {arXiv:1611.06982 [nucl-th]} \BibitemShut
  {NoStop}%
\bibitem [{\citenamefont {Wildenthal}\ \emph {et~al.}(1983)\citenamefont
  {Wildenthal}, \citenamefont {Curtin},\ and\ \citenamefont
  {Brown}}]{Wildenthal:1983SM2}%
  \BibitemOpen
  \bibfield  {author} {\bibinfo {author} {\bibfnamefont {B.~H.}\ \bibnamefont
  {Wildenthal}}, \bibinfo {author} {\bibfnamefont {M.~S.}\ \bibnamefont
  {Curtin}}, \ and\ \bibinfo {author} {\bibfnamefont {B.~A.}\ \bibnamefont
  {Brown}},\ }\href {\doibase 10.1103/PhysRevC.28.1343} {\bibfield  {journal}
  {\bibinfo  {journal} {Phys. Rev. C}\ }\textbf {\bibinfo {volume} {28}},\
  \bibinfo {pages} {1343} (\bibinfo {year} {1983})}\BibitemShut {NoStop}%
\bibitem [{\citenamefont {Poves}\ and\ \citenamefont
  {Retamosa}(1987)}]{Poves:1987SM3}%
  \BibitemOpen
  \bibfield  {author} {\bibinfo {author} {\bibfnamefont {A.}~\bibnamefont
  {Poves}}\ and\ \bibinfo {author} {\bibfnamefont {J.}~\bibnamefont
  {Retamosa}},\ }\href {\doibase 10.1016/0370-2693(87)90171-7} {\bibfield
  {journal} {\bibinfo  {journal} {Phys. Lett. B}\ }\textbf {\bibinfo {volume}
  {184}},\ \bibinfo {pages} {311} (\bibinfo {year} {1987})}\BibitemShut
  {NoStop}%
\bibitem [{\citenamefont {Fukunishi}\ \emph {et~al.}(1992)\citenamefont
  {Fukunishi}, \citenamefont {Otsuka},\ and\ \citenamefont
  {Sebe}}]{Fukunishi:1992SM1}%
  \BibitemOpen
  \bibfield  {author} {\bibinfo {author} {\bibfnamefont {N.}~\bibnamefont
  {Fukunishi}}, \bibinfo {author} {\bibfnamefont {T.}~\bibnamefont {Otsuka}}, \
  and\ \bibinfo {author} {\bibfnamefont {T.}~\bibnamefont {Sebe}},\ }\href
  {\doibase 10.1016/0370-2693(92)91320-9} {\bibfield  {journal} {\bibinfo
  {journal} {Phys. Lett. B}\ }\textbf {\bibinfo {volume} {296}},\ \bibinfo
  {pages} {279} (\bibinfo {year} {1992})}\BibitemShut {NoStop}%
\bibitem [{\citenamefont {Utsuno}\ \emph {et~al.}(2004)\citenamefont {Utsuno},
  \citenamefont {Otsuka}, \citenamefont {Glasmacher}, \citenamefont
  {Mizusaki},\ and\ \citenamefont {Honma}}]{Utsuno:2004SM4}%
  \BibitemOpen
  \bibfield  {author} {\bibinfo {author} {\bibfnamefont {Y.}~\bibnamefont
  {Utsuno}}, \bibinfo {author} {\bibfnamefont {T.}~\bibnamefont {Otsuka}},
  \bibinfo {author} {\bibfnamefont {T.}~\bibnamefont {Glasmacher}}, \bibinfo
  {author} {\bibfnamefont {T.}~\bibnamefont {Mizusaki}}, \ and\ \bibinfo
  {author} {\bibfnamefont {M.}~\bibnamefont {Honma}},\ }\href {\doibase
  10.1103/PhysRevC.70.044307} {\bibfield  {journal} {\bibinfo  {journal} {Phys.
  Rev. C}\ }\textbf {\bibinfo {volume} {70}},\ \bibinfo {pages} {044307}
  (\bibinfo {year} {2004})},\ \Eprint {http://arxiv.org/abs/nucl-th/0407082}
  {arXiv:nucl-th/0407082} \BibitemShut {NoStop}%
\bibitem [{\citenamefont {Caurier}\ \emph {et~al.}(1998)\citenamefont
  {Caurier}, \citenamefont {Nowacki}, \citenamefont {Poves},\ and\
  \citenamefont {Retamosa}}]{Caurier:1998SM5}%
  \BibitemOpen
  \bibfield  {author} {\bibinfo {author} {\bibfnamefont {E.}~\bibnamefont
  {Caurier}}, \bibinfo {author} {\bibfnamefont {F.}~\bibnamefont {Nowacki}},
  \bibinfo {author} {\bibfnamefont {A.}~\bibnamefont {Poves}}, \ and\ \bibinfo
  {author} {\bibfnamefont {J.}~\bibnamefont {Retamosa}},\ }\href {\doibase
  10.1103/PhysRevC.58.2033} {\bibfield  {journal} {\bibinfo  {journal} {Phys.
  Rev. C}\ }\textbf {\bibinfo {volume} {58}},\ \bibinfo {pages} {2033}
  (\bibinfo {year} {1998})}\BibitemShut {NoStop}%
\bibitem [{\citenamefont {Heyde}\ and\ \citenamefont
  {Wood}(2011{\natexlab{a}})}]{Heyde:2011}%
  \BibitemOpen
  \bibfield  {author} {\bibinfo {author} {\bibfnamefont {K.}~\bibnamefont
  {Heyde}}\ and\ \bibinfo {author} {\bibfnamefont {J.~L.}\ \bibnamefont
  {Wood}},\ }\href {\doibase 10.1103/RevModPhys.83.1467} {\bibfield  {journal}
  {\bibinfo  {journal} {Rev. Mod. Phys.}\ }\textbf {\bibinfo {volume} {83}},\
  \bibinfo {pages} {1467} (\bibinfo {year} {2011}{\natexlab{a}})}\BibitemShut
  {NoStop}%
\bibitem [{\citenamefont {Caurier}\ \emph {et~al.}(2014)\citenamefont
  {Caurier}, \citenamefont {Nowacki},\ and\ \citenamefont
  {Poves}}]{Caurier:2013sm6}%
  \BibitemOpen
  \bibfield  {author} {\bibinfo {author} {\bibfnamefont {E.}~\bibnamefont
  {Caurier}}, \bibinfo {author} {\bibfnamefont {F.}~\bibnamefont {Nowacki}}, \
  and\ \bibinfo {author} {\bibfnamefont {A.}~\bibnamefont {Poves}},\ }\href
  {\doibase 10.1103/PhysRevC.90.014302} {\bibfield  {journal} {\bibinfo
  {journal} {Phys. Rev. C}\ }\textbf {\bibinfo {volume} {90}},\ \bibinfo
  {pages} {014302} (\bibinfo {year} {2014})},\ \Eprint
  {http://arxiv.org/abs/1309.6955} {arXiv:1309.6955 [nucl-th]} \BibitemShut
  {NoStop}%
\bibitem [{\citenamefont {Gade}\ and\ \citenamefont
  {Liddick}(2016)}]{Gade:2016}%
  \BibitemOpen
  \bibfield  {author} {\bibinfo {author} {\bibfnamefont {A.}~\bibnamefont
  {Gade}}\ and\ \bibinfo {author} {\bibfnamefont {S.~N.}\ \bibnamefont
  {Liddick}},\ }\href {\doibase 10.1088/0954-3899/43/2/024001} {\bibfield
  {journal} {\bibinfo  {journal} {J. Phys. G}\ }\textbf {\bibinfo {volume}
  {43}},\ \bibinfo {pages} {024001} (\bibinfo {year} {2016})}\BibitemShut
  {NoStop}%
\bibitem [{\citenamefont {Wimmer}\ \emph {et~al.}(2010)\citenamefont {Wimmer}
  \emph {et~al.}}]{Wimmer:2010Mg32prl}%
  \BibitemOpen
  \bibfield  {author} {\bibinfo {author} {\bibfnamefont {K.}~\bibnamefont
  {Wimmer}} \emph {et~al.},\ }\href {\doibase 10.1103/PhysRevLett.105.252501}
  {\bibfield  {journal} {\bibinfo  {journal} {Phys. Rev. Lett.}\ }\textbf
  {\bibinfo {volume} {105}},\ \bibinfo {pages} {252501} (\bibinfo {year}
  {2010})},\ \Eprint {http://arxiv.org/abs/1010.3999} {arXiv:1010.3999
  [nucl-ex]} \BibitemShut {NoStop}%
\bibitem [{\citenamefont {Heyde}\ and\ \citenamefont
  {Wood}(2011{\natexlab{b}})}]{Heyde:2011RMP}%
  \BibitemOpen
  \bibfield  {author} {\bibinfo {author} {\bibfnamefont {K.}~\bibnamefont
  {Heyde}}\ and\ \bibinfo {author} {\bibfnamefont {J.~L.}\ \bibnamefont
  {Wood}},\ }\href {\doibase 10.1103/RevModPhys.83.1467} {\bibfield  {journal}
  {\bibinfo  {journal} {Rev. Mod. Phys.}\ }\textbf {\bibinfo {volume} {83}},\
  \bibinfo {pages} {1467} (\bibinfo {year} {2011}{\natexlab{b}})}\BibitemShut
  {NoStop}%
\bibitem [{\citenamefont {Tsukiyama}\ \emph {et~al.}(2012)\citenamefont
  {Tsukiyama}, \citenamefont {Bogner},\ and\ \citenamefont
  {Schwenk}}]{Tsukiyama:2012PRC}%
  \BibitemOpen
  \bibfield  {author} {\bibinfo {author} {\bibfnamefont {K.}~\bibnamefont
  {Tsukiyama}}, \bibinfo {author} {\bibfnamefont {S.~K.}\ \bibnamefont
  {Bogner}}, \ and\ \bibinfo {author} {\bibfnamefont {A.}~\bibnamefont
  {Schwenk}},\ }\href {\doibase 10.1103/PhysRevC.85.061304} {\bibfield
  {journal} {\bibinfo  {journal} {Phys. Rev. C}\ }\textbf {\bibinfo {volume}
  {85}},\ \bibinfo {pages} {061304} (\bibinfo {year} {2012})},\ \Eprint
  {http://arxiv.org/abs/1203.2515} {arXiv:1203.2515 [nucl-th]} \BibitemShut
  {NoStop}%
\bibitem [{\citenamefont {Bogner}\ \emph {et~al.}(2014)\citenamefont {Bogner},
  \citenamefont {Hergert}, \citenamefont {Holt}, \citenamefont {Schwenk},
  \citenamefont {Binder}, \citenamefont {Calci}, \citenamefont {Langhammer},\
  and\ \citenamefont {Roth}}]{Bogner:2014PRL}%
  \BibitemOpen
  \bibfield  {author} {\bibinfo {author} {\bibfnamefont {S.~K.}\ \bibnamefont
  {Bogner}}, \bibinfo {author} {\bibfnamefont {H.}~\bibnamefont {Hergert}},
  \bibinfo {author} {\bibfnamefont {J.~D.}\ \bibnamefont {Holt}}, \bibinfo
  {author} {\bibfnamefont {A.}~\bibnamefont {Schwenk}}, \bibinfo {author}
  {\bibfnamefont {S.}~\bibnamefont {Binder}}, \bibinfo {author} {\bibfnamefont
  {A.}~\bibnamefont {Calci}}, \bibinfo {author} {\bibfnamefont
  {J.}~\bibnamefont {Langhammer}}, \ and\ \bibinfo {author} {\bibfnamefont
  {R.}~\bibnamefont {Roth}},\ }\href {\doibase 10.1103/PhysRevLett.113.142501}
  {\bibfield  {journal} {\bibinfo  {journal} {Phys. Rev. Lett.}\ }\textbf
  {\bibinfo {volume} {113}},\ \bibinfo {pages} {142501} (\bibinfo {year}
  {2014})},\ \Eprint {http://arxiv.org/abs/1402.1407} {arXiv:1402.1407
  [nucl-th]} \BibitemShut {NoStop}%
\bibitem [{\citenamefont {Stroberg}\ \emph {et~al.}(2019)\citenamefont
  {Stroberg}, \citenamefont {Bogner}, \citenamefont {Hergert},\ and\
  \citenamefont {Holt}}]{Stroberg:2019ARNPS}%
  \BibitemOpen
  \bibfield  {author} {\bibinfo {author} {\bibfnamefont {S.~R.}\ \bibnamefont
  {Stroberg}}, \bibinfo {author} {\bibfnamefont {S.~K.}\ \bibnamefont
  {Bogner}}, \bibinfo {author} {\bibfnamefont {H.}~\bibnamefont {Hergert}}, \
  and\ \bibinfo {author} {\bibfnamefont {J.~D.}\ \bibnamefont {Holt}},\ }\href
  {\doibase 10.1146/annurev-nucl-101917-021120} {\bibfield  {journal} {\bibinfo
   {journal} {Ann. Rev. Nucl. Part. Sci.}\ }\textbf {\bibinfo {volume} {69}},\
  \bibinfo {pages} {307} (\bibinfo {year} {2019})},\ \Eprint
  {http://arxiv.org/abs/1902.06154} {arXiv:1902.06154 [nucl-th]} \BibitemShut
  {NoStop}%
\bibitem [{\citenamefont {Weinberg}(1991)}]{Weinberg:1991}%
  \BibitemOpen
  \bibfield  {author} {\bibinfo {author} {\bibfnamefont {S.}~\bibnamefont
  {Weinberg}},\ }\href {\doibase 10.1016/0550-3213(91)90231-L} {\bibfield
  {journal} {\bibinfo  {journal} {Nucl. Phys. B}\ }\textbf {\bibinfo {volume}
  {363}},\ \bibinfo {pages} {3} (\bibinfo {year} {1991})}\BibitemShut {NoStop}%
\bibitem [{\citenamefont {Epelbaum}\ \emph {et~al.}(2009)\citenamefont
  {Epelbaum}, \citenamefont {Hammer},\ and\ \citenamefont
  {Mei\ss{}ner}}]{Epelbaum:2009RMP}%
  \BibitemOpen
  \bibfield  {author} {\bibinfo {author} {\bibfnamefont {E.}~\bibnamefont
  {Epelbaum}}, \bibinfo {author} {\bibfnamefont {H.-W.}\ \bibnamefont
  {Hammer}}, \ and\ \bibinfo {author} {\bibfnamefont {U.-G.}\ \bibnamefont
  {Mei\ss{}ner}},\ }\href {\doibase 10.1103/RevModPhys.81.1773} {\bibfield
  {journal} {\bibinfo  {journal} {Rev. Mod. Phys.}\ }\textbf {\bibinfo {volume}
  {81}},\ \bibinfo {pages} {1773} (\bibinfo {year} {2009})}\BibitemShut
  {NoStop}%
\bibitem [{\citenamefont {Machleidt}\ and\ \citenamefont
  {Entem}(2011)}]{Machleidt:2011PR}%
  \BibitemOpen
  \bibfield  {author} {\bibinfo {author} {\bibfnamefont {R.}~\bibnamefont
  {Machleidt}}\ and\ \bibinfo {author} {\bibfnamefont {D.~R.}\ \bibnamefont
  {Entem}},\ }\href {\doibase 10.1016/j.physrep.2011.02.001} {\bibfield
  {journal} {\bibinfo  {journal} {Phys. Rept.}\ }\textbf {\bibinfo {volume}
  {503}},\ \bibinfo {pages} {1} (\bibinfo {year} {2011})},\ \Eprint
  {http://arxiv.org/abs/1105.2919} {arXiv:1105.2919 [nucl-th]} \BibitemShut
  {NoStop}%
\bibitem [{\citenamefont {Miyagi}\ \emph {et~al.}(2020)\citenamefont {Miyagi},
  \citenamefont {Stroberg}, \citenamefont {Holt},\ and\ \citenamefont
  {Shimizu}}]{Miyagi:2020PRC}%
  \BibitemOpen
  \bibfield  {author} {\bibinfo {author} {\bibfnamefont {T.}~\bibnamefont
  {Miyagi}}, \bibinfo {author} {\bibfnamefont {S.~R.}\ \bibnamefont
  {Stroberg}}, \bibinfo {author} {\bibfnamefont {J.~D.}\ \bibnamefont {Holt}},
  \ and\ \bibinfo {author} {\bibfnamefont {N.}~\bibnamefont {Shimizu}},\ }\href
  {\doibase 10.1103/PhysRevC.102.034320} {\bibfield  {journal} {\bibinfo
  {journal} {Phys. Rev. C}\ }\textbf {\bibinfo {volume} {102}},\ \bibinfo
  {pages} {034320} (\bibinfo {year} {2020})},\ \Eprint
  {http://arxiv.org/abs/2004.12969} {arXiv:2004.12969 [nucl-th]} \BibitemShut
  {NoStop}%
\bibitem [{\citenamefont {Stroberg}\ \emph {et~al.}(2024)\citenamefont
  {Stroberg}, \citenamefont {Morris},\ and\ \citenamefont
  {He}}]{Stroberg:2024}%
  \BibitemOpen
  \bibfield  {author} {\bibinfo {author} {\bibfnamefont {S.~R.}\ \bibnamefont
  {Stroberg}}, \bibinfo {author} {\bibfnamefont {T.~D.}\ \bibnamefont
  {Morris}}, \ and\ \bibinfo {author} {\bibfnamefont {B.~C.}\ \bibnamefont
  {He}},\ }\href {\doibase 10.1103/PhysRevC.110.044316} {\bibfield  {journal}
  {\bibinfo  {journal} {Phys. Rev. C}\ }\textbf {\bibinfo {volume} {110}},\
  \bibinfo {pages} {044316} (\bibinfo {year} {2024})}\BibitemShut {NoStop}%
\bibitem [{\citenamefont {Hergert}(2020)}]{Hergert:2020}%
  \BibitemOpen
  \bibfield  {author} {\bibinfo {author} {\bibfnamefont {H.}~\bibnamefont
  {Hergert}},\ }\href {\doibase 10.3389/fphy.2020.00379} {\bibfield  {journal}
  {\bibinfo  {journal} {Front. in Phys.}\ }\textbf {\bibinfo {volume} {8}},\
  \bibinfo {pages} {379} (\bibinfo {year} {2020})},\ \Eprint
  {http://arxiv.org/abs/2008.05061} {arXiv:2008.05061 [nucl-th]} \BibitemShut
  {NoStop}%
\bibitem [{\citenamefont {Hagen}\ \emph {et~al.}(2022)\citenamefont {Hagen},
  \citenamefont {Novario}, \citenamefont {Sun}, \citenamefont {Papenbrock},
  \citenamefont {Jansen}, \citenamefont {Lietz}, \citenamefont {Duguet},\ and\
  \citenamefont {Tichai}}]{Hagen:2022PRC}%
  \BibitemOpen
  \bibfield  {author} {\bibinfo {author} {\bibfnamefont {G.}~\bibnamefont
  {Hagen}}, \bibinfo {author} {\bibfnamefont {S.~J.}\ \bibnamefont {Novario}},
  \bibinfo {author} {\bibfnamefont {Z.~H.}\ \bibnamefont {Sun}}, \bibinfo
  {author} {\bibfnamefont {T.}~\bibnamefont {Papenbrock}}, \bibinfo {author}
  {\bibfnamefont {G.~R.}\ \bibnamefont {Jansen}}, \bibinfo {author}
  {\bibfnamefont {J.~G.}\ \bibnamefont {Lietz}}, \bibinfo {author}
  {\bibfnamefont {T.}~\bibnamefont {Duguet}}, \ and\ \bibinfo {author}
  {\bibfnamefont {A.}~\bibnamefont {Tichai}},\ }\href {\doibase
  10.1103/PhysRevC.105.064311} {\bibfield  {journal} {\bibinfo  {journal}
  {Phys. Rev. C}\ }\textbf {\bibinfo {volume} {105}},\ \bibinfo {pages}
  {064311} (\bibinfo {year} {2022})},\ \Eprint
  {http://arxiv.org/abs/2201.07298} {arXiv:2201.07298 [nucl-th]} \BibitemShut
  {NoStop}%
\bibitem [{\citenamefont {Sun}\ \emph {et~al.}(2024)\citenamefont {Sun},
  \citenamefont {Ekstr\"om}, \citenamefont {Forss\'en}, \citenamefont {Hagen},
  \citenamefont {Jansen},\ and\ \citenamefont {Papenbrock}}]{Sun:2024_even}%
  \BibitemOpen
  \bibfield  {author} {\bibinfo {author} {\bibfnamefont {Z.~H.}\ \bibnamefont
  {Sun}}, \bibinfo {author} {\bibfnamefont {A.}~\bibnamefont {Ekstr\"om}},
  \bibinfo {author} {\bibfnamefont {C.}~\bibnamefont {Forss\'en}}, \bibinfo
  {author} {\bibfnamefont {G.}~\bibnamefont {Hagen}}, \bibinfo {author}
  {\bibfnamefont {G.~R.}\ \bibnamefont {Jansen}}, \ and\ \bibinfo {author}
  {\bibfnamefont {T.}~\bibnamefont {Papenbrock}}\ }(\bibinfo {year} {2024})\
  \Eprint {http://arxiv.org/abs/2404.00058} {arXiv:2404.00058 [nucl-th]}
  \BibitemShut {NoStop}%
\bibitem [{\citenamefont {Sun}\ \emph {et~al.}(2025)\citenamefont {Sun},
  \citenamefont {Dj\"arv}, \citenamefont {Hagen}, \citenamefont {Jansen},\ and\
  \citenamefont {Papenbrock}}]{Sun:2025lk}%
  \BibitemOpen
  \bibfield  {author} {\bibinfo {author} {\bibfnamefont {Z.~H.}\ \bibnamefont
  {Sun}}, \bibinfo {author} {\bibfnamefont {T.~R.}\ \bibnamefont {Dj\"arv}},
  \bibinfo {author} {\bibfnamefont {G.}~\bibnamefont {Hagen}}, \bibinfo
  {author} {\bibfnamefont {G.~R.}\ \bibnamefont {Jansen}}, \ and\ \bibinfo
  {author} {\bibfnamefont {T.}~\bibnamefont {Papenbrock}},\ }\href {\doibase
  10.1103/PhysRevC.111.044304} {\bibfield  {journal} {\bibinfo  {journal}
  {Phys. Rev. C}\ }\textbf {\bibinfo {volume} {111}},\ \bibinfo {pages}
  {044304} (\bibinfo {year} {2025})}\BibitemShut {NoStop}%
\bibitem [{\citenamefont {Frosini}\ \emph
  {et~al.}(2022{\natexlab{a}})\citenamefont {Frosini}, \citenamefont {Duguet},
  \citenamefont {Ebran},\ and\ \citenamefont {Som\`a}}]{Frosini:2022_1}%
  \BibitemOpen
  \bibfield  {author} {\bibinfo {author} {\bibfnamefont {M.}~\bibnamefont
  {Frosini}}, \bibinfo {author} {\bibfnamefont {T.}~\bibnamefont {Duguet}},
  \bibinfo {author} {\bibfnamefont {J.-P.}\ \bibnamefont {Ebran}}, \ and\
  \bibinfo {author} {\bibfnamefont {V.}~\bibnamefont {Som\`a}},\ }\href
  {\doibase 10.1140/epja/s10050-022-00692-z} {\bibfield  {journal} {\bibinfo
  {journal} {Eur. Phys. J. A}\ }\textbf {\bibinfo {volume} {58}},\ \bibinfo
  {pages} {62} (\bibinfo {year} {2022}{\natexlab{a}})},\ \Eprint
  {http://arxiv.org/abs/2110.15737} {arXiv:2110.15737 [nucl-th]} \BibitemShut
  {NoStop}%
\bibitem [{\citenamefont {Frosini}\ \emph
  {et~al.}(2022{\natexlab{b}})\citenamefont {Frosini}, \citenamefont {Duguet},
  \citenamefont {Ebran}, \citenamefont {Bally}, \citenamefont {Mongelli},
  \citenamefont {Rodr\'\i{}guez}, \citenamefont {Roth},\ and\ \citenamefont
  {Som\`a}}]{Frosini:2022_2}%
  \BibitemOpen
  \bibfield  {author} {\bibinfo {author} {\bibfnamefont {M.}~\bibnamefont
  {Frosini}}, \bibinfo {author} {\bibfnamefont {T.}~\bibnamefont {Duguet}},
  \bibinfo {author} {\bibfnamefont {J.-P.}\ \bibnamefont {Ebran}}, \bibinfo
  {author} {\bibfnamefont {B.}~\bibnamefont {Bally}}, \bibinfo {author}
  {\bibfnamefont {T.}~\bibnamefont {Mongelli}}, \bibinfo {author}
  {\bibfnamefont {T.~R.}\ \bibnamefont {Rodr\'\i{}guez}}, \bibinfo {author}
  {\bibfnamefont {R.}~\bibnamefont {Roth}}, \ and\ \bibinfo {author}
  {\bibfnamefont {V.}~\bibnamefont {Som\`a}},\ }\href {\doibase
  10.1140/epja/s10050-022-00693-y} {\bibfield  {journal} {\bibinfo  {journal}
  {Eur. Phys. J. A}\ }\textbf {\bibinfo {volume} {58}},\ \bibinfo {pages} {63}
  (\bibinfo {year} {2022}{\natexlab{b}})},\ \Eprint
  {http://arxiv.org/abs/2111.00797} {arXiv:2111.00797 [nucl-th]} \BibitemShut
  {NoStop}%
\bibitem [{\citenamefont {Frosini}\ \emph
  {et~al.}(2022{\natexlab{c}})\citenamefont {Frosini}, \citenamefont {Duguet},
  \citenamefont {Ebran}, \citenamefont {Bally}, \citenamefont {Hergert},
  \citenamefont {Rodr\'\i{}guez}, \citenamefont {Roth}, \citenamefont {Yao},\
  and\ \citenamefont {Som\`a}}]{Frosini:2022_3}%
  \BibitemOpen
  \bibfield  {author} {\bibinfo {author} {\bibfnamefont {M.}~\bibnamefont
  {Frosini}}, \bibinfo {author} {\bibfnamefont {T.}~\bibnamefont {Duguet}},
  \bibinfo {author} {\bibfnamefont {J.-P.}\ \bibnamefont {Ebran}}, \bibinfo
  {author} {\bibfnamefont {B.}~\bibnamefont {Bally}}, \bibinfo {author}
  {\bibfnamefont {H.}~\bibnamefont {Hergert}}, \bibinfo {author} {\bibfnamefont
  {T.~R.}\ \bibnamefont {Rodr\'\i{}guez}}, \bibinfo {author} {\bibfnamefont
  {R.}~\bibnamefont {Roth}}, \bibinfo {author} {\bibfnamefont {J.}~\bibnamefont
  {Yao}}, \ and\ \bibinfo {author} {\bibfnamefont {V.}~\bibnamefont {Som\`a}},\
  }\href {\doibase 10.1140/epja/s10050-022-00694-x} {\bibfield  {journal}
  {\bibinfo  {journal} {Eur. Phys. J. A}\ }\textbf {\bibinfo {volume} {58}},\
  \bibinfo {pages} {64} (\bibinfo {year} {2022}{\natexlab{c}})},\ \Eprint
  {http://arxiv.org/abs/2111.01461} {arXiv:2111.01461 [nucl-th]} \BibitemShut
  {NoStop}%
\bibitem [{\citenamefont {Cao}\ and\ \citenamefont {Jiao}(2025)}]{Cao:2025}%
  \BibitemOpen
  \bibfield  {author} {\bibinfo {author} {\bibfnamefont {X.~C.}\ \bibnamefont
  {Cao}}\ and\ \bibinfo {author} {\bibfnamefont {C.~F.}\ \bibnamefont {Jiao}},\
  }\href {\doibase 10.1016/j.physletb.2025.140034} {\bibfield  {journal}
  {\bibinfo  {journal} {Phys. Lett. B}\ }\textbf {\bibinfo {volume} {871}},\
  \bibinfo {pages} {140034} (\bibinfo {year} {2025})},\ \Eprint
  {http://arxiv.org/abs/2509.04842} {arXiv:2509.04842 [nucl-th]} \BibitemShut
  {NoStop}%
\bibitem [{\citenamefont {Stroberg}\ \emph {et~al.}(2017)\citenamefont
  {Stroberg}, \citenamefont {Calci}, \citenamefont {Hergert}, \citenamefont
  {Holt}, \citenamefont {Bogner}, \citenamefont {Roth},\ and\ \citenamefont
  {Schwenk}}]{Stroberg:2017PRL}%
  \BibitemOpen
  \bibfield  {author} {\bibinfo {author} {\bibfnamefont {S.~R.}\ \bibnamefont
  {Stroberg}}, \bibinfo {author} {\bibfnamefont {A.}~\bibnamefont {Calci}},
  \bibinfo {author} {\bibfnamefont {H.}~\bibnamefont {Hergert}}, \bibinfo
  {author} {\bibfnamefont {J.~D.}\ \bibnamefont {Holt}}, \bibinfo {author}
  {\bibfnamefont {S.~K.}\ \bibnamefont {Bogner}}, \bibinfo {author}
  {\bibfnamefont {R.}~\bibnamefont {Roth}}, \ and\ \bibinfo {author}
  {\bibfnamefont {A.}~\bibnamefont {Schwenk}},\ }\href {\doibase
  10.1103/PhysRevLett.118.032502} {\bibfield  {journal} {\bibinfo  {journal}
  {Phys. Rev. Lett.}\ }\textbf {\bibinfo {volume} {118}},\ \bibinfo {pages}
  {032502} (\bibinfo {year} {2017})}\BibitemShut {NoStop}%
\bibitem [{\citenamefont {Yao}\ \emph {et~al.}(2018)\citenamefont {Yao},
  \citenamefont {Engel}, \citenamefont {Wang}, \citenamefont {Jiao},\ and\
  \citenamefont {Hergert}}]{Yao:2018PRC}%
  \BibitemOpen
  \bibfield  {author} {\bibinfo {author} {\bibfnamefont {J.~M.}\ \bibnamefont
  {Yao}}, \bibinfo {author} {\bibfnamefont {J.}~\bibnamefont {Engel}}, \bibinfo
  {author} {\bibfnamefont {L.~J.}\ \bibnamefont {Wang}}, \bibinfo {author}
  {\bibfnamefont {C.~F.}\ \bibnamefont {Jiao}}, \ and\ \bibinfo {author}
  {\bibfnamefont {H.}~\bibnamefont {Hergert}},\ }\href {\doibase
  10.1103/PhysRevC.98.054311} {\bibfield  {journal} {\bibinfo  {journal} {Phys.
  Rev. C}\ }\textbf {\bibinfo {volume} {98}},\ \bibinfo {pages} {054311}
  (\bibinfo {year} {2018})},\ \Eprint {http://arxiv.org/abs/1807.11053}
  {arXiv:1807.11053 [nucl-th]} \BibitemShut {NoStop}%
\bibitem [{\citenamefont {Hergert}\ \emph {et~al.}(2016)\citenamefont
  {Hergert}, \citenamefont {Bogner}, \citenamefont {Morris}, \citenamefont
  {Schwenk},\ and\ \citenamefont {Tsukiyama}}]{Hergert:2016PR}%
  \BibitemOpen
  \bibfield  {author} {\bibinfo {author} {\bibfnamefont {H.}~\bibnamefont
  {Hergert}}, \bibinfo {author} {\bibfnamefont {S.~K.}\ \bibnamefont {Bogner}},
  \bibinfo {author} {\bibfnamefont {T.~D.}\ \bibnamefont {Morris}}, \bibinfo
  {author} {\bibfnamefont {A.}~\bibnamefont {Schwenk}}, \ and\ \bibinfo
  {author} {\bibfnamefont {K.}~\bibnamefont {Tsukiyama}},\ }\href {\doibase
  10.1016/j.physrep.2015.12.007} {\bibfield  {journal} {\bibinfo  {journal}
  {Phys. Rept.}\ }\textbf {\bibinfo {volume} {621}},\ \bibinfo {pages} {165}
  (\bibinfo {year} {2016})},\ \Eprint {http://arxiv.org/abs/1512.06956}
  {arXiv:1512.06956 [nucl-th]} \BibitemShut {NoStop}%
\bibitem [{\citenamefont {Yao}\ \emph {et~al.}(2020)\citenamefont {Yao},
  \citenamefont {Bally}, \citenamefont {Engel}, \citenamefont {Wirth},
  \citenamefont {Rodr\'{\i}guez},\ and\ \citenamefont {Hergert}}]{Yao:2020PRL}%
  \BibitemOpen
  \bibfield  {author} {\bibinfo {author} {\bibfnamefont {J.~M.}\ \bibnamefont
  {Yao}}, \bibinfo {author} {\bibfnamefont {B.}~\bibnamefont {Bally}}, \bibinfo
  {author} {\bibfnamefont {J.}~\bibnamefont {Engel}}, \bibinfo {author}
  {\bibfnamefont {R.}~\bibnamefont {Wirth}}, \bibinfo {author} {\bibfnamefont
  {T.~R.}\ \bibnamefont {Rodr\'{\i}guez}}, \ and\ \bibinfo {author}
  {\bibfnamefont {H.}~\bibnamefont {Hergert}},\ }\href {\doibase
  10.1103/PhysRevLett.124.232501} {\bibfield  {journal} {\bibinfo  {journal}
  {Phys. Rev. Lett.}\ }\textbf {\bibinfo {volume} {124}},\ \bibinfo {pages}
  {232501} (\bibinfo {year} {2020})}\BibitemShut {NoStop}%
\bibitem [{\citenamefont {Belley}\ \emph {et~al.}(2024)\citenamefont {Belley},
  \citenamefont {Yao}, \citenamefont {Bally}, \citenamefont {Pitcher},
  \citenamefont {Engel}, \citenamefont {Hergert}, \citenamefont {Holt},
  \citenamefont {Miyagi}, \citenamefont {Rodr\'{\i}guez}, \citenamefont
  {Romero}, \citenamefont {Stroberg},\ and\ \citenamefont
  {Zhang}}]{Belley:2024PRL}%
  \BibitemOpen
  \bibfield  {author} {\bibinfo {author} {\bibfnamefont {A.}~\bibnamefont
  {Belley}}, \bibinfo {author} {\bibfnamefont {J.~M.}\ \bibnamefont {Yao}},
  \bibinfo {author} {\bibfnamefont {B.}~\bibnamefont {Bally}}, \bibinfo
  {author} {\bibfnamefont {J.}~\bibnamefont {Pitcher}}, \bibinfo {author}
  {\bibfnamefont {J.}~\bibnamefont {Engel}}, \bibinfo {author} {\bibfnamefont
  {H.}~\bibnamefont {Hergert}}, \bibinfo {author} {\bibfnamefont {J.~D.}\
  \bibnamefont {Holt}}, \bibinfo {author} {\bibfnamefont {T.}~\bibnamefont
  {Miyagi}}, \bibinfo {author} {\bibfnamefont {T.~R.}\ \bibnamefont
  {Rodr\'{\i}guez}}, \bibinfo {author} {\bibfnamefont {A.~M.}\ \bibnamefont
  {Romero}}, \bibinfo {author} {\bibfnamefont {S.~R.}\ \bibnamefont
  {Stroberg}}, \ and\ \bibinfo {author} {\bibfnamefont {X.}~\bibnamefont
  {Zhang}},\ }\href {\doibase 10.1103/PhysRevLett.132.182502} {\bibfield
  {journal} {\bibinfo  {journal} {Phys. Rev. Lett.}\ }\textbf {\bibinfo
  {volume} {132}},\ \bibinfo {pages} {182502} (\bibinfo {year}
  {2024})}\BibitemShut {NoStop}%
\bibitem [{\citenamefont {Zhou}\ \emph {et~al.}(2025)\citenamefont {Zhou},
  \citenamefont {Ding}, \citenamefont {Yao}, \citenamefont {Bally},
  \citenamefont {Hergert}, \citenamefont {Jiao},\ and\ \citenamefont
  {Rodr{\'\i}guez}}]{Zhou:2024_short}%
  \BibitemOpen
  \bibfield  {author} {\bibinfo {author} {\bibfnamefont {E.~F.}\ \bibnamefont
  {Zhou}}, \bibinfo {author} {\bibfnamefont {C.~R.}\ \bibnamefont {Ding}},
  \bibinfo {author} {\bibfnamefont {J.~M.}\ \bibnamefont {Yao}}, \bibinfo
  {author} {\bibfnamefont {B.}~\bibnamefont {Bally}}, \bibinfo {author}
  {\bibfnamefont {H.}~\bibnamefont {Hergert}}, \bibinfo {author} {\bibfnamefont
  {C.~F.}\ \bibnamefont {Jiao}}, \ and\ \bibinfo {author} {\bibfnamefont
  {T.~R.}\ \bibnamefont {Rodr{\'\i}guez}},\ }\href {\doibase
  10.1016/j.physletb.2025.139464} {\bibfield  {journal} {\bibinfo  {journal}
  {Phys. Lett. B}\ }\textbf {\bibinfo {volume} {865}},\ \bibinfo {pages}
  {139464} (\bibinfo {year} {2025})},\ \Eprint
  {http://arxiv.org/abs/2410.23113} {arXiv:2410.23113 [nucl-th]} \BibitemShut
  {NoStop}%
\bibitem [{\citenamefont {Ding}\ \emph {et~al.}(2026)\citenamefont {Ding},
  \citenamefont {Wang}, \citenamefont {Yao}, \citenamefont {Hergert},
  \citenamefont {Liang},\ and\ \citenamefont {Bogner}}]{Ding:2025}%
  \BibitemOpen
  \bibfield  {author} {\bibinfo {author} {\bibfnamefont {C.~R.}\ \bibnamefont
  {Ding}}, \bibinfo {author} {\bibfnamefont {C.~C.}\ \bibnamefont {Wang}},
  \bibinfo {author} {\bibfnamefont {J.~M.}\ \bibnamefont {Yao}}, \bibinfo
  {author} {\bibfnamefont {H.}~\bibnamefont {Hergert}}, \bibinfo {author}
  {\bibfnamefont {H.~Z.}\ \bibnamefont {Liang}}, \ and\ \bibinfo {author}
  {\bibfnamefont {S.~K.}\ \bibnamefont {Bogner}},\ }\href {\doibase
  10.1103/8lzc-j1lx} {\bibfield  {journal} {\bibinfo  {journal} {Phys. Rev.
  Lett.}\ }\textbf {\bibinfo {volume} {136}},\ \bibinfo {pages} {052501}
  (\bibinfo {year} {2026})}\BibitemShut {NoStop}%
\bibitem [{\citenamefont {Lin}\ \emph {et~al.}(2024)\citenamefont {Lin},
  \citenamefont {Zhou}, \citenamefont {Yao},\ and\ \citenamefont
  {Hergert}}]{Lin:2024arl}%
  \BibitemOpen
  \bibfield  {author} {\bibinfo {author} {\bibfnamefont {W.}~\bibnamefont
  {Lin}}, \bibinfo {author} {\bibfnamefont {E.~F.}\ \bibnamefont {Zhou}},
  \bibinfo {author} {\bibfnamefont {J.}~\bibnamefont {Yao}}, \ and\ \bibinfo
  {author} {\bibfnamefont {H.}~\bibnamefont {Hergert}},\ }\href {\doibase
  10.3390/sym16040409} {\bibfield  {journal} {\bibinfo  {journal} {Symmetry}\
  }\textbf {\bibinfo {volume} {16}},\ \bibinfo {pages} {409} (\bibinfo {year}
  {2024})},\ \Eprint {http://arxiv.org/abs/2403.01177} {arXiv:2403.01177
  [nucl-th]} \BibitemShut {NoStop}%
\bibitem [{\citenamefont {Bogner}\ \emph {et~al.}(2010)\citenamefont {Bogner},
  \citenamefont {Furnstahl},\ and\ \citenamefont {Schwenk}}]{Bogner:2010PPNP}%
  \BibitemOpen
  \bibfield  {author} {\bibinfo {author} {\bibfnamefont {S.~K.}\ \bibnamefont
  {Bogner}}, \bibinfo {author} {\bibfnamefont {R.~J.}\ \bibnamefont
  {Furnstahl}}, \ and\ \bibinfo {author} {\bibfnamefont {A.}~\bibnamefont
  {Schwenk}},\ }\href {\doibase 10.1016/j.ppnp.2010.03.001} {\bibfield
  {journal} {\bibinfo  {journal} {Prog. Part. Nucl. Phys.}\ }\textbf {\bibinfo
  {volume} {65}},\ \bibinfo {pages} {94} (\bibinfo {year} {2010})},\ \Eprint
  {http://arxiv.org/abs/0912.3688} {arXiv:0912.3688 [nucl-th]} \BibitemShut
  {NoStop}%
\bibitem [{\citenamefont {Hebeler}\ \emph {et~al.}(2011)\citenamefont
  {Hebeler}, \citenamefont {Bogner}, \citenamefont {Furnstahl}, \citenamefont
  {Nogga},\ and\ \citenamefont {Schwenk}}]{Hebeler:2011PRC}%
  \BibitemOpen
  \bibfield  {author} {\bibinfo {author} {\bibfnamefont {K.}~\bibnamefont
  {Hebeler}}, \bibinfo {author} {\bibfnamefont {S.~K.}\ \bibnamefont {Bogner}},
  \bibinfo {author} {\bibfnamefont {R.~J.}\ \bibnamefont {Furnstahl}}, \bibinfo
  {author} {\bibfnamefont {A.}~\bibnamefont {Nogga}}, \ and\ \bibinfo {author}
  {\bibfnamefont {A.}~\bibnamefont {Schwenk}},\ }\href {\doibase
  10.1103/PhysRevC.83.031301} {\bibfield  {journal} {\bibinfo  {journal} {Phys.
  Rev. C}\ }\textbf {\bibinfo {volume} {83}},\ \bibinfo {pages} {031301}
  (\bibinfo {year} {2011})}\BibitemShut {NoStop}%
\bibitem [{\citenamefont {Entem}\ and\ \citenamefont
  {Machleidt}(2003)}]{Entem:2003PRC}%
  \BibitemOpen
  \bibfield  {author} {\bibinfo {author} {\bibfnamefont {D.~R.}\ \bibnamefont
  {Entem}}\ and\ \bibinfo {author} {\bibfnamefont {R.}~\bibnamefont
  {Machleidt}},\ }\href {\doibase 10.1103/PhysRevC.68.041001} {\bibfield
  {journal} {\bibinfo  {journal} {Phys. Rev. C}\ }\textbf {\bibinfo {volume}
  {68}},\ \bibinfo {pages} {041001} (\bibinfo {year} {2003})}\BibitemShut
  {NoStop}%
\bibitem [{\citenamefont {Hergert}\ \emph {et~al.}(2013)\citenamefont
  {Hergert}, \citenamefont {Binder}, \citenamefont {Calci}, \citenamefont
  {Langhammer},\ and\ \citenamefont {Roth}}]{Hergert:2013PRL}%
  \BibitemOpen
  \bibfield  {author} {\bibinfo {author} {\bibfnamefont {H.}~\bibnamefont
  {Hergert}}, \bibinfo {author} {\bibfnamefont {S.}~\bibnamefont {Binder}},
  \bibinfo {author} {\bibfnamefont {A.}~\bibnamefont {Calci}}, \bibinfo
  {author} {\bibfnamefont {J.}~\bibnamefont {Langhammer}}, \ and\ \bibinfo
  {author} {\bibfnamefont {R.}~\bibnamefont {Roth}},\ }\href {\doibase
  10.1103/PhysRevLett.110.242501} {\bibfield  {journal} {\bibinfo  {journal}
  {Phys. Rev. Lett.}\ }\textbf {\bibinfo {volume} {110}},\ \bibinfo {pages}
  {242501} (\bibinfo {year} {2013})},\ \Eprint {http://arxiv.org/abs/1302.7294}
  {arXiv:1302.7294 [nucl-th]} \BibitemShut {NoStop}%
\bibitem [{\citenamefont {Gebrerufael}\ \emph {et~al.}(2017)\citenamefont
  {Gebrerufael}, \citenamefont {Vobig}, \citenamefont {Hergert},\ and\
  \citenamefont {Roth}}]{Gebrerufael:2017PRL}%
  \BibitemOpen
  \bibfield  {author} {\bibinfo {author} {\bibfnamefont {E.}~\bibnamefont
  {Gebrerufael}}, \bibinfo {author} {\bibfnamefont {K.}~\bibnamefont {Vobig}},
  \bibinfo {author} {\bibfnamefont {H.}~\bibnamefont {Hergert}}, \ and\
  \bibinfo {author} {\bibfnamefont {R.}~\bibnamefont {Roth}},\ }\href {\doibase
  10.1103/PhysRevLett.118.152503} {\bibfield  {journal} {\bibinfo  {journal}
  {Phys. Rev. Lett.}\ }\textbf {\bibinfo {volume} {118}},\ \bibinfo {pages}
  {152503} (\bibinfo {year} {2017})},\ \Eprint
  {http://arxiv.org/abs/1610.05254} {arXiv:1610.05254 [nucl-th]} \BibitemShut
  {NoStop}%
\bibitem [{\citenamefont {Belley}\ \emph {et~al.}(2023)\citenamefont {Belley}
  \emph {et~al.}}]{Belley:2023}%
  \BibitemOpen
  \bibfield  {author} {\bibinfo {author} {\bibfnamefont {A.}~\bibnamefont
  {Belley}} \emph {et~al.},\ }\href@noop {} {\  (\bibinfo {year} {2023})},\
  \Eprint {http://arxiv.org/abs/2308.15634} {arXiv:2308.15634 [nucl-th]}
  \BibitemShut {NoStop}%
\bibitem [{\citenamefont {Tsukiyama}\ \emph {et~al.}(2011)\citenamefont
  {Tsukiyama}, \citenamefont {Bogner},\ and\ \citenamefont
  {Schwenk}}]{Tsukiyama:2011PRL}%
  \BibitemOpen
  \bibfield  {author} {\bibinfo {author} {\bibfnamefont {K.}~\bibnamefont
  {Tsukiyama}}, \bibinfo {author} {\bibfnamefont {S.~K.}\ \bibnamefont
  {Bogner}}, \ and\ \bibinfo {author} {\bibfnamefont {A.}~\bibnamefont
  {Schwenk}},\ }\href {\doibase 10.1103/PhysRevLett.106.222502} {\bibfield
  {journal} {\bibinfo  {journal} {Phys. Rev. Lett.}\ }\textbf {\bibinfo
  {volume} {106}},\ \bibinfo {pages} {222502} (\bibinfo {year} {2011})},\
  \Eprint {http://arxiv.org/abs/1006.3639} {arXiv:1006.3639 [nucl-th]}
  \BibitemShut {NoStop}%
\bibitem [{\citenamefont {Magnus}(1954)}]{Magnus:1954}%
  \BibitemOpen
  \bibfield  {author} {\bibinfo {author} {\bibfnamefont {W.}~\bibnamefont
  {Magnus}},\ }\href {\doibase 10.1002/cpa.3160070404} {\bibfield  {journal}
  {\bibinfo  {journal} {Commun. Pure Appl. Math.}\ }\textbf {\bibinfo {volume}
  {7}},\ \bibinfo {pages} {649} (\bibinfo {year} {1954})}\BibitemShut {NoStop}%
\bibitem [{\citenamefont {Morris}\ \emph {et~al.}(2015)\citenamefont {Morris},
  \citenamefont {Parzuchowski},\ and\ \citenamefont {Bogner}}]{Morris:2015}%
  \BibitemOpen
  \bibfield  {author} {\bibinfo {author} {\bibfnamefont {T.~D.}\ \bibnamefont
  {Morris}}, \bibinfo {author} {\bibfnamefont {N.}~\bibnamefont
  {Parzuchowski}}, \ and\ \bibinfo {author} {\bibfnamefont {S.~K.}\
  \bibnamefont {Bogner}},\ }\href {\doibase 10.1103/PhysRevC.92.034331}
  {\bibfield  {journal} {\bibinfo  {journal} {Phys. Rev. C}\ }\textbf {\bibinfo
  {volume} {92}},\ \bibinfo {pages} {034331} (\bibinfo {year} {2015})},\
  \Eprint {http://arxiv.org/abs/1507.06725} {arXiv:1507.06725 [nucl-th]}
  \BibitemShut {NoStop}%
\bibitem [{Ber()}]{Bernoulli}%
  \BibitemOpen
  \href@noop {} {\enquote {\bibinfo {title} {The bernoulli numbers page},}\
  }\bibinfo {howpublished} {\url{https://www.bernoulli.org/}},\ \bibinfo {note}
  {online resource for Bernoulli numbers and related mathematical
  information}\BibitemShut {NoStop}%
\bibitem [{\citenamefont {Hergert}(2017)}]{Hergert:2017PS}%
  \BibitemOpen
  \bibfield  {author} {\bibinfo {author} {\bibfnamefont {H.}~\bibnamefont
  {Hergert}},\ }\href {\doibase 10.1088/1402-4896/92/2/023002} {\bibfield
  {journal} {\bibinfo  {journal} {Phys. Scripta}\ }\textbf {\bibinfo {volume}
  {92}},\ \bibinfo {pages} {023002} (\bibinfo {year} {2017})},\ \Eprint
  {http://arxiv.org/abs/1607.06882} {arXiv:1607.06882 [nucl-th]} \BibitemShut
  {NoStop}%
\bibitem [{\citenamefont {Hergert}\ \emph {et~al.}(2017)\citenamefont
  {Hergert}, \citenamefont {Bogner}, \citenamefont {Lietz}, \citenamefont
  {Morris}, \citenamefont {Novario}, \citenamefont {Parzuchowski},\ and\
  \citenamefont {Yuan}}]{Hergert:2017bc}%
  \BibitemOpen
  \bibfield  {author} {\bibinfo {author} {\bibfnamefont {H.}~\bibnamefont
  {Hergert}}, \bibinfo {author} {\bibfnamefont {S.~K.}\ \bibnamefont {Bogner}},
  \bibinfo {author} {\bibfnamefont {J.~G.}\ \bibnamefont {Lietz}}, \bibinfo
  {author} {\bibfnamefont {T.~D.}\ \bibnamefont {Morris}}, \bibinfo {author}
  {\bibfnamefont {S.~J.}\ \bibnamefont {Novario}}, \bibinfo {author}
  {\bibfnamefont {N.~M.}\ \bibnamefont {Parzuchowski}}, \ and\ \bibinfo
  {author} {\bibfnamefont {F.}~\bibnamefont {Yuan}},\ }\enquote {\bibinfo
  {title} {In-medium similarity renormalization group approach to the nuclear
  many-body problem},}\ in\ \href {\doibase 10.1007/978-3-319-53336-0{\_}10}
  {\emph {\bibinfo {booktitle} {An Advanced Course in Computational Nuclear
  Physics: Bridging the Scales from Quarks to Neutron Stars}}},\ \bibinfo
  {editor} {edited by\ \bibinfo {editor} {\bibfnamefont {M.}~\bibnamefont
  {Hjorth-Jensen}}, \bibinfo {editor} {\bibfnamefont {M.~P.}\ \bibnamefont
  {Lombardo}}, \ and\ \bibinfo {editor} {\bibfnamefont {U.}~\bibnamefont {van
  Kolck}}}\ (\bibinfo  {publisher} {Springer International Publishing},\
  \bibinfo {address} {Cham},\ \bibinfo {year} {2017})\ pp.\ \bibinfo {pages}
  {477--570}\BibitemShut {NoStop}%
\bibitem [{\citenamefont {Parzuchowski}\ \emph
  {et~al.}(2017{\natexlab{a}})\citenamefont {Parzuchowski}, \citenamefont
  {Morris},\ and\ \citenamefont {Bogner}}]{Parzuchowski:2017yq}%
  \BibitemOpen
  \bibfield  {author} {\bibinfo {author} {\bibfnamefont {N.~M.}\ \bibnamefont
  {Parzuchowski}}, \bibinfo {author} {\bibfnamefont {T.~D.}\ \bibnamefont
  {Morris}}, \ and\ \bibinfo {author} {\bibfnamefont {S.~K.}\ \bibnamefont
  {Bogner}},\ }\href {\doibase 10.1103/PhysRevC.95.044304} {\bibfield
  {journal} {\bibinfo  {journal} {Phys. Rev. C}\ }\textbf {\bibinfo {volume}
  {95}},\ \bibinfo {pages} {044304} (\bibinfo {year}
  {2017}{\natexlab{a}})}\BibitemShut {NoStop}%
\bibitem [{\citenamefont {Ring}\ and\ \citenamefont
  {Schuck}(1980)}]{Ring:1980}%
  \BibitemOpen
  \bibfield  {author} {\bibinfo {author} {\bibfnamefont {P.}~\bibnamefont
  {Ring}}\ and\ \bibinfo {author} {\bibfnamefont {P.}~\bibnamefont {Schuck}},\
  }\href@noop {} {\emph {\bibinfo {title} {The nuclear many-body problem}}}\
  (\bibinfo  {publisher} {Springer-Verlag},\ \bibinfo {address} {New York},\
  \bibinfo {year} {1980})\BibitemShut {NoStop}%
\bibitem [{\citenamefont {Bally}\ \emph {et~al.}(2021)\citenamefont {Bally},
  \citenamefont {S\'anchez-Fern\'andez},\ and\ \citenamefont
  {Rodr\'\i{}guez}}]{Bally:2021_EPJA}%
  \BibitemOpen
  \bibfield  {author} {\bibinfo {author} {\bibfnamefont {B.}~\bibnamefont
  {Bally}}, \bibinfo {author} {\bibfnamefont {A.}~\bibnamefont
  {S\'anchez-Fern\'andez}}, \ and\ \bibinfo {author} {\bibfnamefont {T.~R.}\
  \bibnamefont {Rodr\'\i{}guez}},\ }\href {\doibase
  10.1140/epja/s10050-021-00369-z} {\bibfield  {journal} {\bibinfo  {journal}
  {Eur. Phys. J. A}\ }\textbf {\bibinfo {volume} {57}},\ \bibinfo {pages} {69}
  (\bibinfo {year} {2021})},\ \bibinfo {note} {[Erratum: Eur.Phys.J.A 57, 124
  (2021)]},\ \Eprint {http://arxiv.org/abs/2010.14169} {arXiv:2010.14169
  [nucl-th]} \BibitemShut {NoStop}%
\bibitem [{\citenamefont {Hill}\ and\ \citenamefont
  {Wheeler}(1953)}]{Hill:1953}%
  \BibitemOpen
  \bibfield  {author} {\bibinfo {author} {\bibfnamefont {D.~L.}\ \bibnamefont
  {Hill}}\ and\ \bibinfo {author} {\bibfnamefont {J.~A.}\ \bibnamefont
  {Wheeler}},\ }\href {\doibase 10.1103/PhysRev.89.1102} {\bibfield  {journal}
  {\bibinfo  {journal} {Phys. Rev.}\ }\textbf {\bibinfo {volume} {89}},\
  \bibinfo {pages} {1102} (\bibinfo {year} {1953})}\BibitemShut {NoStop}%
\bibitem [{\citenamefont {Avez}\ and\ \citenamefont
  {Bender}(2012)}]{Avez:2012PRC}%
  \BibitemOpen
  \bibfield  {author} {\bibinfo {author} {\bibfnamefont {B.}~\bibnamefont
  {Avez}}\ and\ \bibinfo {author} {\bibfnamefont {M.}~\bibnamefont {Bender}},\
  }\href {\doibase 10.1103/PhysRevC.85.034325} {\bibfield  {journal} {\bibinfo
  {journal} {Phys. Rev. C}\ }\textbf {\bibinfo {volume} {85}},\ \bibinfo
  {pages} {034325} (\bibinfo {year} {2012})}\BibitemShut {NoStop}%
\bibitem [{\citenamefont {Yao}\ \emph {et~al.}(2010)\citenamefont {Yao},
  \citenamefont {Meng}, \citenamefont {Ring},\ and\ \citenamefont
  {Vretenar}}]{Yao:2010}%
  \BibitemOpen
  \bibfield  {author} {\bibinfo {author} {\bibfnamefont {J.~M.}\ \bibnamefont
  {Yao}}, \bibinfo {author} {\bibfnamefont {J.}~\bibnamefont {Meng}}, \bibinfo
  {author} {\bibfnamefont {P.}~\bibnamefont {Ring}}, \ and\ \bibinfo {author}
  {\bibfnamefont {D.}~\bibnamefont {Vretenar}},\ }\href {\doibase
  10.1103/PhysRevC.81.044311} {\bibfield  {journal} {\bibinfo  {journal} {Phys.
  Rev. C}\ }\textbf {\bibinfo {volume} {81}},\ \bibinfo {pages} {044311}
  (\bibinfo {year} {2010})},\ \Eprint {http://arxiv.org/abs/0912.2650}
  {arXiv:0912.2650 [nucl-th]} \BibitemShut {NoStop}%
\bibitem [{\citenamefont {Yao}\ \emph {et~al.}(2015)\citenamefont {Yao},
  \citenamefont {Bender},\ and\ \citenamefont {Heenen}}]{Yao:2015TD}%
  \BibitemOpen
  \bibfield  {author} {\bibinfo {author} {\bibfnamefont {J.~M.}\ \bibnamefont
  {Yao}}, \bibinfo {author} {\bibfnamefont {M.}~\bibnamefont {Bender}}, \ and\
  \bibinfo {author} {\bibfnamefont {P.-H.}\ \bibnamefont {Heenen}},\ }\href
  {\doibase 10.1103/PhysRevC.91.024301} {\bibfield  {journal} {\bibinfo
  {journal} {Phys. Rev. C}\ }\textbf {\bibinfo {volume} {91}},\ \bibinfo
  {pages} {024301} (\bibinfo {year} {2015})}\BibitemShut {NoStop}%
\bibitem [{\citenamefont {Gao}\ \emph {et~al.}(2023)\citenamefont {Gao},
  \citenamefont {Chen},\ and\ \citenamefont {Wang}}]{Gao:PRC2023}%
  \BibitemOpen
  \bibfield  {author} {\bibinfo {author} {\bibfnamefont {F.}~\bibnamefont
  {Gao}}, \bibinfo {author} {\bibfnamefont {Z.-R.}\ \bibnamefont {Chen}}, \
  and\ \bibinfo {author} {\bibfnamefont {L.-J.}\ \bibnamefont {Wang}},\ }\href
  {\doibase 10.1103/PhysRevC.108.054313} {\bibfield  {journal} {\bibinfo
  {journal} {Phys. Rev. C}\ }\textbf {\bibinfo {volume} {108}},\ \bibinfo
  {pages} {054313} (\bibinfo {year} {2023})}\BibitemShut {NoStop}%
\bibitem [{\citenamefont {Miyagi}\ \emph {et~al.}(2024)\citenamefont {Miyagi},
  \citenamefont {Cao}, \citenamefont {Seutin}, \citenamefont {Bacca},
  \citenamefont {Garcia~Ruiz}, \citenamefont {Hebeler}, \citenamefont {Holt},\
  and\ \citenamefont {Schwenk}}]{Miyagi:2024}%
  \BibitemOpen
  \bibfield  {author} {\bibinfo {author} {\bibfnamefont {T.}~\bibnamefont
  {Miyagi}}, \bibinfo {author} {\bibfnamefont {X.}~\bibnamefont {Cao}},
  \bibinfo {author} {\bibfnamefont {R.}~\bibnamefont {Seutin}}, \bibinfo
  {author} {\bibfnamefont {S.}~\bibnamefont {Bacca}}, \bibinfo {author}
  {\bibfnamefont {R.~F.}\ \bibnamefont {Garcia~Ruiz}}, \bibinfo {author}
  {\bibfnamefont {K.}~\bibnamefont {Hebeler}}, \bibinfo {author} {\bibfnamefont
  {J.~D.}\ \bibnamefont {Holt}}, \ and\ \bibinfo {author} {\bibfnamefont
  {A.}~\bibnamefont {Schwenk}},\ }\href {\doibase
  10.1103/PhysRevLett.132.232503} {\bibfield  {journal} {\bibinfo  {journal}
  {Phys. Rev. Lett.}\ }\textbf {\bibinfo {volume} {132}},\ \bibinfo {pages}
  {232503} (\bibinfo {year} {2024})},\ \Eprint
  {http://arxiv.org/abs/2311.14383} {arXiv:2311.14383 [nucl-th]} \BibitemShut
  {NoStop}%
\bibitem [{\citenamefont {Stevenson}\ \emph {et~al.}(2002)\citenamefont
  {Stevenson}, \citenamefont {Rikovska Stone},\ and\ \citenamefont
  {Strayer}}]{Stevenson:2002}%
  \BibitemOpen
  \bibfield  {author} {\bibinfo {author} {\bibfnamefont {P.}~\bibnamefont
  {Stevenson}}, \bibinfo {author} {\bibfnamefont {J.}~\bibnamefont
  {Rikovska Stone}}, \ and\ \bibinfo {author} {\bibfnamefont {M.}~\bibnamefont
  {Strayer}},\ }\href {\doibase https://doi.org/10.1016/S0370-2693(02)02634-5}
  {\bibfield  {journal} {\bibinfo  {journal} {Phys. Lett. B}\ }\textbf
  {\bibinfo {volume} {545}},\ \bibinfo {pages} {291} (\bibinfo {year}
  {2002})}\BibitemShut {NoStop}%
\bibitem [{\citenamefont {Kimura}\ and\ \citenamefont
  {Horiuchi}(2002)}]{Kimura:2002}%
  \BibitemOpen
  \bibfield  {author} {\bibinfo {author} {\bibfnamefont {M.}~\bibnamefont
  {Kimura}}\ and\ \bibinfo {author} {\bibfnamefont {H.}~\bibnamefont
  {Horiuchi}},\ }\href {\doibase 10.1143/PTP.107.33} {\bibfield  {journal}
  {\bibinfo  {journal} {Progress of Theoretical Physics}\ }\textbf {\bibinfo
  {volume} {107}},\ \bibinfo {pages} {33} (\bibinfo {year} {2002})},\ \Eprint
  {http://arxiv.org/abs/https://academic.oup.com/ptp/article-pdf/107/1/33/5213174/107-1-33.pdf}
  {https://academic.oup.com/ptp/article-pdf/107/1/33/5213174/107-1-33.pdf}
  \BibitemShut {NoStop}%
\bibitem [{\citenamefont {Doornenbal}\ \emph
  {et~al.}(2016{\natexlab{b}})\citenamefont {Doornenbal} \emph
  {et~al.}}]{Ne30E2Doornenbal:2016}%
  \BibitemOpen
  \bibfield  {author} {\bibinfo {author} {\bibfnamefont {P.}~\bibnamefont
  {Doornenbal}} \emph {et~al.},\ }\href {\doibase 10.1103/PhysRevC.93.044306}
  {\bibfield  {journal} {\bibinfo  {journal} {Phys. Rev. C}\ }\textbf {\bibinfo
  {volume} {93}},\ \bibinfo {pages} {044306} (\bibinfo {year}
  {2016}{\natexlab{b}})}\BibitemShut {NoStop}%
\bibitem [{\citenamefont {{National Nuclear Data Center}}(2020)}]{NNDC}%
  \BibitemOpen
  \bibfield  {author} {\bibinfo {author} {\bibnamefont {{National Nuclear Data
  Center}}},\ }\href {https://www.nndc.bnl.gov/nudat2} {\enquote {\bibinfo
  {title} {{NuDat 2 Database}},}\ } (\bibinfo {year} {2020}),\ \bibinfo {note}
  {\url{https://www.nndc.bnl.gov/nudat2}}\BibitemShut {NoStop}%
\bibitem [{\citenamefont {Shamsuzzoha~Basunia}(2010)}]{sheeta=30}%
  \BibitemOpen
  \bibfield  {author} {\bibinfo {author} {\bibfnamefont {M.}~\bibnamefont
  {Shamsuzzoha~Basunia}},\ }\href {\doibase 10.1016/j.nds.2010.09.001}
  {\bibfield  {journal} {\bibinfo  {journal} {Nucl. Data Sheets}\ }\textbf
  {\bibinfo {volume} {111}},\ \bibinfo {pages} {2331} (\bibinfo {year}
  {2010})}\BibitemShut {NoStop}%
\bibitem [{\citenamefont {Rotaru}\ \emph {et~al.}(2012)\citenamefont {Rotaru}
  \emph {et~al.}}]{Rotaru:2012Si34isomer}%
  \BibitemOpen
  \bibfield  {author} {\bibinfo {author} {\bibfnamefont {F.}~\bibnamefont
  {Rotaru}} \emph {et~al.},\ }\href {\doibase 10.1103/PhysRevLett.109.092503}
  {\bibfield  {journal} {\bibinfo  {journal} {Phys. Rev. Lett.}\ }\textbf
  {\bibinfo {volume} {109}},\ \bibinfo {pages} {092503} (\bibinfo {year}
  {2012})},\ \Eprint {http://arxiv.org/abs/1206.6572} {arXiv:1206.6572
  [nucl-ex]} \BibitemShut {NoStop}%
\bibitem [{\citenamefont {Bazin}\ \emph {et~al.}(2021)\citenamefont {Bazin}
  \emph {et~al.}}]{Bazin:2021nti}%
  \BibitemOpen
  \bibfield  {author} {\bibinfo {author} {\bibfnamefont {D.}~\bibnamefont
  {Bazin}} \emph {et~al.},\ }\href {\doibase 10.1103/PhysRevC.103.064318}
  {\bibfield  {journal} {\bibinfo  {journal} {Phys. Rev. C}\ }\textbf {\bibinfo
  {volume} {103}},\ \bibinfo {pages} {064318} (\bibinfo {year} {2021})},\
  \Eprint {http://arxiv.org/abs/2101.12182} {arXiv:2101.12182 [nucl-ex]}
  \BibitemShut {NoStop}%
\bibitem [{\citenamefont {Yao}(2022)}]{Yao:2022_HB}%
  \BibitemOpen
  \bibfield  {author} {\bibinfo {author} {\bibfnamefont {J.~M.}\ \bibnamefont
  {Yao}},\ }\enquote {\bibinfo {title} {{Symmetry Restoration Methods}},}\ in\
  \href {\doibase 10.1007/978-981-15-8818-1_18-1} {\emph {\bibinfo {booktitle}
  {{Handbook of Nuclear Physics}}}},\ \bibinfo {editor} {edited by\ \bibinfo
  {editor} {\bibfnamefont {I.}~\bibnamefont {Tanihata}}, \bibinfo {editor}
  {\bibfnamefont {H.}~\bibnamefont {Toki}}, \ and\ \bibinfo {editor}
  {\bibfnamefont {T.}~\bibnamefont {Kajino}}}\ (\bibinfo  {publisher} {Springer
  Nature Singapore},\ \bibinfo {year} {2022})\ pp.\ \bibinfo {pages} {1--36},\
  \Eprint {http://arxiv.org/abs/2204.12126} {arXiv:2204.12126 [nucl-th]}
  \BibitemShut {NoStop}%
\bibitem [{\citenamefont {Doornenbal}\ \emph {et~al.}(2017)\citenamefont
  {Doornenbal} \emph {et~al.}}]{Doornenbal:2017}%
  \BibitemOpen
  \bibfield  {author} {\bibinfo {author} {\bibfnamefont {P.}~\bibnamefont
  {Doornenbal}} \emph {et~al.},\ }\href {\doibase 10.1103/PhysRevC.95.041301}
  {\bibfield  {journal} {\bibinfo  {journal} {Phys. Rev. C}\ }\textbf {\bibinfo
  {volume} {95}},\ \bibinfo {pages} {041301} (\bibinfo {year}
  {2017})}\BibitemShut {NoStop}%
\bibitem [{\citenamefont {Kahlbow}(2024)}]{Kahlbow:2024}%
  \BibitemOpen
  \bibfield  {author} {\bibinfo {author} {\bibfnamefont {J.}~\bibnamefont
  {Kahlbow}}\ }(\bibinfo {year} {2024})\ \Eprint
  {http://arxiv.org/abs/2412.16799} {arXiv:2412.16799 [nucl-ex]} \BibitemShut
  {NoStop}%
\bibitem [{\citenamefont {Kimura}(2007)}]{Kimura:2007}%
  \BibitemOpen
  \bibfield  {author} {\bibinfo {author} {\bibfnamefont {M.}~\bibnamefont
  {Kimura}},\ }\href {\doibase 10.1103/PhysRevC.75.041302} {\bibfield
  {journal} {\bibinfo  {journal} {Phys. Rev. C}\ }\textbf {\bibinfo {volume}
  {75}},\ \bibinfo {pages} {041302} (\bibinfo {year} {2007})},\ \Eprint
  {http://arxiv.org/abs/nucl-th/0702012} {arXiv:nucl-th/0702012} \BibitemShut
  {NoStop}%
\bibitem [{\citenamefont {Sahoo}\ \emph {et~al.}(2023)\citenamefont {Sahoo},
  \citenamefont {Srivastava},\ and\ \citenamefont {Suzuki}}]{Naisoptop2023}%
  \BibitemOpen
  \bibfield  {author} {\bibinfo {author} {\bibfnamefont {S.}~\bibnamefont
  {Sahoo}}, \bibinfo {author} {\bibfnamefont {P.~C.}\ \bibnamefont
  {Srivastava}}, \ and\ \bibinfo {author} {\bibfnamefont {T.}~\bibnamefont
  {Suzuki}},\ }\href {\doibase 10.1016/j.nuclphysa.2023.122618} {\bibfield
  {journal} {\bibinfo  {journal} {Nucl. Phys. A}\ }\textbf {\bibinfo {volume}
  {1032}},\ \bibinfo {pages} {122618} (\bibinfo {year} {2023})},\ \Eprint
  {http://arxiv.org/abs/2211.04133} {arXiv:2211.04133 [nucl-th]} \BibitemShut
  {NoStop}%
\bibitem [{\citenamefont {Geithner}\ \emph {et~al.}(2000)\citenamefont
  {Geithner}, \citenamefont {Georg}, \citenamefont {Kappertz}, \citenamefont
  {Keim}, \citenamefont {Klein}, \citenamefont {Lievens}, \citenamefont
  {Neugart}, \citenamefont {Neuroth}, \citenamefont {Vermeeren},\ and\
  \citenamefont {Wilbert}}]{Na31qs}%
  \BibitemOpen
  \bibfield  {author} {\bibinfo {author} {\bibfnamefont {W.}~\bibnamefont
  {Geithner}}, \bibinfo {author} {\bibfnamefont {U.}~\bibnamefont {Georg}},
  \bibinfo {author} {\bibfnamefont {S.}~\bibnamefont {Kappertz}}, \bibinfo
  {author} {\bibfnamefont {M.}~\bibnamefont {Keim}}, \bibinfo {author}
  {\bibfnamefont {A.}~\bibnamefont {Klein}}, \bibinfo {author} {\bibfnamefont
  {P.}~\bibnamefont {Lievens}}, \bibinfo {author} {\bibfnamefont
  {R.}~\bibnamefont {Neugart}}, \bibinfo {author} {\bibfnamefont
  {M.}~\bibnamefont {Neuroth}}, \bibinfo {author} {\bibfnamefont
  {L.}~\bibnamefont {Vermeeren}}, \ and\ \bibinfo {author} {\bibfnamefont
  {S.}~\bibnamefont {Wilbert}},\ }\href {\doibase 10.1023/A:1012642707147}
  {\bibfield  {journal} {\bibinfo  {journal} {Hyperfine Interactions}\ }\textbf
  {\bibinfo {volume} {129}},\ \bibinfo {pages} {271} (\bibinfo {year}
  {2000})}\BibitemShut {NoStop}%
\bibitem [{\citenamefont {Ouellet}\ and\ \citenamefont
  {Singh}(2013)}]{SheetA=31}%
  \BibitemOpen
  \bibfield  {author} {\bibinfo {author} {\bibfnamefont {C.}~\bibnamefont
  {Ouellet}}\ and\ \bibinfo {author} {\bibfnamefont {B.}~\bibnamefont
  {Singh}},\ }\href {\doibase 10.1016/j.nds.2013.03.001} {\bibfield  {journal}
  {\bibinfo  {journal} {Nucl. Data Sheets}\ }\textbf {\bibinfo {volume}
  {114}},\ \bibinfo {pages} {209} (\bibinfo {year} {2013})}\BibitemShut
  {NoStop}%
\bibitem [{\citenamefont {Shamsuzzoha~Basunia}(2012)}]{sheeta=29}%
  \BibitemOpen
  \bibfield  {author} {\bibinfo {author} {\bibfnamefont {M.}~\bibnamefont
  {Shamsuzzoha~Basunia}},\ }\href {\doibase 10.1016/j.nds.2012.04.001}
  {\bibfield  {journal} {\bibinfo  {journal} {Nucl. Data Sheets}\ }\textbf
  {\bibinfo {volume} {113}},\ \bibinfo {pages} {909} (\bibinfo {year}
  {2012})}\BibitemShut {NoStop}%
\bibitem [{\citenamefont {Chen}\ and\ \citenamefont {Singh}(2011)}]{sheetA=33}%
  \BibitemOpen
  \bibfield  {author} {\bibinfo {author} {\bibfnamefont {J.}~\bibnamefont
  {Chen}}\ and\ \bibinfo {author} {\bibfnamefont {B.}~\bibnamefont {Singh}},\
  }\href {\doibase 10.1016/j.nds.2011.04.003} {\bibfield  {journal} {\bibinfo
  {journal} {Nucl. Data Sheets}\ }\textbf {\bibinfo {volume} {112}},\ \bibinfo
  {pages} {1393} (\bibinfo {year} {2011})}\BibitemShut {NoStop}%
\bibitem [{\citenamefont {Chen}\ \emph {et~al.}(2011)\citenamefont {Chen},
  \citenamefont {Cameron},\ and\ \citenamefont {Singh}}]{sheeta=35}%
  \BibitemOpen
  \bibfield  {author} {\bibinfo {author} {\bibfnamefont {J.}~\bibnamefont
  {Chen}}, \bibinfo {author} {\bibfnamefont {J.}~\bibnamefont {Cameron}}, \
  and\ \bibinfo {author} {\bibfnamefont {B.}~\bibnamefont {Singh}},\ }\href
  {\doibase 10.1016/j.nds.2011.10.001} {\bibfield  {journal} {\bibinfo
  {journal} {Nucl. Data Sheets}\ }\textbf {\bibinfo {volume} {112}},\ \bibinfo
  {pages} {2715} (\bibinfo {year} {2011})}\BibitemShut {NoStop}%
\bibitem [{\citenamefont {Shimada}\ \emph {et~al.}(2012)\citenamefont {Shimada}
  \emph {et~al.}}]{qsAl33}%
  \BibitemOpen
  \bibfield  {author} {\bibinfo {author} {\bibfnamefont {K.}~\bibnamefont
  {Shimada}} \emph {et~al.},\ }\href {\doibase 10.1016/j.physletb.2012.07.030}
  {\bibfield  {journal} {\bibinfo  {journal} {Phys. Lett. B}\ }\textbf
  {\bibinfo {volume} {714}},\ \bibinfo {pages} {246} (\bibinfo {year}
  {2012})}\BibitemShut {NoStop}%
\bibitem [{\citenamefont {Yordanov}\ \emph {et~al.}(2019)\citenamefont
  {Yordanov} \emph {et~al.}}]{qsMg33}%
  \BibitemOpen
  \bibfield  {author} {\bibinfo {author} {\bibfnamefont {D.~T.}\ \bibnamefont
  {Yordanov}} \emph {et~al.},\ }\href {\doibase 10.1007/s10751-019-1609-4}
  {\bibfield  {journal} {\bibinfo  {journal} {Hyperfine Interact.}\ }\textbf
  {\bibinfo {volume} {240}},\ \bibinfo {pages} {67} (\bibinfo {year} {2019})},\
  \Eprint {http://arxiv.org/abs/1905.05580} {arXiv:1905.05580 [nucl-ex]}
  \BibitemShut {NoStop}%
\bibitem [{\citenamefont {Varshalovich}\ \emph {et~al.}(1988)\citenamefont
  {Varshalovich}, \citenamefont {Moskalev},\ and\ \citenamefont
  {Khersonskii}}]{Varshalovich:1988}%
  \BibitemOpen
  \bibfield  {author} {\bibinfo {author} {\bibfnamefont {D.~A.}\ \bibnamefont
  {Varshalovich}}, \bibinfo {author} {\bibfnamefont {A.~N.}\ \bibnamefont
  {Moskalev}}, \ and\ \bibinfo {author} {\bibfnamefont {V.~K.}\ \bibnamefont
  {Khersonskii}},\ }\href {\doibase 10.1142/0270} {\emph {\bibinfo {title}
  {{Quantum Theory of Angular Momentum}: {Irreducible Tensors, Spherical
  Harmonics, Vector Coupling Coefficients, 3nj Symbols}}}}\ (\bibinfo
  {publisher} {World Scientific Publishing Company},\ \bibinfo {year}
  {1988})\BibitemShut {NoStop}%
\bibitem [{\citenamefont {Salinas}\ \emph {et~al.}(2026)\citenamefont
  {Salinas}, \citenamefont {Iwasaki}, \citenamefont {Revel}, \citenamefont
  {Brown}, \citenamefont {Ash}, \citenamefont {Bazin}, \citenamefont {Chen},
  \citenamefont {Elder}, \citenamefont {Farris}, \citenamefont {Gade},
  \citenamefont {Grinder}, \citenamefont {Kobayashi}, \citenamefont {Li},
  \citenamefont {Longfellow}, \citenamefont {Mijatovi\ifmmode~\acute{c}\else
  \'{c}\fi{}}, \citenamefont {Pereira}, \citenamefont {Sanchez}, \citenamefont
  {Spieker}, \citenamefont {Utsuno}, \citenamefont {Weisshaar},\ and\
  \citenamefont {Wu}}]{Salinas:2026}%
  \BibitemOpen
  \bibfield  {author} {\bibinfo {author} {\bibfnamefont {R.}~\bibnamefont
  {Salinas}}, \bibinfo {author} {\bibfnamefont {H.}~\bibnamefont {Iwasaki}},
  \bibinfo {author} {\bibfnamefont {A.}~\bibnamefont {Revel}}, \bibinfo
  {author} {\bibfnamefont {B.~A.}\ \bibnamefont {Brown}}, \bibinfo {author}
  {\bibfnamefont {J.}~\bibnamefont {Ash}}, \bibinfo {author} {\bibfnamefont
  {D.}~\bibnamefont {Bazin}}, \bibinfo {author} {\bibfnamefont
  {J.}~\bibnamefont {Chen}}, \bibinfo {author} {\bibfnamefont {R.}~\bibnamefont
  {Elder}}, \bibinfo {author} {\bibfnamefont {P.}~\bibnamefont {Farris}},
  \bibinfo {author} {\bibfnamefont {A.}~\bibnamefont {Gade}}, \bibinfo {author}
  {\bibfnamefont {M.}~\bibnamefont {Grinder}}, \bibinfo {author} {\bibfnamefont
  {N.}~\bibnamefont {Kobayashi}}, \bibinfo {author} {\bibfnamefont
  {J.}~\bibnamefont {Li}}, \bibinfo {author} {\bibfnamefont {B.}~\bibnamefont
  {Longfellow}}, \bibinfo {author} {\bibfnamefont {T.}~\bibnamefont
  {Mijatovi\ifmmode~\acute{c}\else \'{c}\fi{}}}, \bibinfo {author}
  {\bibfnamefont {J.}~\bibnamefont {Pereira}}, \bibinfo {author} {\bibfnamefont
  {A.}~\bibnamefont {Sanchez}}, \bibinfo {author} {\bibfnamefont
  {M.}~\bibnamefont {Spieker}}, \bibinfo {author} {\bibfnamefont
  {Y.}~\bibnamefont {Utsuno}}, \bibinfo {author} {\bibfnamefont
  {D.}~\bibnamefont {Weisshaar}}, \ and\ \bibinfo {author} {\bibfnamefont
  {J.}~\bibnamefont {Wu}},\ }\href {\doibase 10.1103/zp75-1ctq} {\bibfield
  {journal} {\bibinfo  {journal} {Phys. Rev. C}\ }\textbf {\bibinfo {volume}
  {113}},\ \bibinfo {pages} {014330} (\bibinfo {year} {2026})}\BibitemShut
  {NoStop}%
\bibitem [{\citenamefont {Wang}\ \emph {et~al.}(2021)\citenamefont {Wang},
  \citenamefont {Huang}, \citenamefont {Kondev}, \citenamefont {Audi},\ and\
  \citenamefont {Naimi}}]{sheetformass}%
  \BibitemOpen
  \bibfield  {author} {\bibinfo {author} {\bibfnamefont {M.}~\bibnamefont
  {Wang}}, \bibinfo {author} {\bibfnamefont {W.~J.}\ \bibnamefont {Huang}},
  \bibinfo {author} {\bibfnamefont {F.~G.}\ \bibnamefont {Kondev}}, \bibinfo
  {author} {\bibfnamefont {G.}~\bibnamefont {Audi}}, \ and\ \bibinfo {author}
  {\bibfnamefont {S.}~\bibnamefont {Naimi}},\ }\href {\doibase
  10.1088/1674-1137/abddaf} {\bibfield  {journal} {\bibinfo  {journal} {Chin.
  Phys. C}\ }\textbf {\bibinfo {volume} {45}},\ \bibinfo {pages} {030003}
  (\bibinfo {year} {2021})}\BibitemShut {NoStop}%
\bibitem [{\citenamefont {Angeli}\ and\ \citenamefont
  {Marinova}(2013)}]{radii}%
  \BibitemOpen
  \bibfield  {author} {\bibinfo {author} {\bibfnamefont {I.}~\bibnamefont
  {Angeli}}\ and\ \bibinfo {author} {\bibfnamefont {K.~P.}\ \bibnamefont
  {Marinova}},\ }\href {\doibase 10.1016/j.adt.2011.12.006} {\bibfield
  {journal} {\bibinfo  {journal} {Atom. Data Nucl. Data Tabl.}\ }\textbf
  {\bibinfo {volume} {99}},\ \bibinfo {pages} {69} (\bibinfo {year}
  {2013})}\BibitemShut {NoStop}%
\bibitem [{\citenamefont {Yordanov}\ \emph {et~al.}(2012)\citenamefont
  {Yordanov}, \citenamefont {Bissell}, \citenamefont {Blaum}, \citenamefont
  {De~Rydt}, \citenamefont {Geppert}, \citenamefont {Kowalska}, \citenamefont
  {Kr\"amer}, \citenamefont {Kreim}, \citenamefont {Krieger}, \citenamefont
  {Lievens}, \citenamefont {Neff}, \citenamefont {Neugart}, \citenamefont
  {Neyens}, \citenamefont {N\"ortersh\"auser}, \citenamefont {S\'anchez},\ and\
  \citenamefont {Vingerhoets}}]{Yordanov:2012}%
  \BibitemOpen
  \bibfield  {author} {\bibinfo {author} {\bibfnamefont {D.~T.}\ \bibnamefont
  {Yordanov}}, \bibinfo {author} {\bibfnamefont {M.~L.}\ \bibnamefont
  {Bissell}}, \bibinfo {author} {\bibfnamefont {K.}~\bibnamefont {Blaum}},
  \bibinfo {author} {\bibfnamefont {M.}~\bibnamefont {De~Rydt}}, \bibinfo
  {author} {\bibfnamefont {C.}~\bibnamefont {Geppert}}, \bibinfo {author}
  {\bibfnamefont {M.}~\bibnamefont {Kowalska}}, \bibinfo {author}
  {\bibfnamefont {J.}~\bibnamefont {Kr\"amer}}, \bibinfo {author}
  {\bibfnamefont {K.}~\bibnamefont {Kreim}}, \bibinfo {author} {\bibfnamefont
  {A.}~\bibnamefont {Krieger}}, \bibinfo {author} {\bibfnamefont
  {P.}~\bibnamefont {Lievens}}, \bibinfo {author} {\bibfnamefont
  {T.}~\bibnamefont {Neff}}, \bibinfo {author} {\bibfnamefont {R.}~\bibnamefont
  {Neugart}}, \bibinfo {author} {\bibfnamefont {G.}~\bibnamefont {Neyens}},
  \bibinfo {author} {\bibfnamefont {W.}~\bibnamefont {N\"ortersh\"auser}},
  \bibinfo {author} {\bibfnamefont {R.}~\bibnamefont {S\'anchez}}, \ and\
  \bibinfo {author} {\bibfnamefont {P.}~\bibnamefont {Vingerhoets}},\ }\href
  {\doibase 10.1103/PhysRevLett.108.042504} {\bibfield  {journal} {\bibinfo
  {journal} {Phys. Rev. Lett.}\ }\textbf {\bibinfo {volume} {108}},\ \bibinfo
  {pages} {042504} (\bibinfo {year} {2012})}\BibitemShut {NoStop}%
\bibitem [{\citenamefont {Ekstr\"om}\ \emph {et~al.}(2015)\citenamefont
  {Ekstr\"om}, \citenamefont {Jansen}, \citenamefont {Wendt}, \citenamefont
  {Hagen}, \citenamefont {Papenbrock}, \citenamefont {Carlsson}, \citenamefont
  {Forss\'en}, \citenamefont {Hjorth-Jensen}, \citenamefont {Navr\'atil},\ and\
  \citenamefont {Nazarewicz}}]{radiiexp}%
  \BibitemOpen
  \bibfield  {author} {\bibinfo {author} {\bibfnamefont {A.}~\bibnamefont
  {Ekstr\"om}}, \bibinfo {author} {\bibfnamefont {G.~R.}\ \bibnamefont
  {Jansen}}, \bibinfo {author} {\bibfnamefont {K.~A.}\ \bibnamefont {Wendt}},
  \bibinfo {author} {\bibfnamefont {G.}~\bibnamefont {Hagen}}, \bibinfo
  {author} {\bibfnamefont {T.}~\bibnamefont {Papenbrock}}, \bibinfo {author}
  {\bibfnamefont {B.~D.}\ \bibnamefont {Carlsson}}, \bibinfo {author}
  {\bibfnamefont {C.}~\bibnamefont {Forss\'en}}, \bibinfo {author}
  {\bibfnamefont {M.}~\bibnamefont {Hjorth-Jensen}}, \bibinfo {author}
  {\bibfnamefont {P.}~\bibnamefont {Navr\'atil}}, \ and\ \bibinfo {author}
  {\bibfnamefont {W.}~\bibnamefont {Nazarewicz}},\ }\href {\doibase
  10.1103/PhysRevC.109.059901} {\bibfield  {journal} {\bibinfo  {journal}
  {Phys. Rev. C}\ }\textbf {\bibinfo {volume} {91}},\ \bibinfo {pages} {051301}
  (\bibinfo {year} {2015})},\ \bibinfo {note} {[Erratum: Phys.Rev.C 109, 059901
  (2024)]},\ \Eprint {http://arxiv.org/abs/1502.04682} {arXiv:1502.04682
  [nucl-th]} \BibitemShut {NoStop}%
\bibitem [{\citenamefont {Lykiardopoulou}\ \emph {et~al.}(2025)\citenamefont
  {Lykiardopoulou} \emph {et~al.}}]{Lykiardopoulou:2025}%
  \BibitemOpen
  \bibfield  {author} {\bibinfo {author} {\bibfnamefont {E.~M.}\ \bibnamefont
  {Lykiardopoulou}} \emph {et~al.},\ }\href {\doibase
  10.1103/PhysRevLett.134.052503} {\bibfield  {journal} {\bibinfo  {journal}
  {Phys. Rev. Lett.}\ }\textbf {\bibinfo {volume} {134}},\ \bibinfo {pages}
  {052503} (\bibinfo {year} {2025})},\ \Eprint
  {http://arxiv.org/abs/2502.06720} {arXiv:2502.06720 [nucl-ex]} \BibitemShut
  {NoStop}%
\bibitem [{\citenamefont {Novario}\ \emph {et~al.}(2020)\citenamefont
  {Novario}, \citenamefont {Hagen}, \citenamefont {Jansen},\ and\ \citenamefont
  {Papenbrock}}]{Novario:2020}%
  \BibitemOpen
  \bibfield  {author} {\bibinfo {author} {\bibfnamefont {S.~J.}\ \bibnamefont
  {Novario}}, \bibinfo {author} {\bibfnamefont {G.}~\bibnamefont {Hagen}},
  \bibinfo {author} {\bibfnamefont {G.~R.}\ \bibnamefont {Jansen}}, \ and\
  \bibinfo {author} {\bibfnamefont {T.}~\bibnamefont {Papenbrock}},\ }\href
  {\doibase 10.1103/PhysRevC.102.051303} {\bibfield  {journal} {\bibinfo
  {journal} {Phys. Rev. C}\ }\textbf {\bibinfo {volume} {102}},\ \bibinfo
  {pages} {051303} (\bibinfo {year} {2020})}\BibitemShut {NoStop}%
\bibitem [{\citenamefont {Jiang}\ \emph {et~al.}(2020)\citenamefont {Jiang},
  \citenamefont {Ekstr\"om}, \citenamefont {Forss\'en}, \citenamefont {Hagen},
  \citenamefont {Jansen},\ and\ \citenamefont {Papenbrock}}]{Jiang:2020}%
  \BibitemOpen
  \bibfield  {author} {\bibinfo {author} {\bibfnamefont {W.~G.}\ \bibnamefont
  {Jiang}}, \bibinfo {author} {\bibfnamefont {A.}~\bibnamefont {Ekstr\"om}},
  \bibinfo {author} {\bibfnamefont {C.}~\bibnamefont {Forss\'en}}, \bibinfo
  {author} {\bibfnamefont {G.}~\bibnamefont {Hagen}}, \bibinfo {author}
  {\bibfnamefont {G.~R.}\ \bibnamefont {Jansen}}, \ and\ \bibinfo {author}
  {\bibfnamefont {T.}~\bibnamefont {Papenbrock}},\ }\href {\doibase
  10.1103/PhysRevC.102.054301} {\bibfield  {journal} {\bibinfo  {journal}
  {Phys. Rev. C}\ }\textbf {\bibinfo {volume} {102}},\ \bibinfo {pages}
  {054301} (\bibinfo {year} {2020})}\BibitemShut {NoStop}%
\bibitem [{\citenamefont {Zhang}\ \emph {et~al.}(2025)\citenamefont {Zhang},
  \citenamefont {Wang}, \citenamefont {Ding},\ and\ \citenamefont
  {Yao}}]{Zhang:2024nqr}%
  \BibitemOpen
  \bibfield  {author} {\bibinfo {author} {\bibfnamefont {X.}~\bibnamefont
  {Zhang}}, \bibinfo {author} {\bibfnamefont {C.~C.}\ \bibnamefont {Wang}},
  \bibinfo {author} {\bibfnamefont {C.~R.}\ \bibnamefont {Ding}}, \ and\
  \bibinfo {author} {\bibfnamefont {J.~M.}\ \bibnamefont {Yao}},\ }\href
  {\doibase 10.1103/4cnl-5dnm} {\bibfield  {journal} {\bibinfo  {journal}
  {Phys. Rev. C}\ }\textbf {\bibinfo {volume} {112}},\ \bibinfo {pages}
  {L021302} (\bibinfo {year} {2025})},\ \Eprint
  {http://arxiv.org/abs/2408.00691} {arXiv:2408.00691 [nucl-th]} \BibitemShut
  {NoStop}%
\bibitem [{\citenamefont {Zhou}\ \emph {et~al.}(2026)\citenamefont {Zhou},
  \citenamefont {Ding}, \citenamefont {Luo}, \citenamefont {Yao},\ and\
  \citenamefont {Hergert}}]{Data}%
  \BibitemOpen
  \bibfield  {author} {\bibinfo {author} {\bibfnamefont {E.~F.}\ \bibnamefont
  {Zhou}}, \bibinfo {author} {\bibfnamefont {C.~R.}\ \bibnamefont {Ding}},
  \bibinfo {author} {\bibfnamefont {Q.~Y.}\ \bibnamefont {Luo}}, \bibinfo
  {author} {\bibfnamefont {J.~M.}\ \bibnamefont {Yao}}, \ and\ \bibinfo
  {author} {\bibfnamefont {H.}~\bibnamefont {Hergert}},\ }\href {\doibase
  10.5281/zenodo.22031953} {\bibfield  {journal} {\bibinfo  {journal} {Zenodo}\
  } (\bibinfo {year} {2026}),\ 10.5281/zenodo.22031953}\BibitemShut {NoStop}%
\bibitem [{\citenamefont {Parzuchowski}\ \emph
  {et~al.}(2017{\natexlab{b}})\citenamefont {Parzuchowski}, \citenamefont
  {Stroberg}, \citenamefont {Navr\'atil}, \citenamefont {Hergert},\ and\
  \citenamefont {Bogner}}]{Parzuchowski:2017}%
  \BibitemOpen
  \bibfield  {author} {\bibinfo {author} {\bibfnamefont {N.~M.}\ \bibnamefont
  {Parzuchowski}}, \bibinfo {author} {\bibfnamefont {S.~R.}\ \bibnamefont
  {Stroberg}}, \bibinfo {author} {\bibfnamefont {P.}~\bibnamefont
  {Navr\'atil}}, \bibinfo {author} {\bibfnamefont {H.}~\bibnamefont {Hergert}},
  \ and\ \bibinfo {author} {\bibfnamefont {S.~K.}\ \bibnamefont {Bogner}},\
  }\href {\doibase 10.1103/PhysRevC.96.034324} {\bibfield  {journal} {\bibinfo
  {journal} {Phys. Rev. C}\ }\textbf {\bibinfo {volume} {96}},\ \bibinfo
  {pages} {034324} (\bibinfo {year} {2017}{\natexlab{b}})}\BibitemShut
  {NoStop}%
\end{thebibliography}

%

\end{CJK*}
\end{document}